\documentclass[useAMS,usenatbib,twocolumn,fleqn]{mn2e}
\usepackage{natbib, graphics, epsfig}

\usepackage{times}
\usepackage{amsmath}
\usepackage{amssymb}
\usepackage{textcomp}
\usepackage{verbatim}
\newcommand{\sm}{\small}

\newcommand{\msun}{{\,\rm M_\odot}}
\newcommand{\kms}{\,{\rm km}\,{\rm s}^{-1}}

\newcommand{\kpc}{\,{\rm kpc}}

\title[Direct detection of self-interacting dark matter]
      {Direct detection of self-interacting dark matter}
      \author[M. Vogelsberger \& J. Zavala] {\parbox{18.5cm}{
          Mark Vogelsberger$^{1}$\thanks{Hubble Fellow, mvogelsb@cfa.harvard.edu} and
          Jesus Zavala$^{2,3}$\thanks{CITA National Fellow},
        }\vspace{0.3cm}\\
        $^1$Harvard-Smithsonian Center for Astrophysics, 60 Garden Street, Cambridge, MA 02138, USA\\
        $^2$Department of Physics and Astronomy, University of Waterloo, Waterloo, Ontario, N2L 3G1, Canada\\
        $^3$Perimeter Institute for Theoretical Physics, 31 Caroline St. N., Waterloo, ON, N2L 2Y5, Canada}

\begin{document}
\date{Accepted ???. Received ???; in original form ???}

\pagerange{\pageref{firstpage}--\pageref{lastpage}} \pubyear{2012}

\maketitle

\label{firstpage}

\begin{abstract}
Self-interacting dark matter offers an interesting alternative to collisionless
dark matter because of its ability to preserve the large-scale success of the cold dark
matter model, while seemingly solving its challenges on small scales. We
present here the first study of the expected dark matter detection signal in a
fully cosmological context taking into account different self-scattering models
for dark matter. We demonstrate that models with constant and velocity
dependent cross sections, which are consistent with observational constraints,
lead to distinct signatures in the velocity distribution, because
non-thermalised features found in the cold dark matter distribution are
thermalised through particle scattering. Depending on the model,
self-interaction can lead to a $10\%$ reduction of the recoil rates at high
energies, corresponding to a minimum speed that can cause recoil larger than $300\kms$, 
compared to the cold dark matter case. At lower energies these
differences are smaller than $5\%$ for all models. The amplitude of the annual
modulation signal can increase by up to $25\%$, and the day of maximum
amplitude can shift by about two weeks with respect to the cold dark matter
expectation. Furthermore, the exact day of phase reversal of the modulation
signal can also differ by about a week between the different models. In general,
models with velocity dependent cross sections peaking at the typical velocities
of dwarf galaxies lead only to minor changes in the detection signals, whereas
allowed constant cross section models lead to significant changes. We conclude
that different self-interacting dark matter scenarios might be distinguished
from each other through the details of direct detection signals.  Furthermore,
detailed constraints on the intrinsic properties of dark matter based on null
detections, should take into account the possibility of self-scattering and the
resulting effects on the detector signal.  
\end{abstract}


\section{Introduction}

The substantial evidence for the existence of dark matter (DM) accumulated over
the last decades, has been purely gravitational in nature. 
Unambiguous proof of its existence as a new particle  beyond the Standard Model
(e.g. a supersymmetric particle like the neutralino) relies on a detection
through a non-gravitational signature. This is currently one of the most significant
challenges for modern astrophysics and particle physics, and
it is the reason why many experiments are currently looking for such a signal,
either directly trying to measure the recoil of nuclei after a collision with
the DM particle, or indirectly by measuring the byproducts of its
self-annihilation \citep[see][for a review]{Bertone2005}.

The interaction rate of DM particles with experiment targets at laboratories on
Earth not only depends on the mass and interaction cross section of DM but also
on its local phase-space distribution at the scale of the apparatus. Any hope
of extracting information about the nature of DM relies therefore on detailed
knowledge of the latter.  Numerical simulations that follow the gravitational
collapse of the primordial DM density perturbations have been very successful
in reproducing the large-scale structure of the Universe
\citep[e.g.][]{Springel2005_Nat} and provide detailed predictions on the
internal structure of DM haloes \citep[e.g.][]{Springel2008, Diemand2008,
Stadel2009, Wu2012}. These simulations are the most reliable approach to study
structure formation \citep[see][for a recent review]{Kuhlen2012} and its
importance for direct detection experiments has been pointed out recently, by
finding departures from the traditional assumptions: a smooth density
distribution and a Maxwellian velocity distribution. It was found that the
phase-space distribution of relaxed DM haloes is not featureless but contains
imprints of its formation history in the energy distribution
\citep[e.g.][]{Vogelsberger2009,Kuhlen2010, Kuhlen2012b}. Over the last years it
also became possible to study the fine-grained phase-space structure of DM by
extending classical N-body methods \citep[][]{Vogelsberger2008, White2009,
Vogelsberger2009caustic, Vogelsberger2011, Abel2011, Neyrinck2012}.

An important point to be made about these simulations is that most of them,
certainly the ones used for estimating direct detection rates, assumed a Cold
DM (CDM) cosmology. In this model,  DM particles are thought to be cold (i.e.,
the thermal motions of DM particles were essentially negligible at the time of
matter-radiation equality) and collisionless (i.e., negligible
self-interaction).  The possibility remains, however, that one or both of these
hypothesis are not true on all scales. In fact, the evidence that supports them
clearly allows for significant DM self-scattering \citep[e.g.][]{Peter2012} and
for the possibility of DM being warm \citep[e.g.][]{Boyarsky2009}. Indeed, the
alternative self-interacting DM (SIDM) and Warm DM (WDM) models are attractive
possibilities to solve some of the current challenges to CDM at the scale of
dwarf galaxies without spoiling its successes at larger scales
\citep[e.g.][]{Colin2000,Bode2001,Zavala2009,SpergelSteinhardt2000,LoebWeiner2011,Vogelsberger2012,Rocha2012}.
The SIDM possibility is particularly exciting since it is not in conflict with
current astrophysical constraints, such as the shape of DM haloes and the
survivability of satellite haloes \citep{LoebWeiner2011,Peter2012}, and is able
to produce central density cores that are seemingly consistent with
observations of dwarf galaxies at different scales (see \cite{Rocha2012} and
\cite{Vogelsberger2012}, hereafter VZL). The scattering cross section is assumed
to be either constant or velocity dependent. In the constant cross section case,
tight constraints have been derived using halo shapes based on X-ray and lensing
data from ellipticals and galaxy clusters: the momentum-transfer
weighted cross section $\sigma_T/m_\chi$ is likely $<1\,{\rm cm}^2\,{\rm g}^{-1}$ by current estimates
\citep[e.g.][]{Peter2012}. Until very recently, it was not clear if
it was possible to solve all outstanding small-scale CDM problems
in a constant cross section scenario with a value lower than currently allowed. 
For example, \cite{Yoshida2000} argued that for $\sigma_T/m_\chi \sim 0.1\,{\rm cm}^2\,{\rm g}^{-1}$, the 
collision rate would be too low to create sizable cores in dwarf galaxies while \cite{Rocha2012} 
suggests that values in this ballpark would be sufficient, but both of these contradictory claims are
based on simulations that do not resolve the scales in dispute.
In \citet{Zavala2012} we used the simulations presented in this paper to clarify this issue showing that a constant 
scattering cross section is a viable explanation for the low dark matter densities inferred in the Milky Way (MW) dwarf spheroidals
only within the very narrow range: $0.6\,{\rm cm}^2\,{\rm g}^{-1}\leq\sigma_T/m_\chi\leq1.0\,{\rm cm}^2\,{\rm g}^{-1}$.
On the other hand, a model with a velocity dependent cross section can easily circumvent cluster
constraints, but preserve the SIDM effects on small scales by adopting an
appropriate velocity dependence.

The magnitude of the self-interacting cross section for ${\rm GeV}$-scale DM particles needed to produce a significant
impact at the scale of dwarf galaxies is of the order of the strong interaction
cross section for neutron-neutron or neutron-proton scattering. This does not at all 
imply that DM should interact with nuclei through the strong nuclear force.
If that were the case, DM particles would not reach ground-based detectors
that are looking for Weakly Interacting Massive Particles (WIMPs), 
since they would scatter in the atmosphere and reach the ground with energies
below the threshold of current experiments. Also, any kind of shielding would 
further prevent detection of strongly interacting DM. Significant constraints on this type
of DM are therefore obtained with space-based instruments \citep{Erickcek2007}.
Instead, an example of the particle physics model we imagine here is that presented
in \citet{Hooper2012}. In this type of hidden-sector DM, a new gauge boson
that has a small kinetic mixing with the photon can result in elastic scattering
of DM with protons at the level of what is hypothetically seen by the DAMA
experiment \citep{Bernabei2012}. The same gauge boson can act between the DM particles 
to enhance self-scattering \citep[e.g.][]{Buckley2010,LoebWeiner2011,Aarssen2012}. Another example
of a similar model is given by \citet{Fornengo2011}, while another distinct
possibility is that of the Atomic Dark Matter model \citep[see][]{Cyr-Racine2012}.

If DM is indeed self-interacting, then direct detection experiments should assume 
the appropriate DM phase-space distribution and not
the one based on CDM simulations. The purpose of this paper is to study the
SIDM local density and velocity distribution using high resolution cosmological
simulations and to quantify the difference between currently allowed SIDM
models and CDM. In particular, we estimate the impact of self-scattering on the
expected interaction rates with nuclei and its signature in direct detection
experiments.

The paper is organised as follows. In Section 2 we present the SIDM models we
consider and details of our simulation suite.  Section 3 presents general
results on the coarse-grained phase-space structure in the inner halo. In
Section 4 we focus on the velocity structure of the different models and
contrast them to the CDM case. Here we also show that our results are valid for
generic MW like haloes, i.e. they are not sensitive to cosmic
variance.  The differences in the velocity distribution directly impact the
direct detection signal as we demonstrate in Section 5.  Finally, we give our
conclusions in Section 6. 

\section{SIDM models and simulations}

\begin{table}
\begin{center}
\begin{tabular}{cccc}
\hline
Name         & $\sigma_T^{\rm max}/m_\chi [{\rm cm}^2\,{\rm g}^{-1}]$     & $v_{\rm max} [{\rm km}\,{\rm s}^{-1}]$                     \\
\hline
\hline
CDM          & /                               & /                                 \\  \hline
SIDM10       & $10$                            & /                                 \\  \hline
SIDM1        & $1$                             & /                                 \\  \hline
SIDM0.1      & $0.1$                           & /                                 \\  \hline
vdSIDMa      & $3.5$                           & $30$                              \\  \hline
vdSIDMb      & $35$                            & $10$                              \\  \hline
\hline
\end{tabular}
\end{center}
\caption{DM models considered in this paper. CDM is the standard collisionless model
without any self-interaction. SIDM10 is a reference model with a constant cross
section much larger than allowed by current observational constraints. vdSIDMa 
and vdSIDMb have a velocity-dependent cross section motivated by the particle
physics model presented in \protect\cite{LoebWeiner2011}, are allowed by all
astrophysical constraints, and solve the ``too big to fail'' problem
\protect\citep[see][]{Boylan2011a} as demonstrated in VZL. SIDM1 and SIDM0.1 were
not discussed in VZL, and are constant cross section scenarios with $10/100$
times smaller transfer cross section than SIDM10. They were considered recently
by \protect\cite{Rocha2012} and \protect\cite{Peter2012}.}
\label{table:ref_points} 
\end{table}

\begin{table}
\begin{center}
\begin{tabular}{lllll}
\hline
Name & $m_{\rm p} [\mathrm{M}_\odot]$  & $\epsilon [\mathrm{pc}]$ & $M_{\rm 200} [\mathrm{M}_\odot]$ & $r_{\rm 200} [\mathrm{kpc}]$ \\
\hline
Aq-A-3   &  $4.911\times 10^4$  & 120.5  &   $1.836\times 10^{12}$ &  245.64 \\
Aq-A-4   &  $3.929\times 10^5$  & 342.5  &   $1.838\times 10^{12}$ &  245.70 \\
\hline
Aq-B-4   &  $2.242\times 10^5$  & 342.5  &   $8.345\times 10^{11}$ &  188.85 \\
\hline
Aq-C-4   &  $3.213\times 10^5$  & 342.5  &   $1.793\times 10^{12}$ &  243.68 \\
\hline
Aq-D-4   &  $2.677\times 10^5$  & 342.5  &   $1.791\times 10^{12}$ &  243.60 \\
\hline
Aq-E-4   &  $2.604\times 10^5$  & 342.5 &    $1.208\times 10^{12}$ &  213.63 \\
\hline
Aq-F-4   &  $1.831\times 10^5$  & 342.5 &    $1.100\times 10^{12}$ &  206.60 \\
\hline
\end{tabular}
\end{center}
\caption{Parameters of the Aquarius simulations. $m_{\rm p}$ is the DM particle
mass, $\epsilon$ is the  Plummer equivalent gravitational softening length.
$M_{\rm 200}$ is the virial mass of the halo, defined as the mass enclosed in a
sphere with mean density 200 times the critical value and $r_{\rm 200}$ gives
the corresponding virial radius (both for the CDM case). We mainly focus on Aq-A-3 and study how the
different DM models affect its phase-space structure. The level-4
haloes are only used to estimate how cosmic variance affects the results.}
\label{table:sims} \end{table}

In VZL we followed the formation and evolution of a single MW-size halo
(simulated at different resolutions) in the context of a few benchmark points
within elastic velocity-dependent SIDM models. In these models, linear momentum
and energy are conserved during a scatter event, but the direction of the
relative velocity vector is randomly redistributed resulting in isotropic
scattering. We refer the reader to VZL for details and testing of our Monte
Carlo based self-scattering algorithm, and its implementation within the {\sm
GADGET-3} code for cosmological simulations (last described in
\cite{Springel2005}). For this paper, we extend the simulation suite
presented in VZL by adding two more SIDM benchmark models  with a constant
cross section (see Table~\ref{table:ref_points}) and more individual haloes to
study the effect of cosmic variance allowing us to have a range of MW-size
haloes with different formation histories, subhalo abundance, environments,
etc. The initial conditions for our simulations are taken from the Aquarius
Project \citep{Springel2008} with the following cosmological parameters:
$\Omega_m=0.25$, $\Omega_{\Lambda}=0.75$, $h=0.73$, $\sigma_8=0.9$ and
$n_{s}=1$; where $\Omega_m$ and $\Omega_{\Lambda}$ are the contribution from
matter and cosmological constant to the mass/energy density of the Universe,
respectively, $h$ is the dimensionless Hubble constant parameter at redshift
zero, $n_s$ is the spectral index of the primordial power spectrum, and
$\sigma_8$ is the rms amplitude of linear mass fluctuations in $8~h^{-1}\,{\rm
Mpc}$ spheres at redshift zero. 

\begin{figure}
\centering
\includegraphics[width=0.475\textwidth]{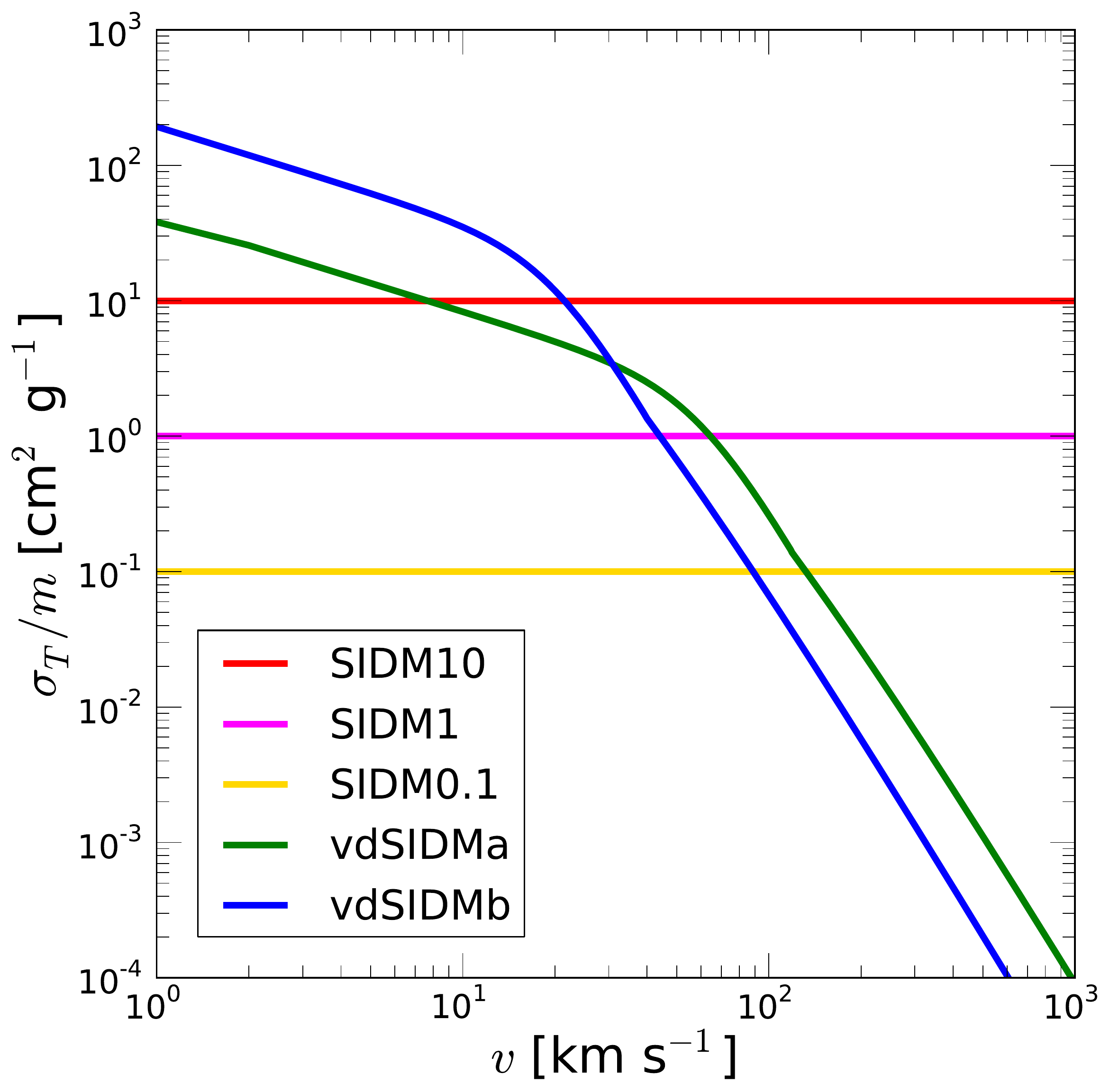}
\caption{Dependence of the transfer cross section on the relative velocity for
the different SIDM models (see Table~\ref{table:ref_points}). SIDM10, SIDM1,
SIDM0.1 are models with constant cross sections differing by factors of ten
ranging from $\sigma_T/m_\chi=0.1\,{\rm cm}^2\,{\rm g}^{-1}$ (SIDM0.1) to $\sigma_T/m_\chi=10\,{\rm cm}^2\,{\rm
g}^{-1}$ (SIDM10). The vdSIDMa and vdSIDMb models have a velocity-dependent cross
section with slightly different peak values and velocities where the
cross section is maximal. vdSIDMa, vdSIDMb and SIDM0.1 are compatible with
astrophysical constraints. SIDM1 is likely too large to be consistent with
cluster observations, whereas SIDM10 is clearly ruled out and mainly serves here
as a test scenario to demonstrate the various effects. 
The vdSIDM models fall off as $v^{-4}$ for higher relative velocities which allows
for large cross sections towards lower velocities.  
Cluster constraints can then be easily circumvented
for such models, whereas they impose a clear upper limit for the constant cross
section models.} 
\label{fig:cross_section} 
\end{figure}

In Table~\ref{table:ref_points} we list the different SIDM models we consider
in this paper. CDM is the vanilla
collisionless case and SIDM10 only serves as a test case to demonstrate the effects of a
large constant constant cross section (already ruled out by various
astrophysical constraints). vdSIDMa and vdSIDMb on the other hand are
velocity-dependent SIDM (vdSIDM) cases, that do not violate any astrophysical
constraints and are well-motivated from a particle physics point of view
\citep[e.g.][]{Arkani-Hamed2009,LoebWeiner2011}. They produce
$\mathcal{O}(1\kpc)$ density cores in the MW subhaloes as discussed in VZL.  We
extended our sample here by adding SIDM1 which has a constant cross section ten
times smaller than SIDM10, bringing it closer to current constraints although it
is still above them as recently pointed out by \cite{Peter2012}.  SIDM0.1 has an
even smaller constant cross section of $\sigma_T/m_\chi=0.1\,{\rm cm}^2\,{\rm g}^{-1}$, which
is consistent with all astrophysical observations including cluster scale
constraints \citep[for further details, see][]{Peter2012}.  In
Fig.~\ref{fig:cross_section} we show the transfer cross sections as a function
of relative velocity for all the benchmark models. The cross section of the
vdSIDM models falls off as $v^{-4}$ towards larger relative velocities. They
can therefore have a quite large cross section at smaller velocities without
violating cluster constraints. For example, at $v=10\kms$ the vdSIDM models
have cross sections even larger than SIDM10. 

In VZL we simulated the elastic models CDM, SIDM10, vdSIDMa and vdSIDMb for the
Aq-A halo at resolution levels 3, 4 and 5, corresponding to an approximate
spatial resolution (given by $2.8\epsilon$, where $\epsilon$ is the Plummer
equivalent gravitational softening length) of about $2.0\kpc$, $1.0\kpc$, and
$0.3\kpc$, respectively. The exact mass resolution depends on the particular
halo, but is of the order of $10^6\msun$, $10^5\msun$, and $10^4\msun$ for
level 5, 4, and 3, respectively. The simulation details are summarised in
Table~\ref{table:sims}, where we also specify the virial properties of the
different haloes.

\begin{figure*}
\centering
\includegraphics[width=0.33\textwidth]{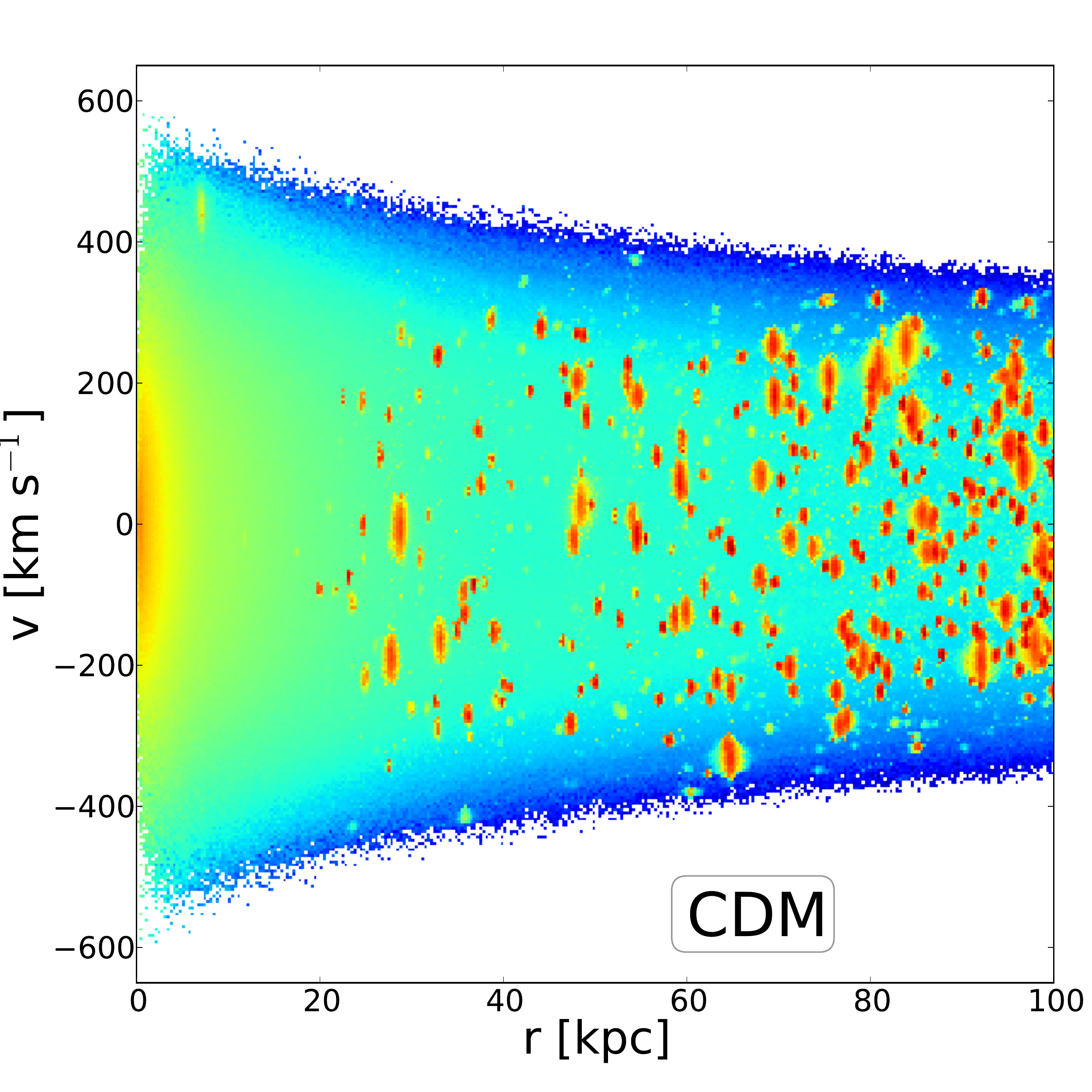}
\includegraphics[width=0.33\textwidth]{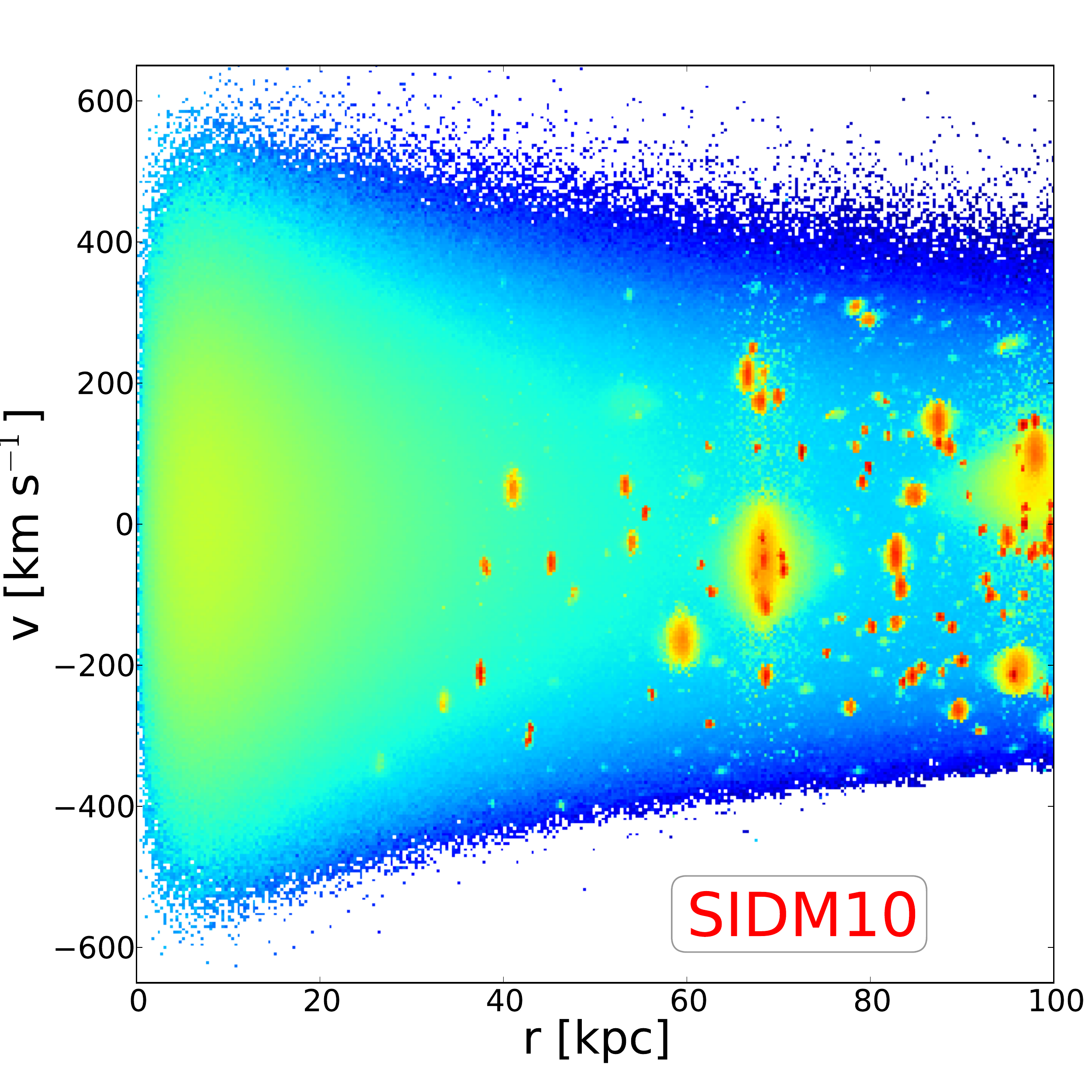}
\includegraphics[width=0.33\textwidth]{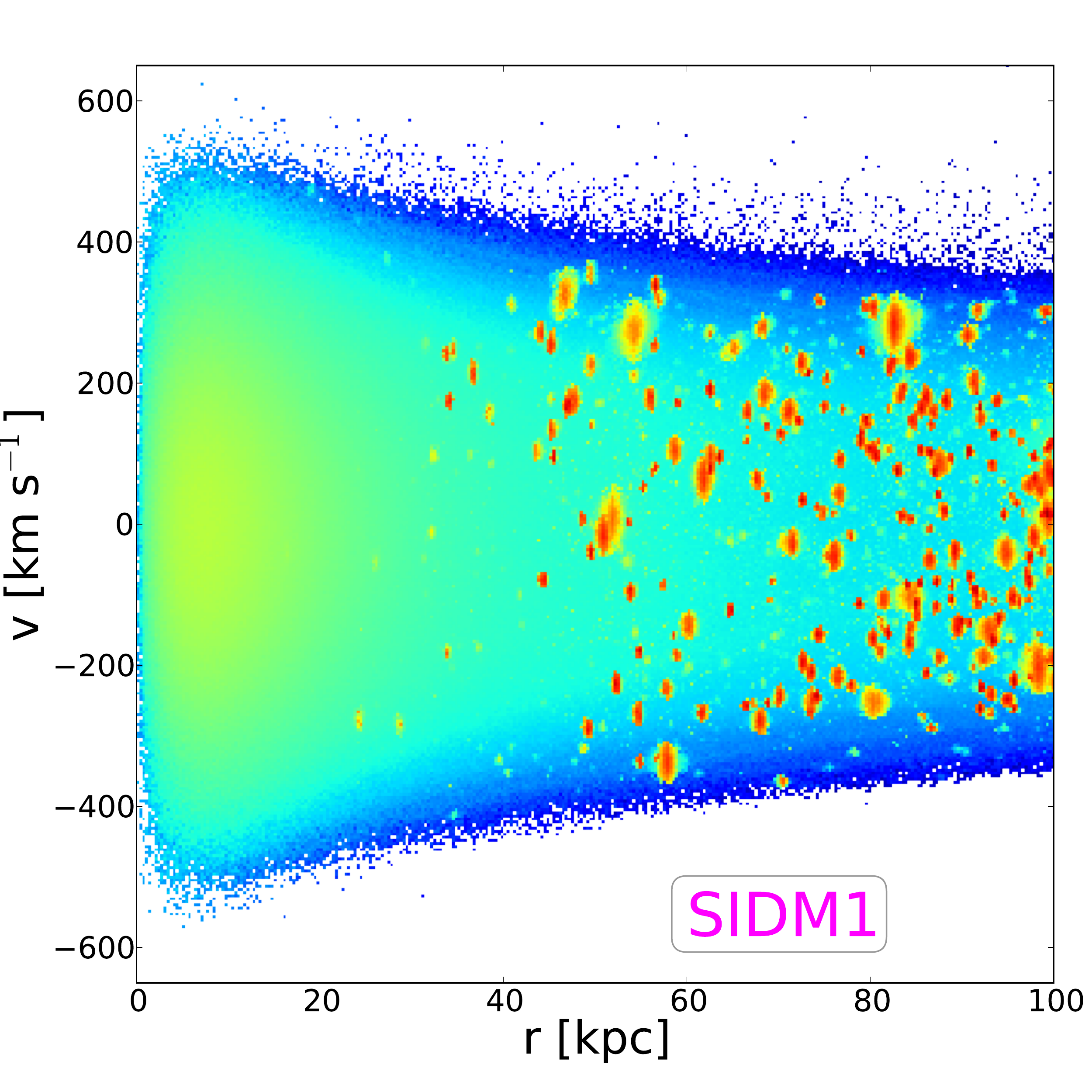}
\includegraphics[width=0.33\textwidth]{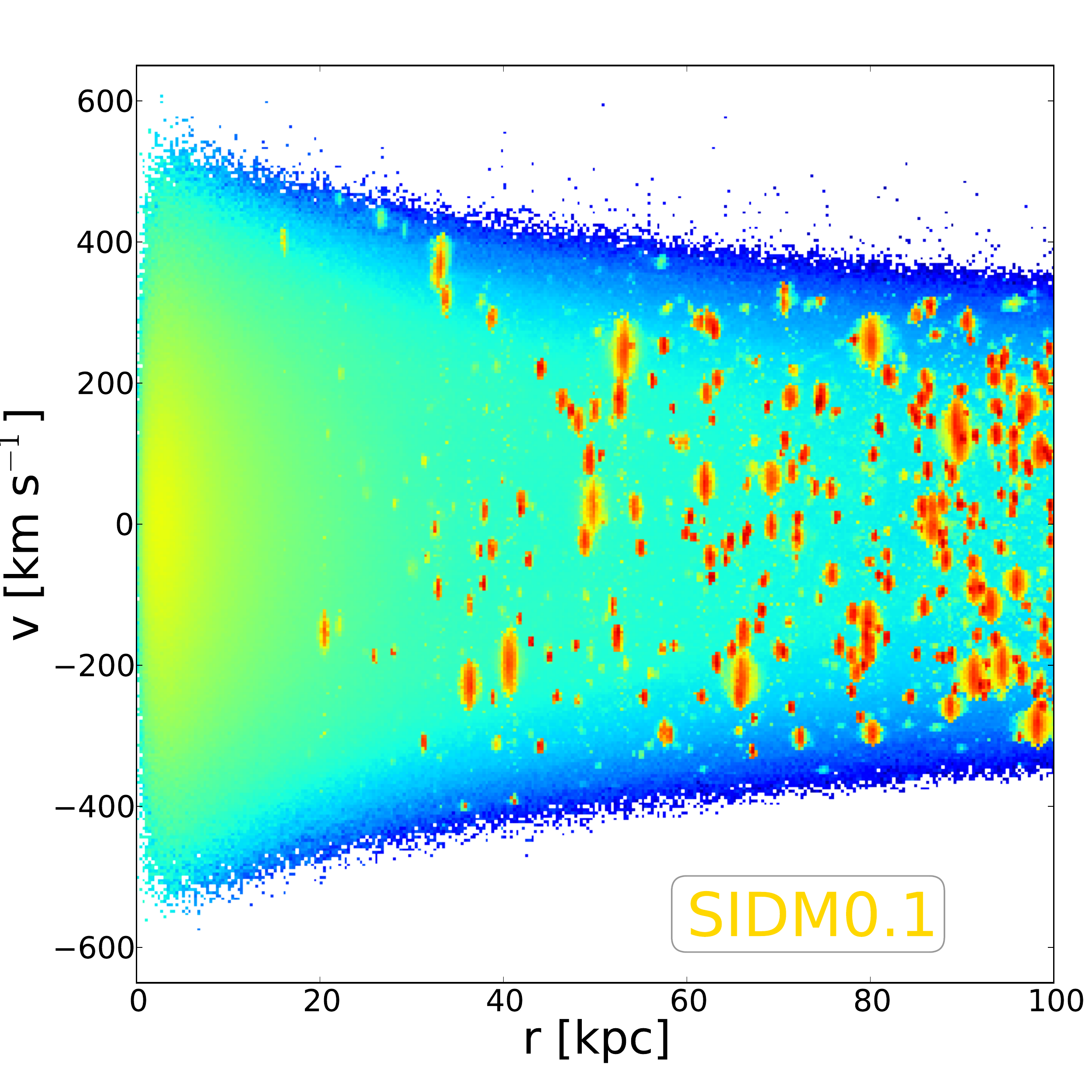}
\includegraphics[width=0.33\textwidth]{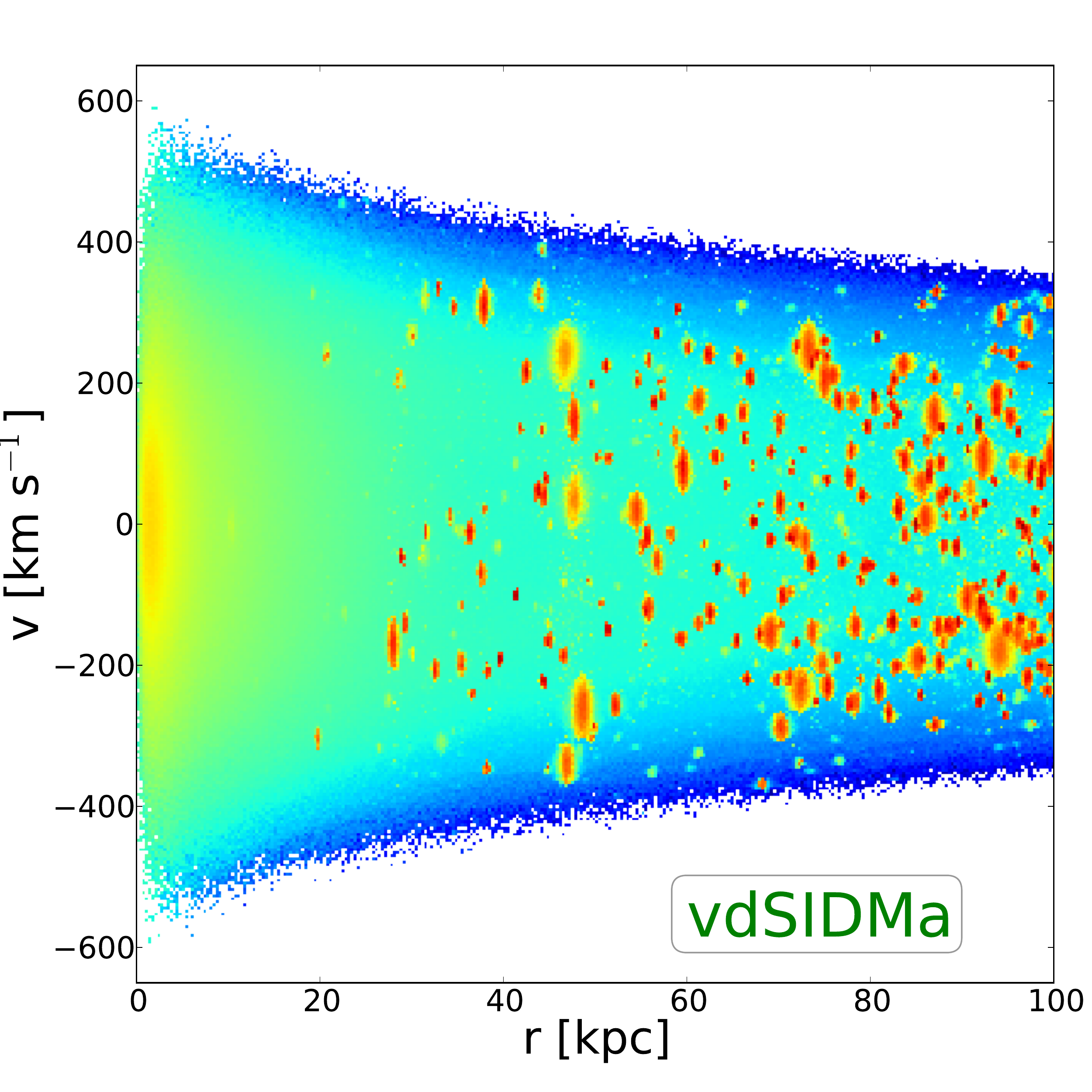}
\includegraphics[width=0.33\textwidth]{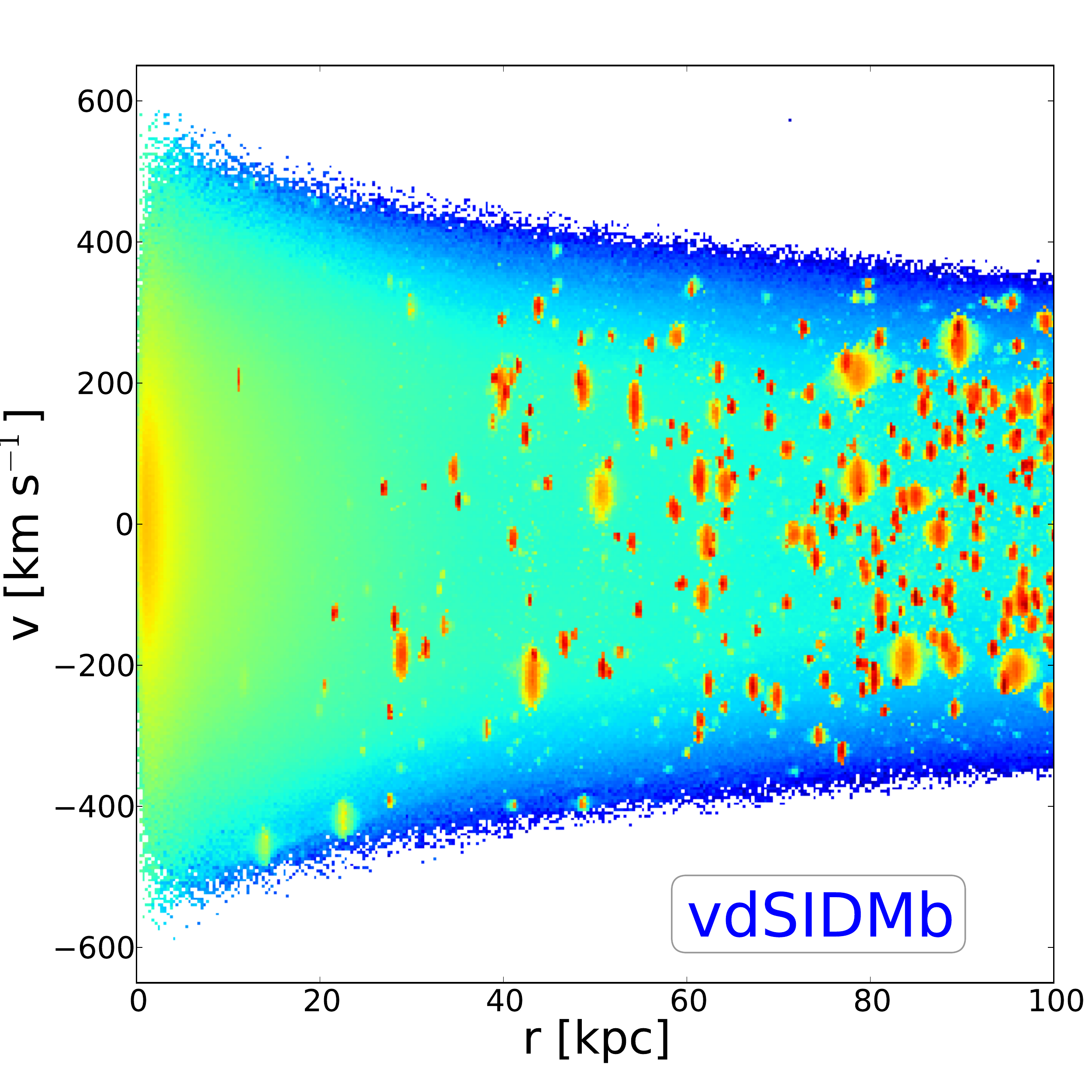}
\caption{Phase-space histograms of all DM models considered in this work. The
contribution of each particle is weighted by its local coarse-grained phase-space density
$\rho/\sigma^3$. SIDM10 leads to a significant suppression of substructure, and
also to a strong reduction of the central main halo phase-space density.
Although SIDM1 has a ten times smaller constant cross section it still shows a
strong reduction of the central main halo phase-space density, but the
abundance of substructure is similar to the CDM case. The former effect is
nearly absent for vdSIDMa and vdSIDMb because the cross section is
velocity-dependent in these cases, which compensates for the rising
configuration space density towards the center. In these cases, it is also
apparent that the phase-space density in substructures is lower than in CDM.
SIDM0.1, with the smallest constant cross section, still looks different from the
vdSIDM models in the center of the main halo even though the substructure
distribution looks very similar. } 
\label{fig:phasespace_density} 
\end{figure*}

We extend the original simulation suite and simulate the remaining SIDM models
(SIDM1, SIDM0.1) at resolution level 3.  We also explore here five other Aquarius
haloes (labelled Aq-B to Aq-F) at level 4 resolution for all the SIDM cases.
The full set of Aquarius haloes covers a wide range of formation histories and
virial masses of MW-type haloes, allowing us to bracket the uncertainties in
the local phase-space distribution due to halo-to-halo scatter, see
\cite{Springel2008} for details. Resolution level 4 is sufficient to
accomplish this.

\section{Phase space structure}

\begin{figure*}
\centering
\vspace{-0.1cm}\hspace{-0.5cm}\includegraphics[width=0.245\textwidth]{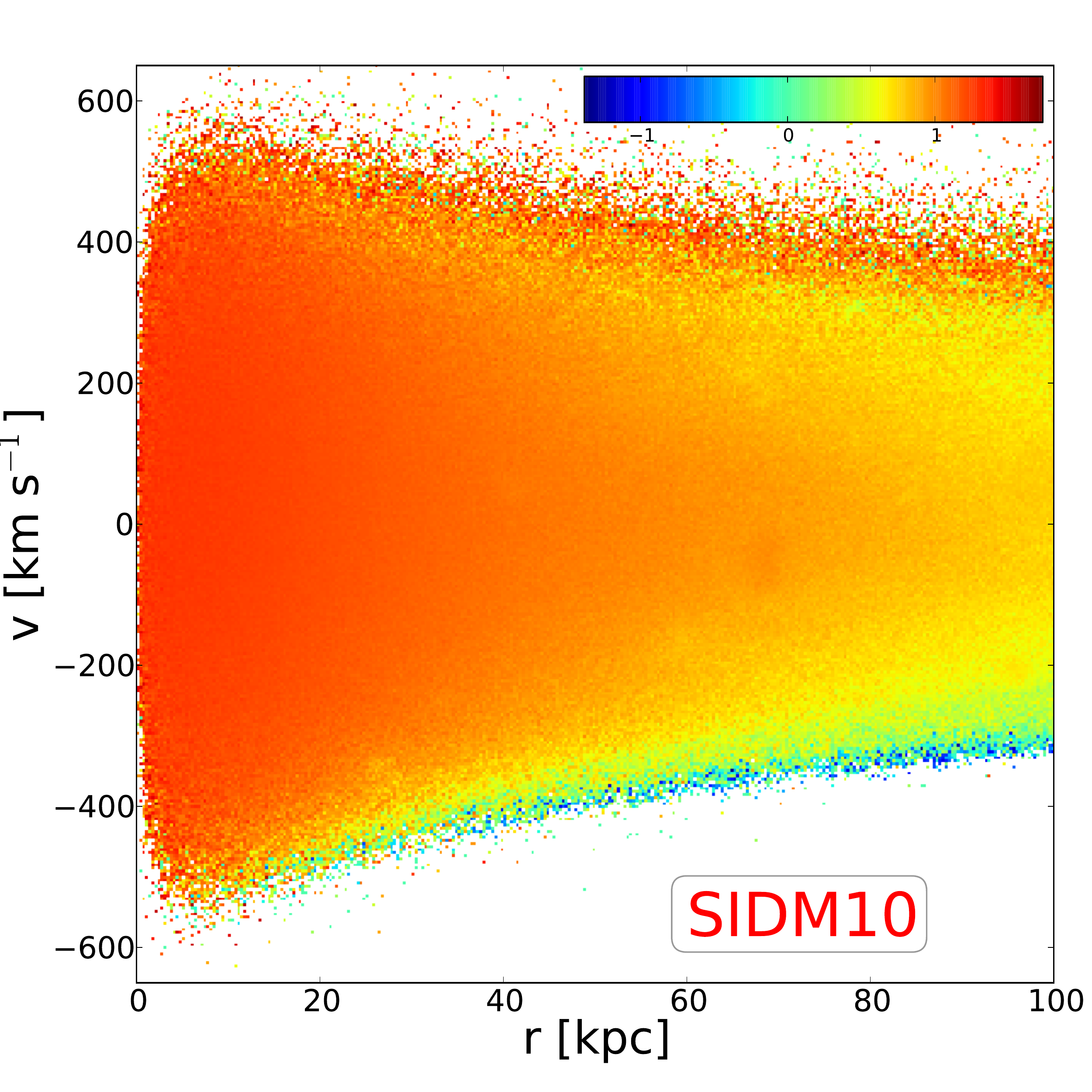}
\hspace{-0.2cm}\includegraphics[width=0.245\textwidth]{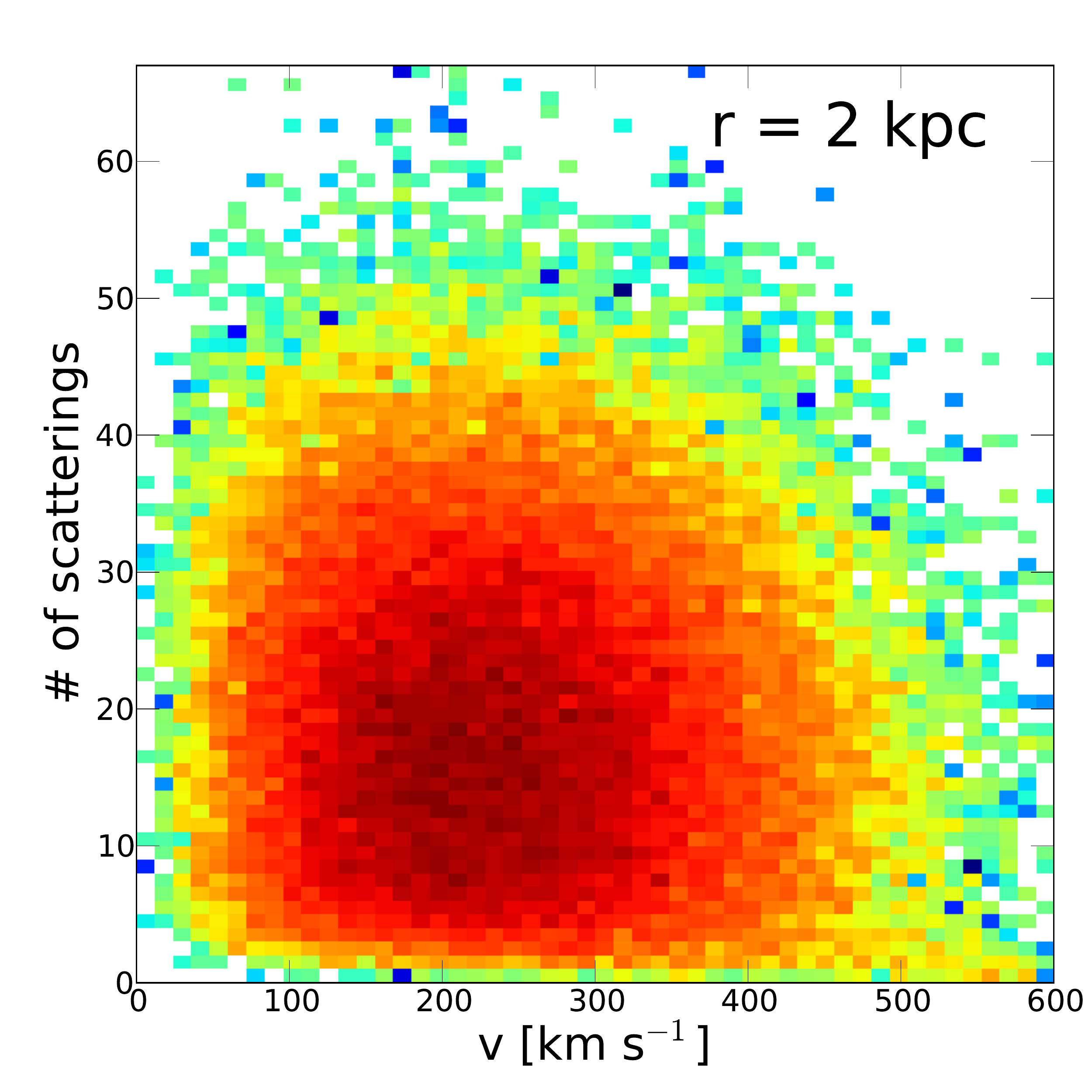}
\hspace{-0.2cm}\includegraphics[width=0.245\textwidth]{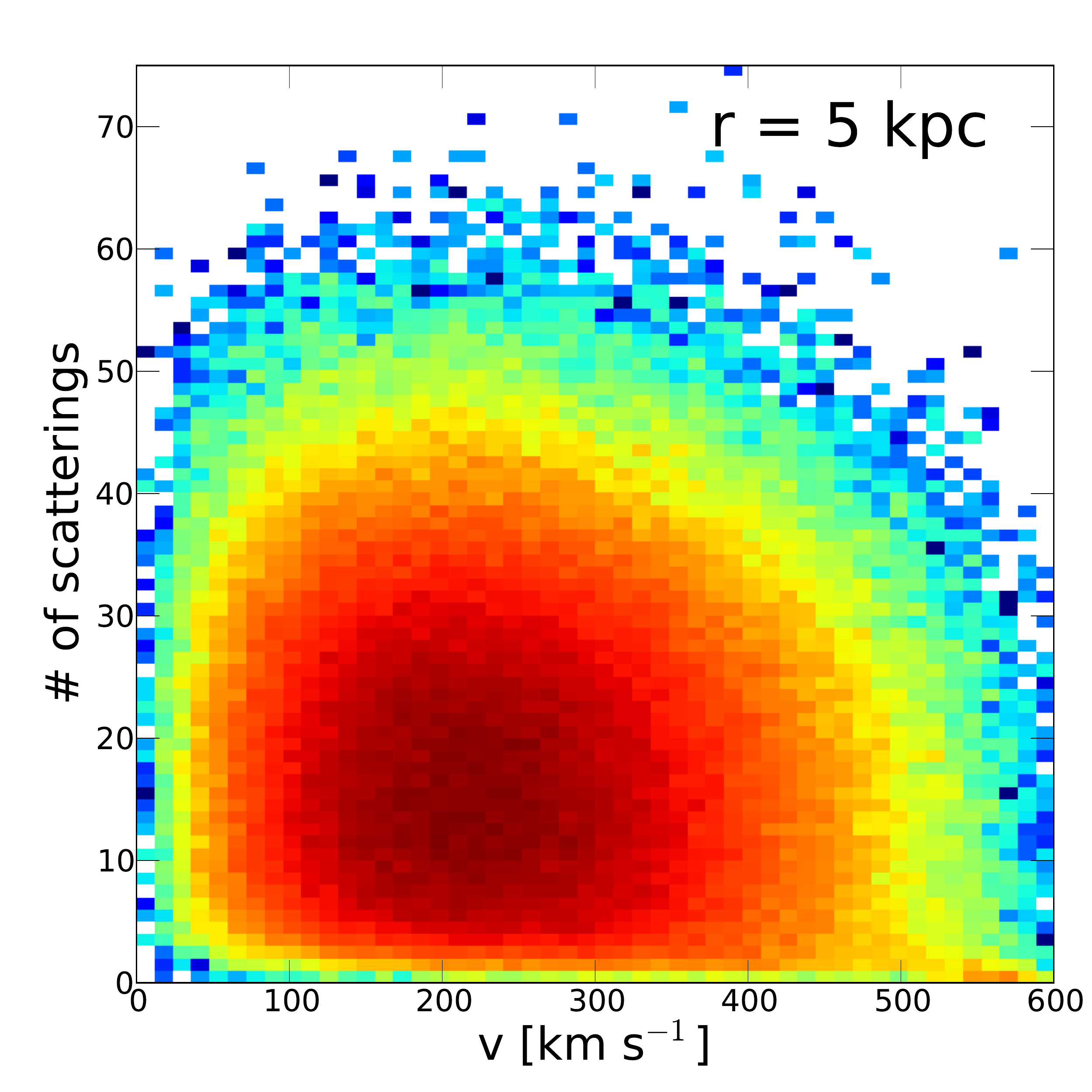}
\hspace{-0.2cm}\includegraphics[width=0.245\textwidth]{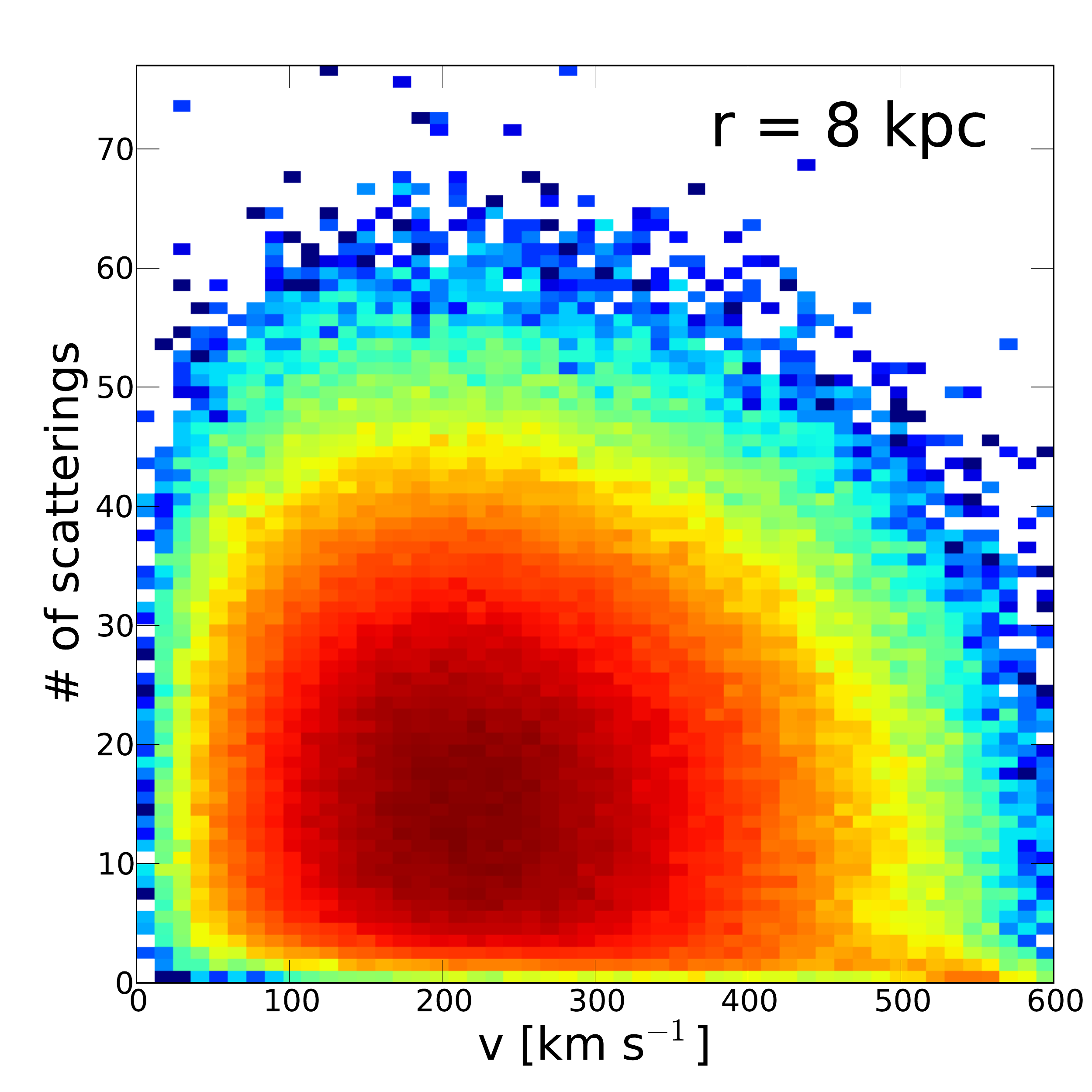}\\
\vspace{-0.1cm}\hspace{-0.5cm}\includegraphics[width=0.245\textwidth]{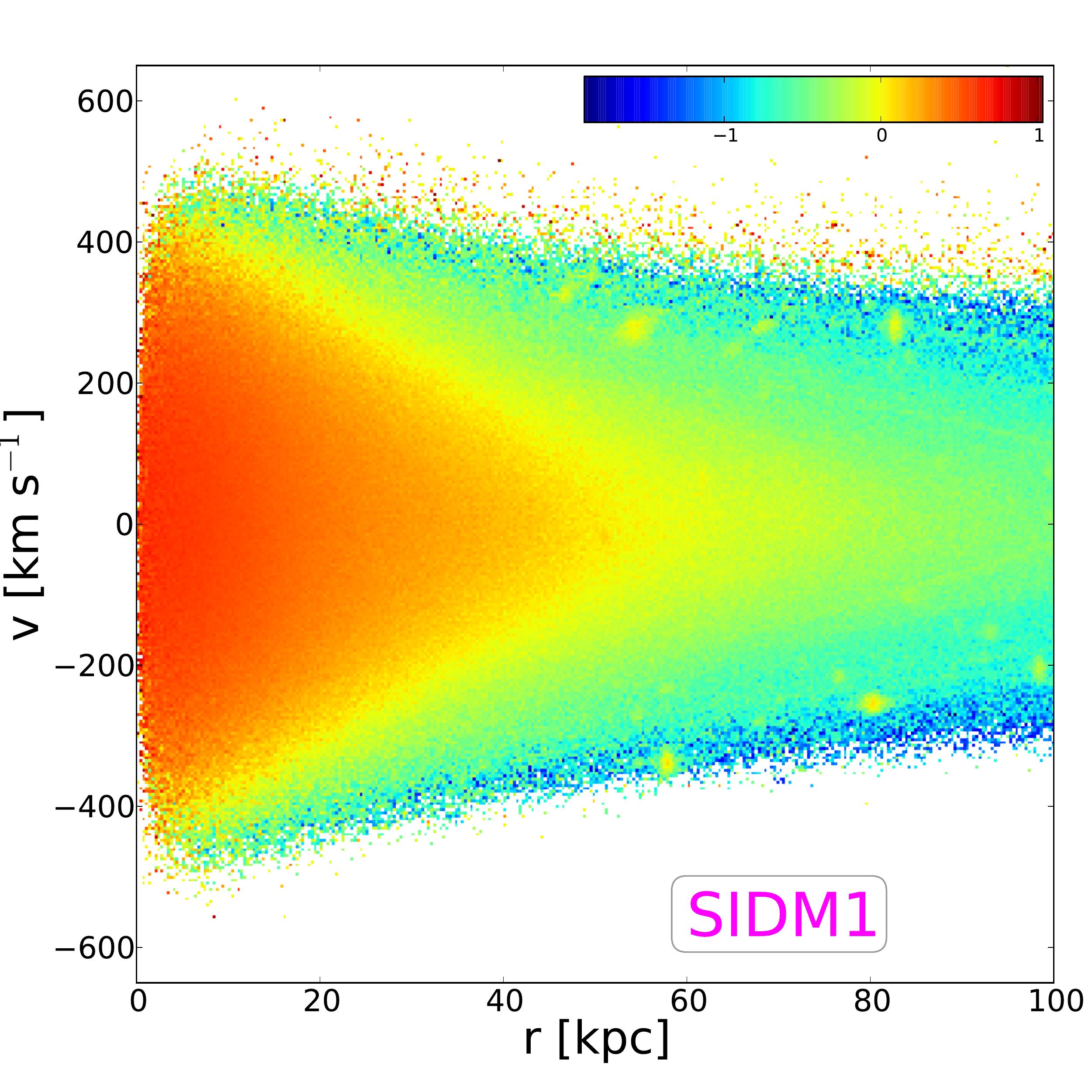}
\hspace{-0.2cm}\includegraphics[width=0.245\textwidth]{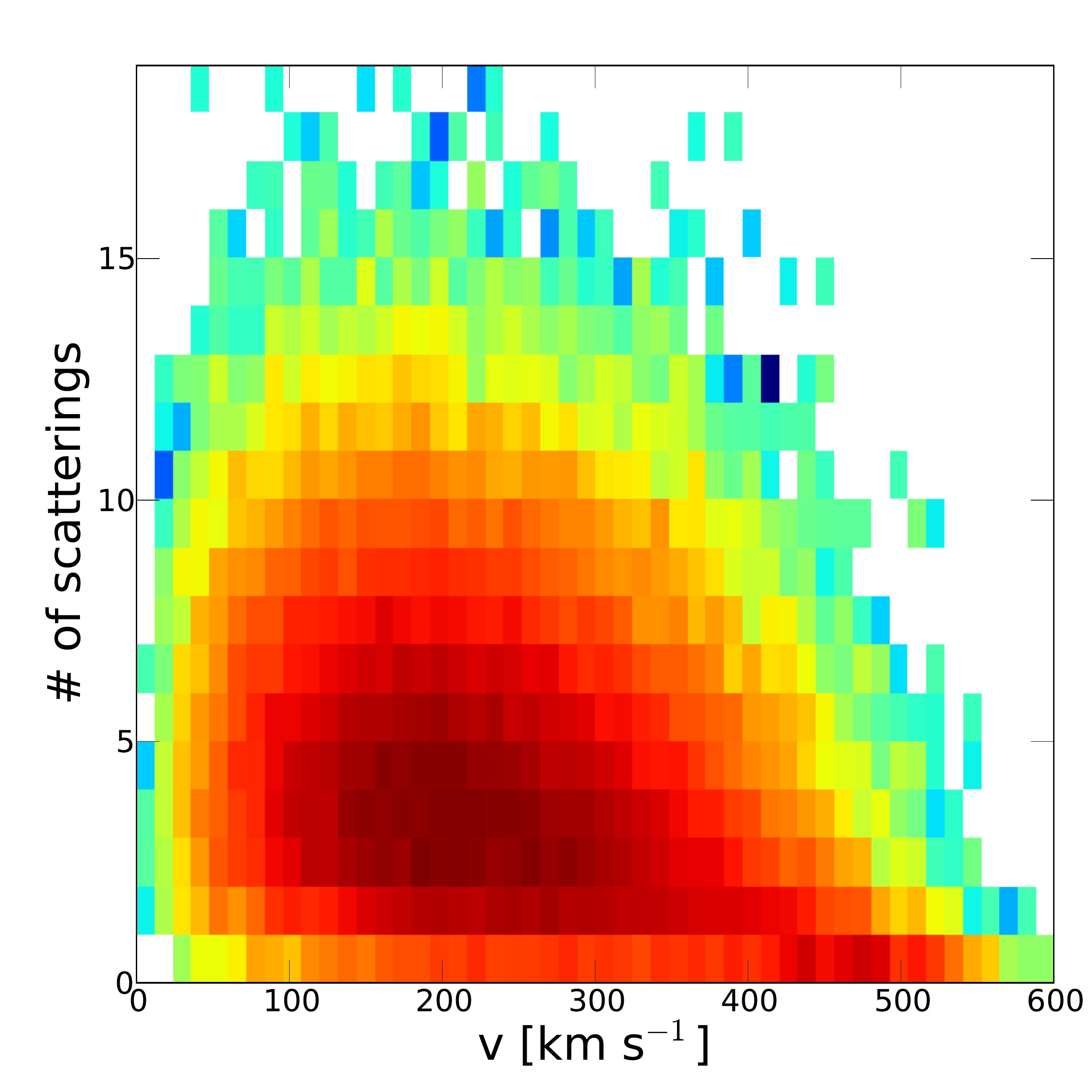}
\hspace{-0.2cm}\includegraphics[width=0.245\textwidth]{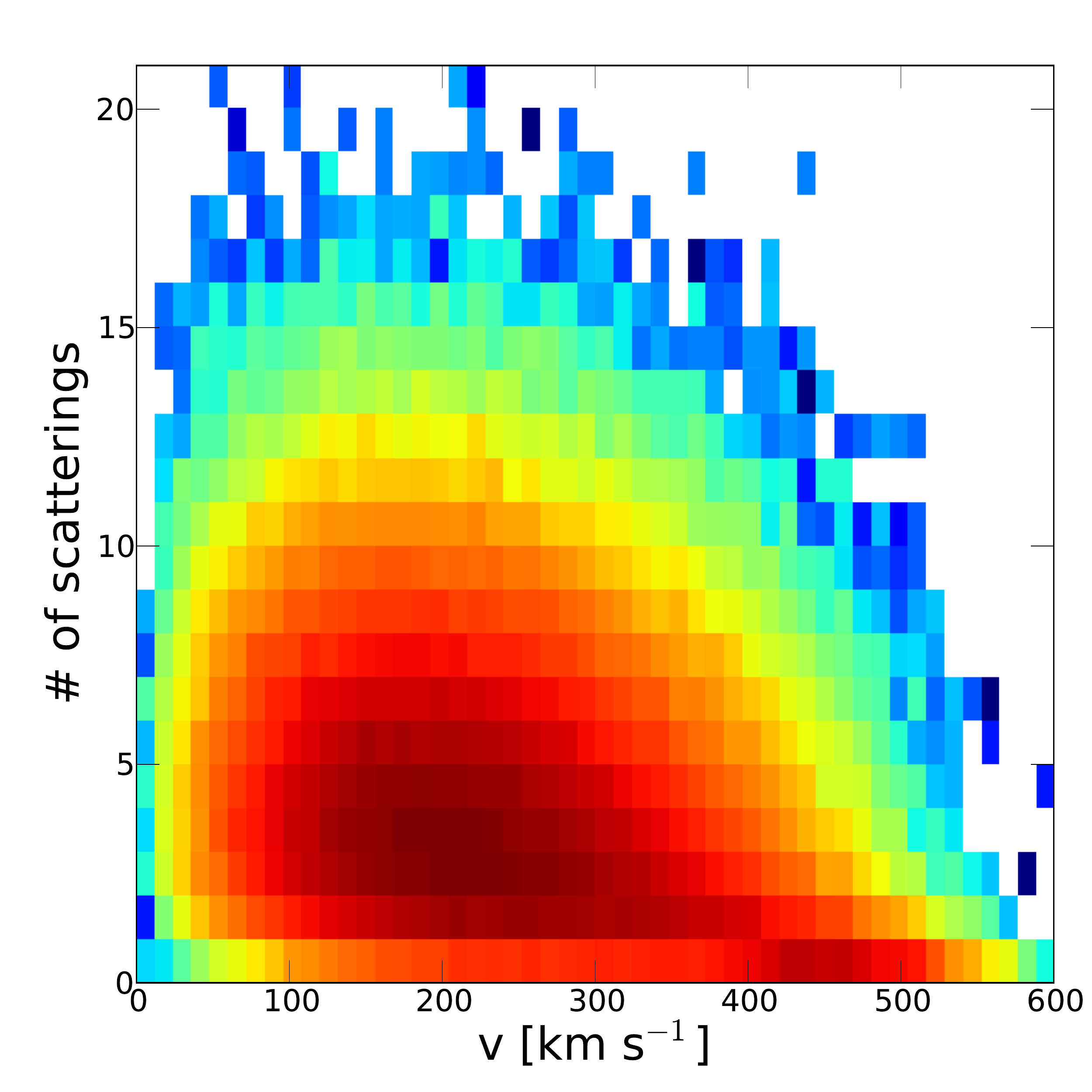}
\hspace{-0.2cm}\includegraphics[width=0.245\textwidth]{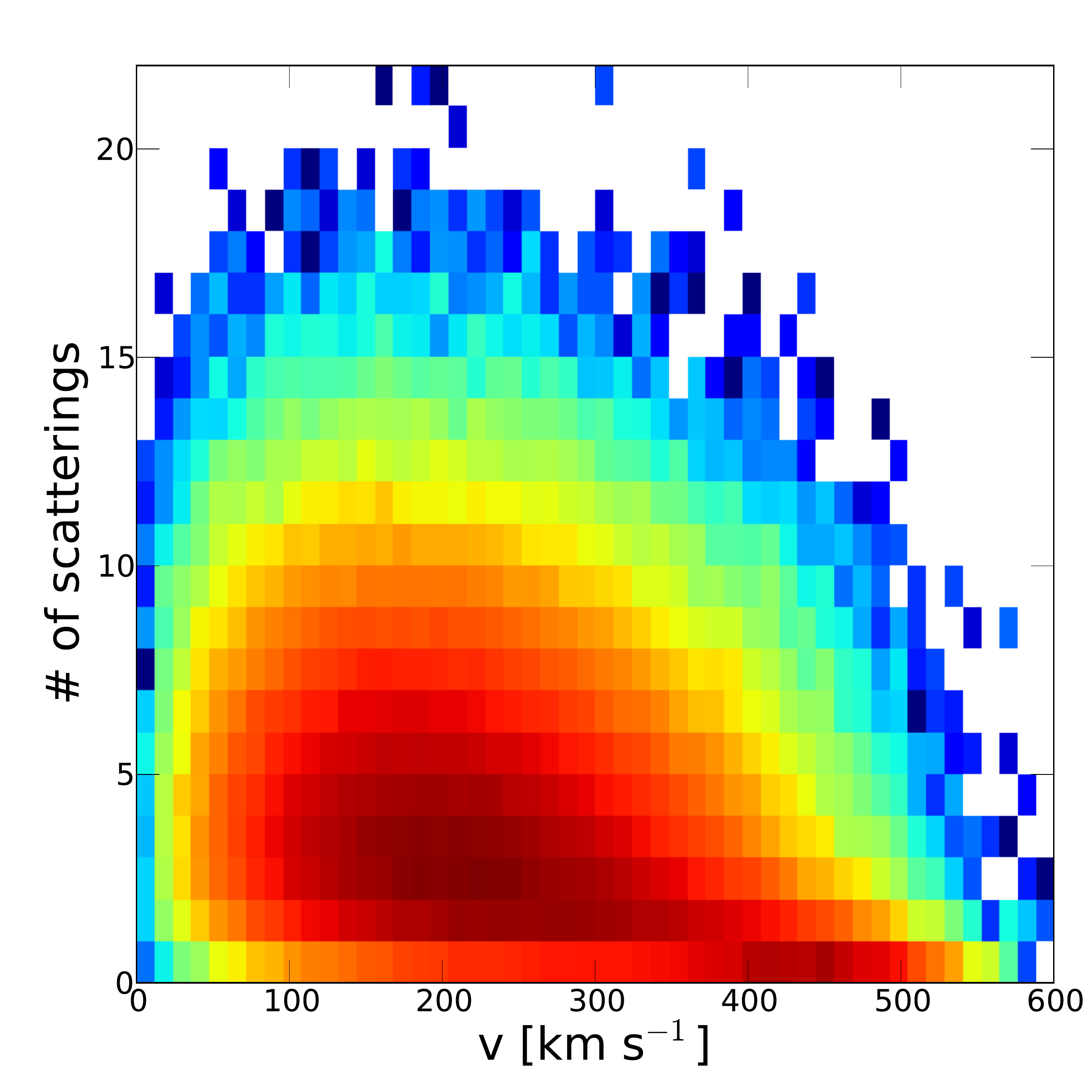}\\
\vspace{-0.1cm}\hspace{-0.5cm}\includegraphics[width=0.245\textwidth]{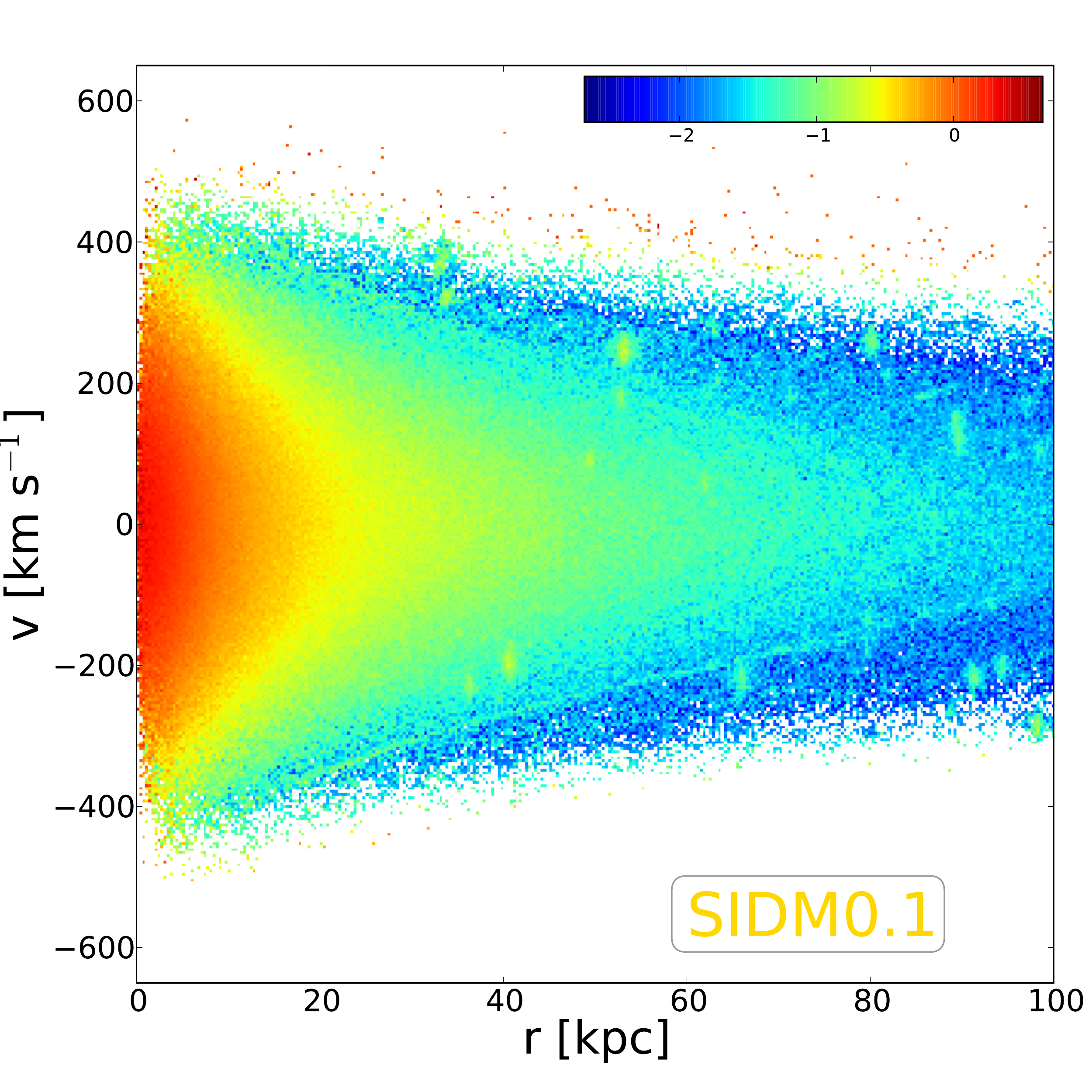}
\hspace{-0.2cm}\includegraphics[width=0.245\textwidth]{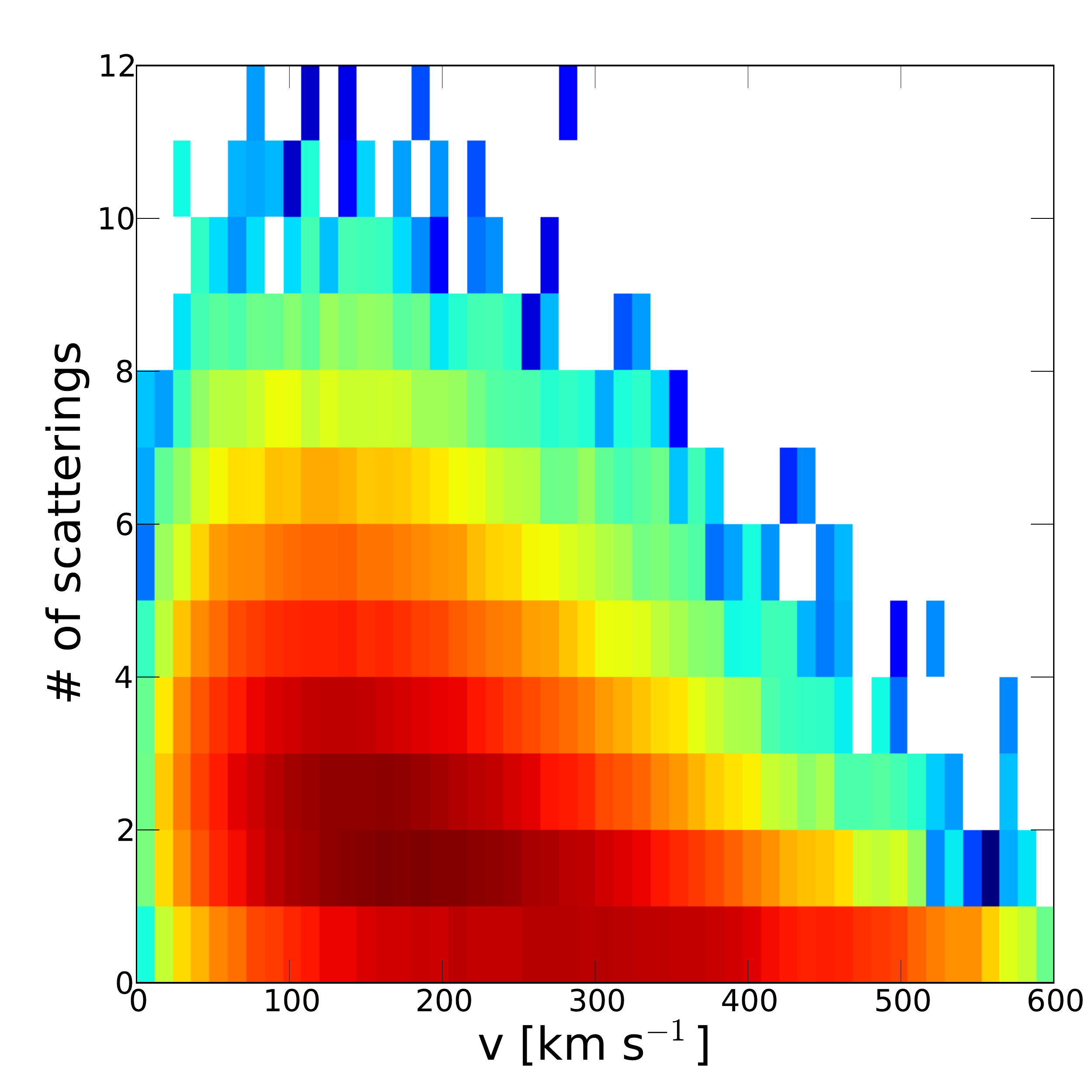}
\hspace{-0.2cm}\includegraphics[width=0.245\textwidth]{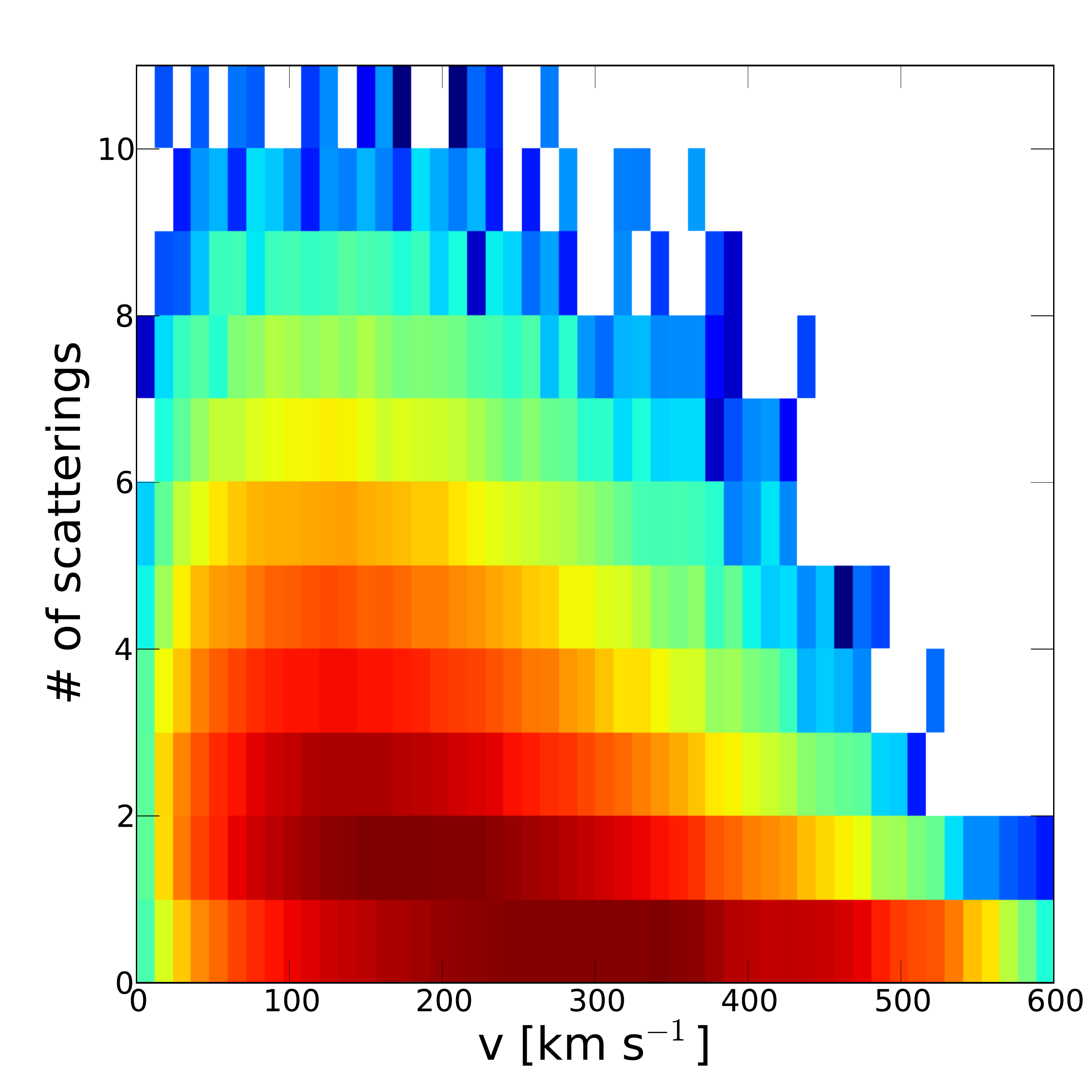}
\hspace{-0.2cm}\includegraphics[width=0.245\textwidth]{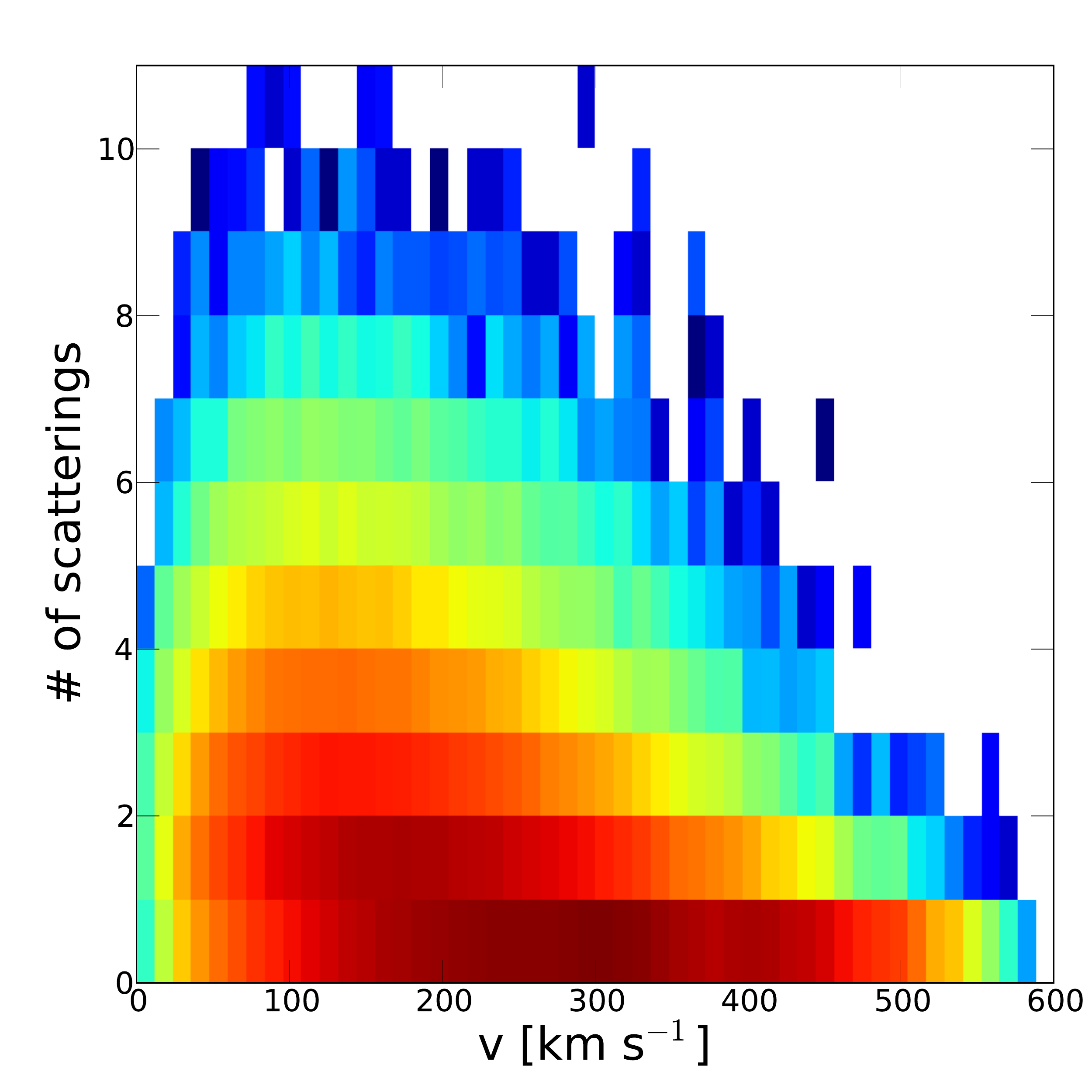}\\
\vspace{-0.1cm}\hspace{-0.5cm}\includegraphics[width=0.245\textwidth]{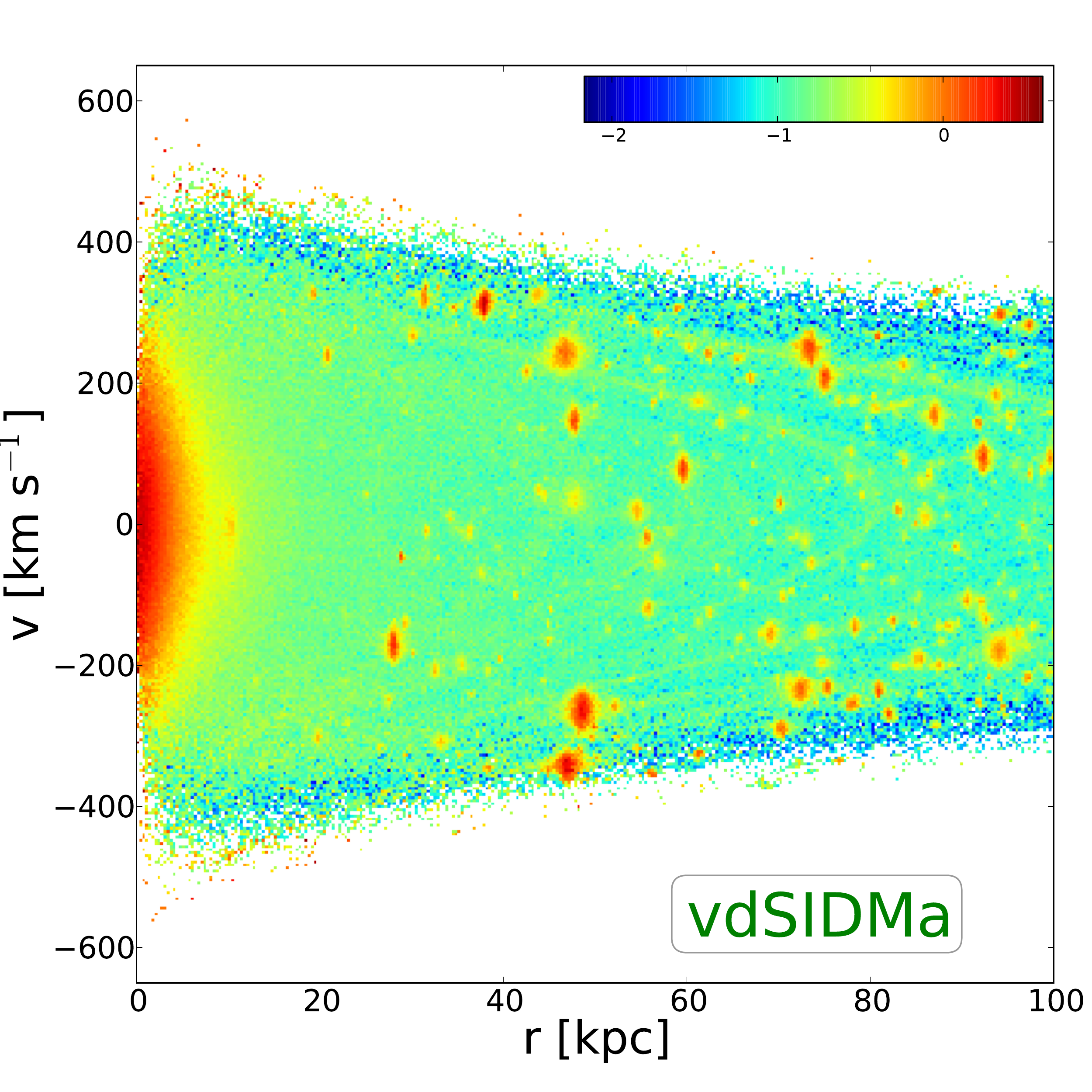}
\hspace{-0.2cm}\includegraphics[width=0.245\textwidth]{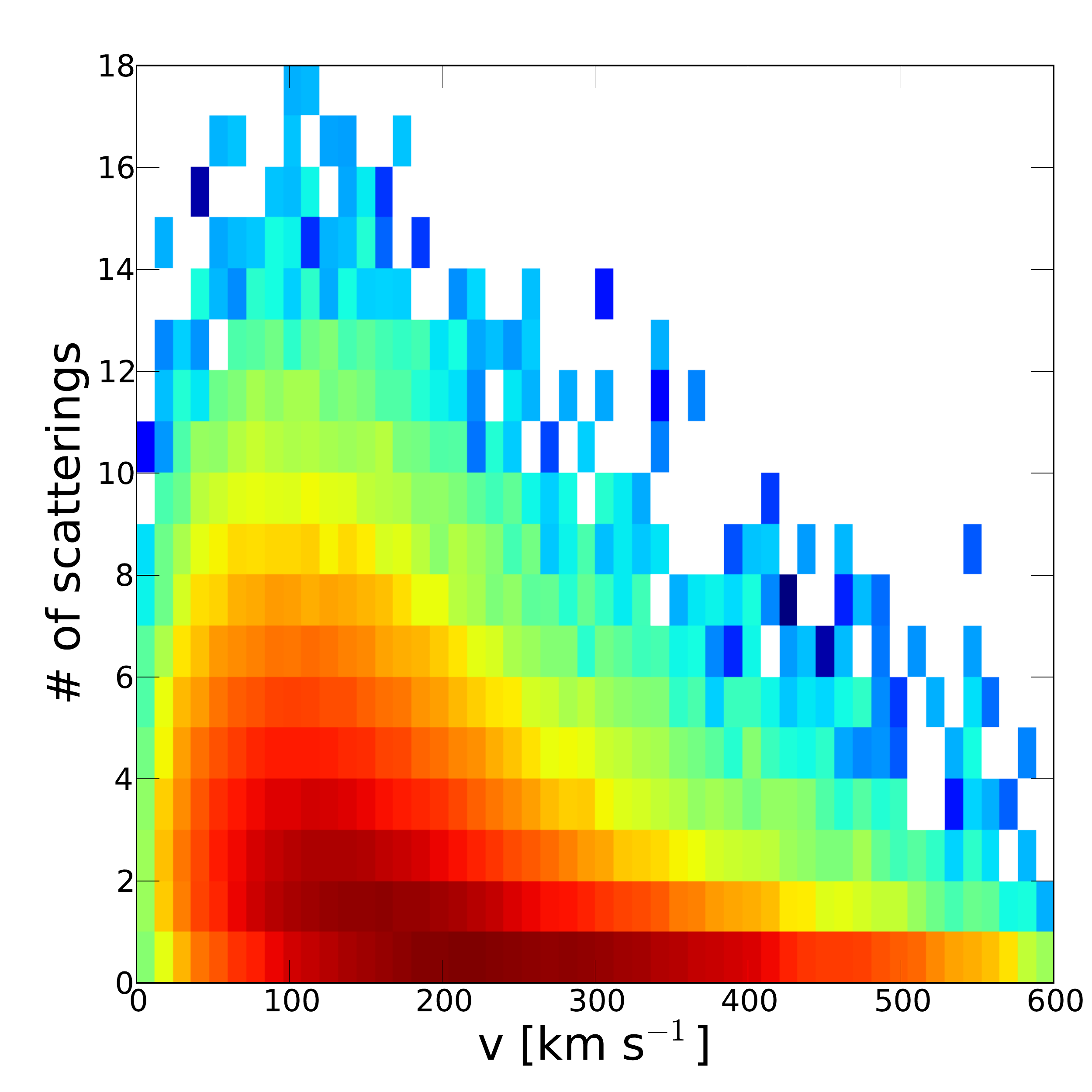}
\hspace{-0.2cm}\includegraphics[width=0.245\textwidth]{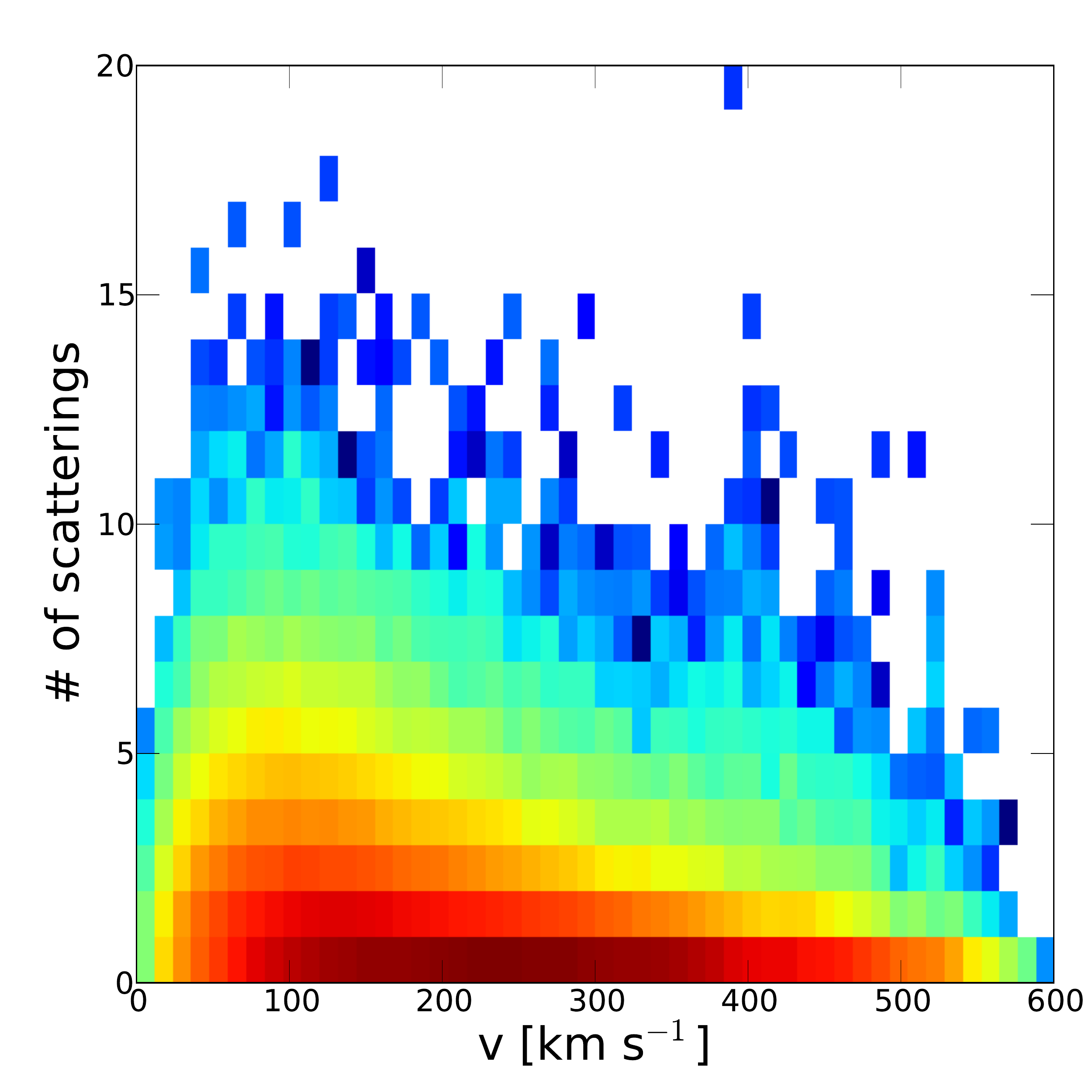}
\hspace{-0.2cm}\includegraphics[width=0.245\textwidth]{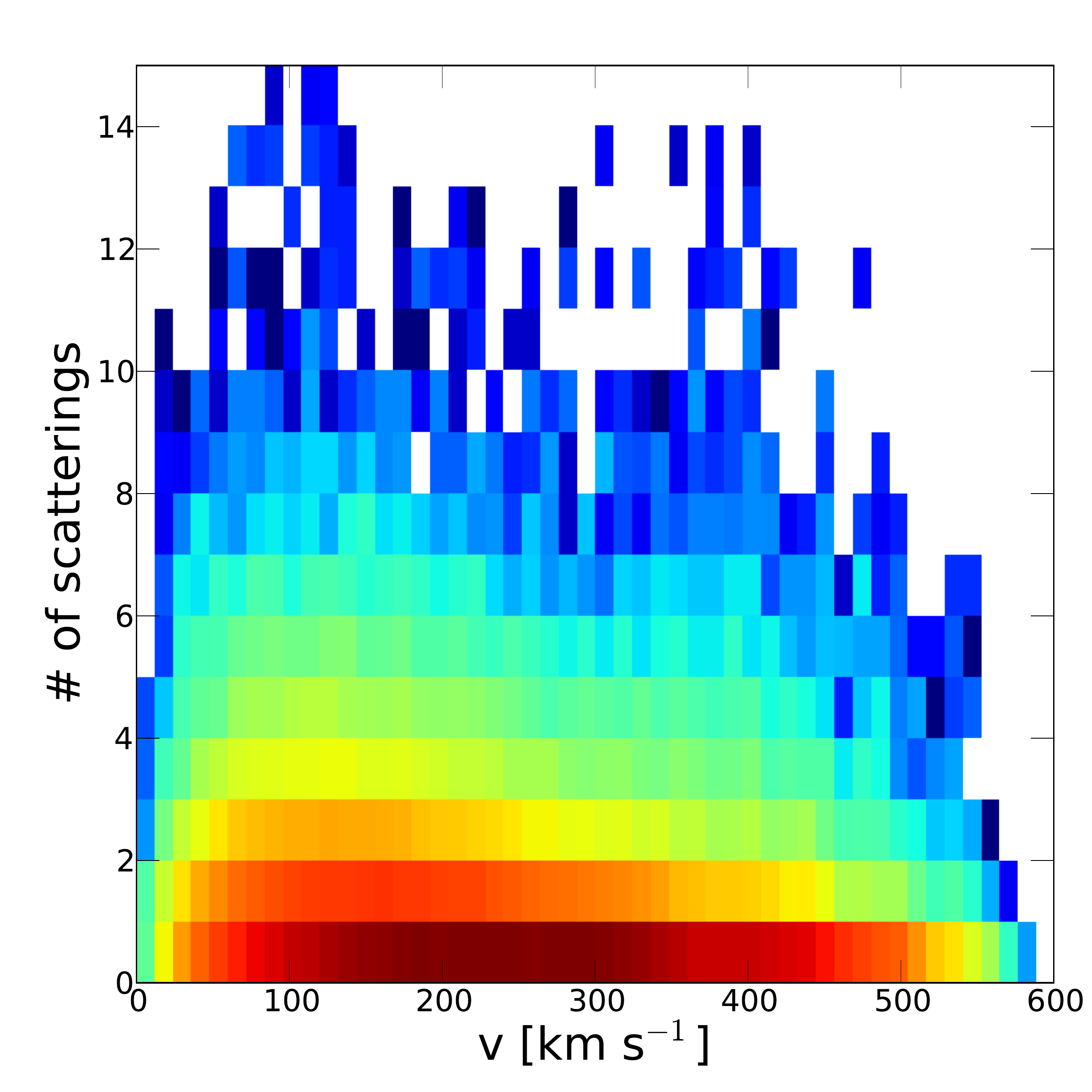}\\
\vspace{-0.1cm}\hspace{-0.5cm}\includegraphics[width=0.245\textwidth]{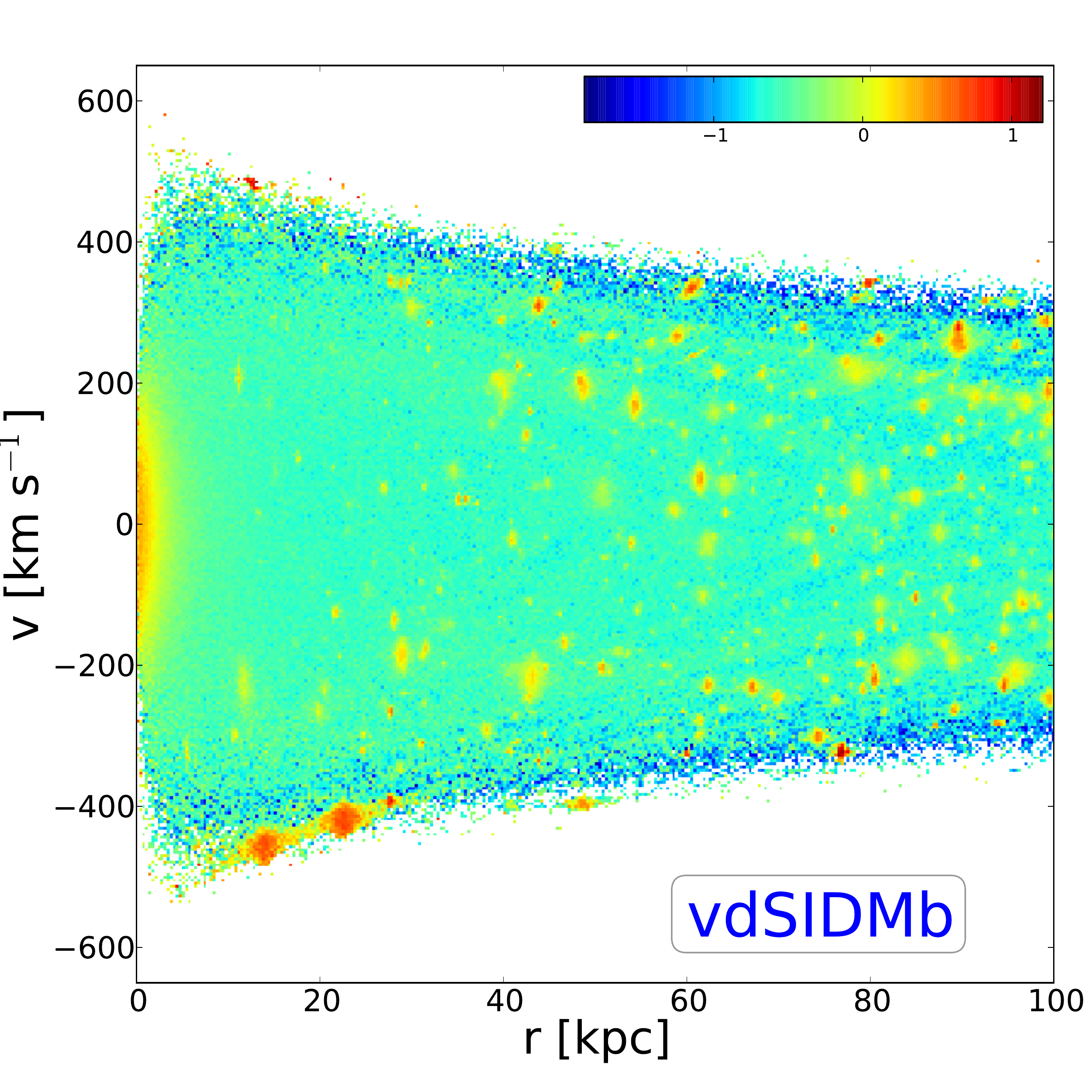}
\hspace{-0.2cm}\includegraphics[width=0.245\textwidth]{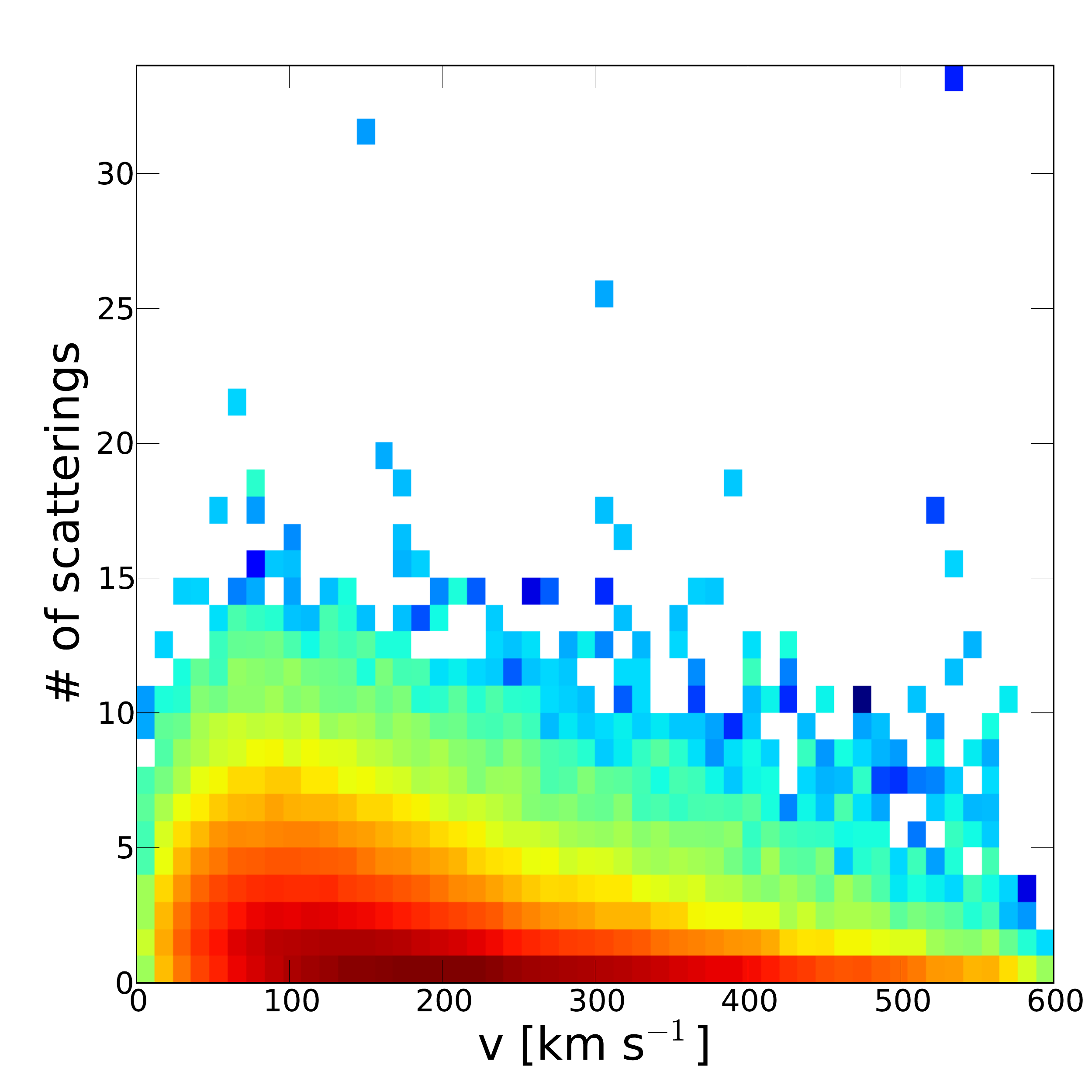}
\hspace{-0.2cm}\includegraphics[width=0.245\textwidth]{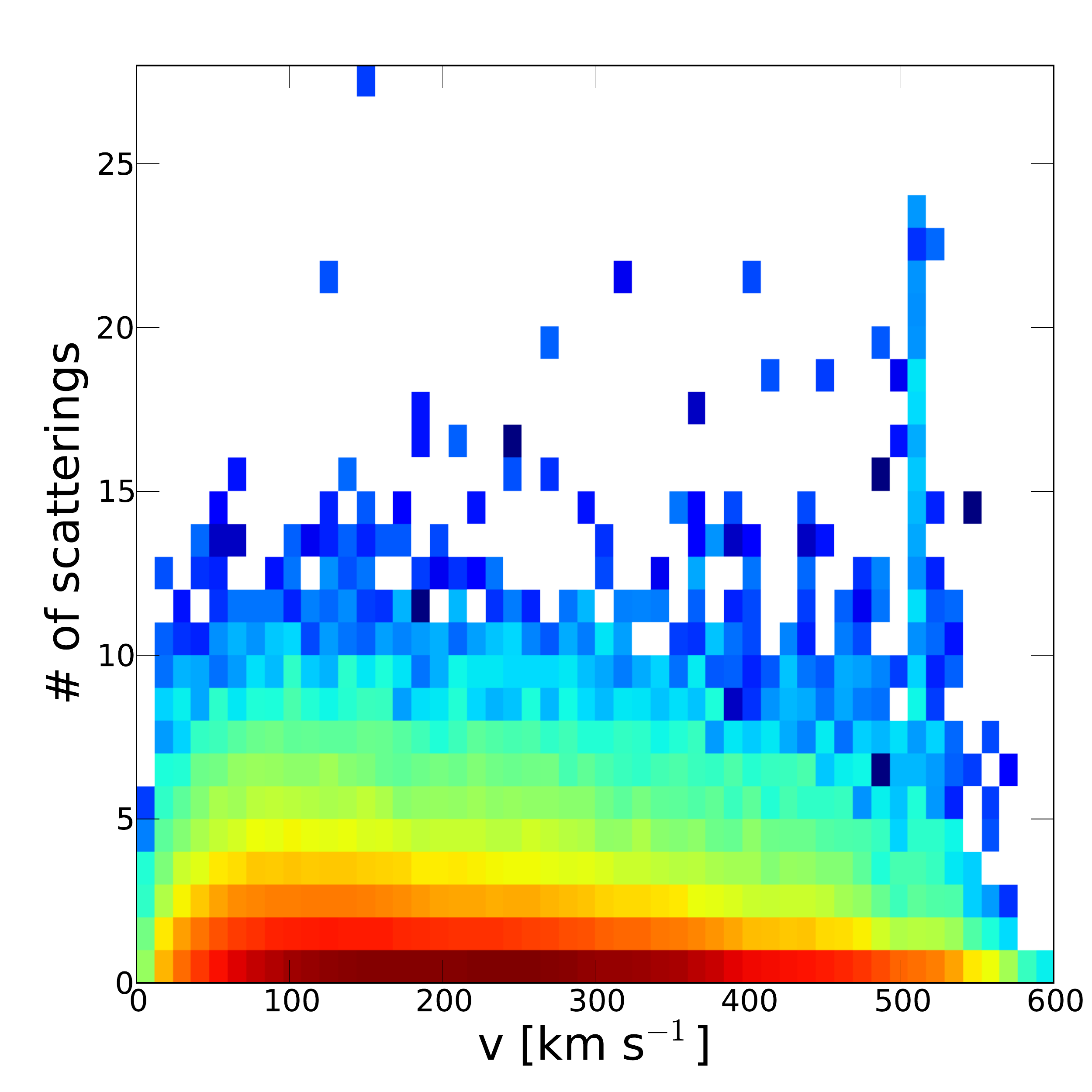}
\hspace{-0.2cm}\includegraphics[width=0.245\textwidth]{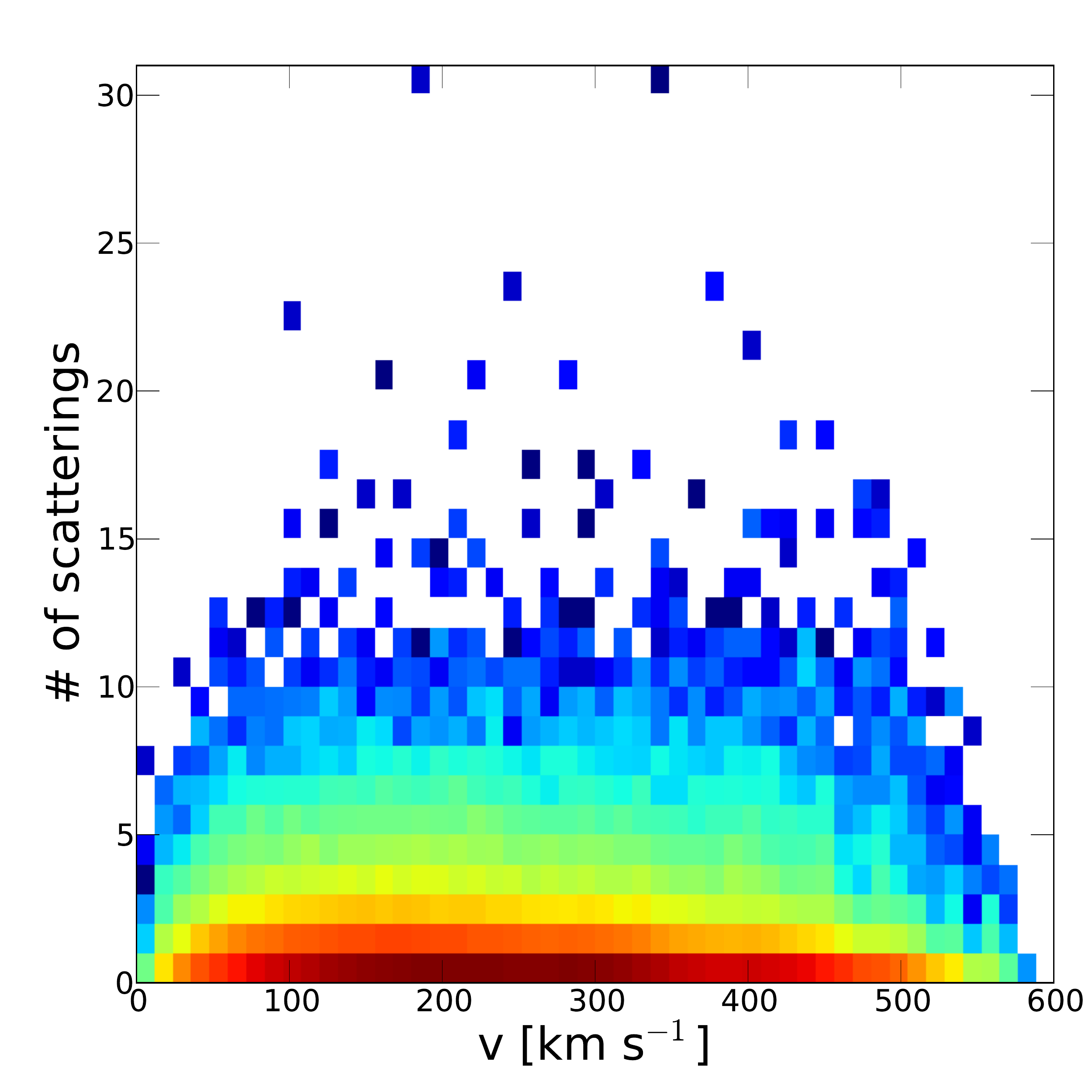}\\
\caption{Left column: Phase-space histograms where each pixel represents the
mean mass-weighted number of scatterings that particles in that bin
underwent until $z=0$ (see colorbar). Remaining
columns: Number of particle scatterings ($y$-axis) as a function of velocity modulus ($x$-axis). 
Colours indicate the mass in each bin. We show the distributions for
three different distances (second column: $2\kpc$; third column: $5\kpc$;
fourth column: $8\kpc$) and in all cases for $1000$ observers. The distribution is measured within spheres of $1\kpc$ radius. 
The number of particles with large number of scatterings increases
strongly towards the main halo center for all models. The constant cross
section cases look similar, although SIDM10 shows clearly a very large number of
scattering events until $z=0$. On the other hand, the vdSIDM models have a
steeper distribution of the scattering events towards the center of the
(sub)haloes, where DM is colder and denser. Notice how subhaloes are clearly
highlighted in the left column for these cases.}
\label{fig:phasespace_scattering} 
\end{figure*}
We examine here first the inner coarse-grained phase-space distributions of the
Aquarius Aq-A-3 halo for the different SIDM models, and contrast them to the
CDM case.  In Fig.~\ref{fig:phasespace_density}, we show the phase-space
structure weighting the simulation particles by their local coarse-grained
phase-space density $\rho/\sigma^3$, where $\rho$ and $\sigma$ are calculated
based on SPH estimates using $64$ neighbours.  Except for SIDM10, where
(sub)haloes are significantly more spherical and less dense (as demonstrated in
VZL), the phase-space structure of the different models looks rather similar to
the CDM case, albeit all of them produce density cores.  The core density of
the main halo is lower for the constant cross section models than for the
vdSIDM models.  Although SIDM1 (SIDM0.1) has a constant cross section that is ten
(a hundred) times smaller than in SIDM10, there is still a noticeable reduction
of the central phase-space density of the main halo.  This effect is nearly
absent for the vdSIDMa and vdSIDMb models due to the velocity-dependence of their
cross sections, which compensates for the rising configuration space density
towards the center. This then leads to a reduced scatter rate in the center and
therefore the central phase-space density is not significantly reduced for the
models with velocity-dependent cross section. We note that except for SIDM10,
all models have a subhalo abundance that is very similar to the one in CDM.

\begin{figure*}
\centering
\hspace{-0.5cm}\includegraphics[width=0.5\textwidth]{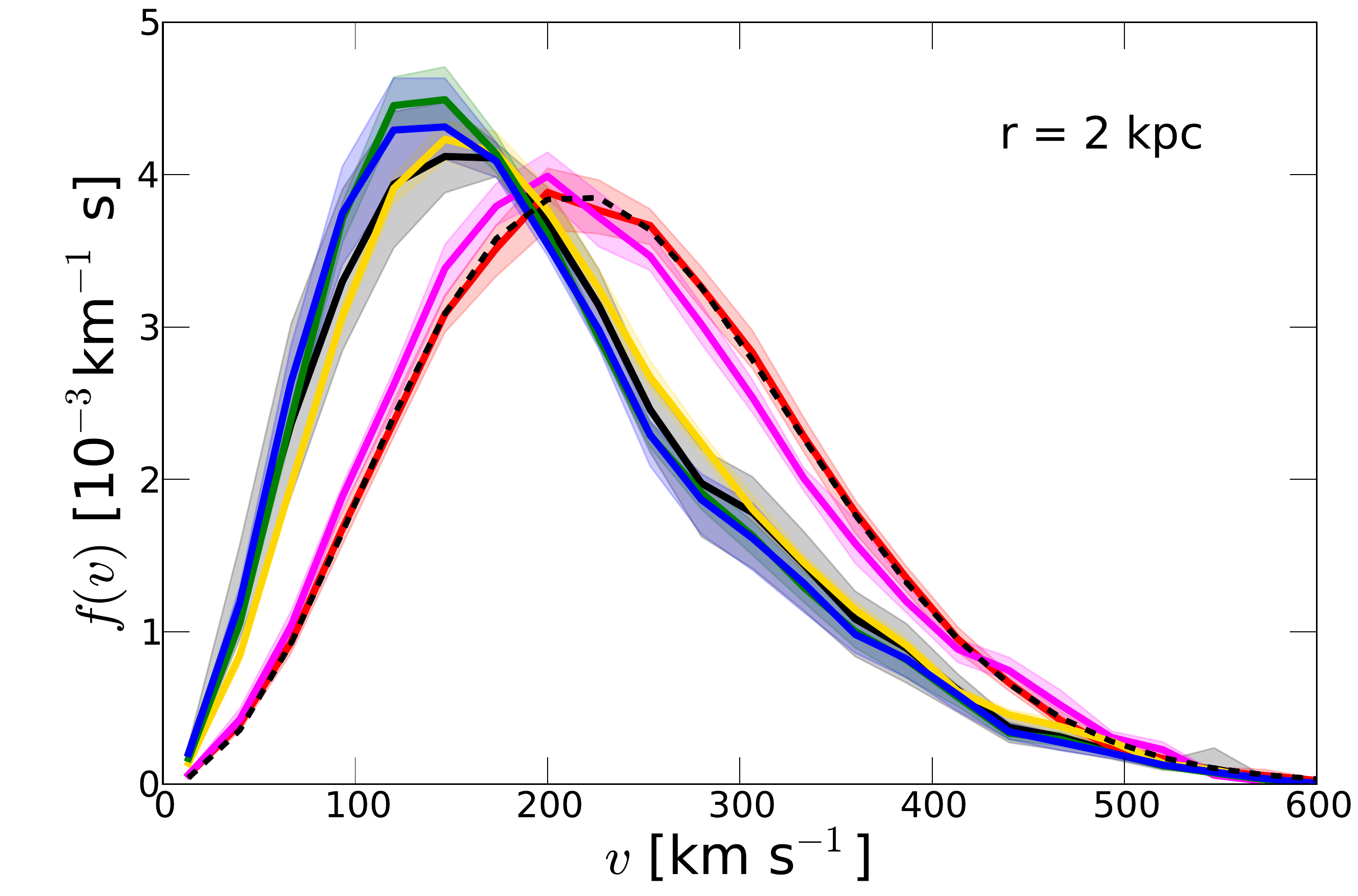}
\includegraphics[width=0.5\textwidth]{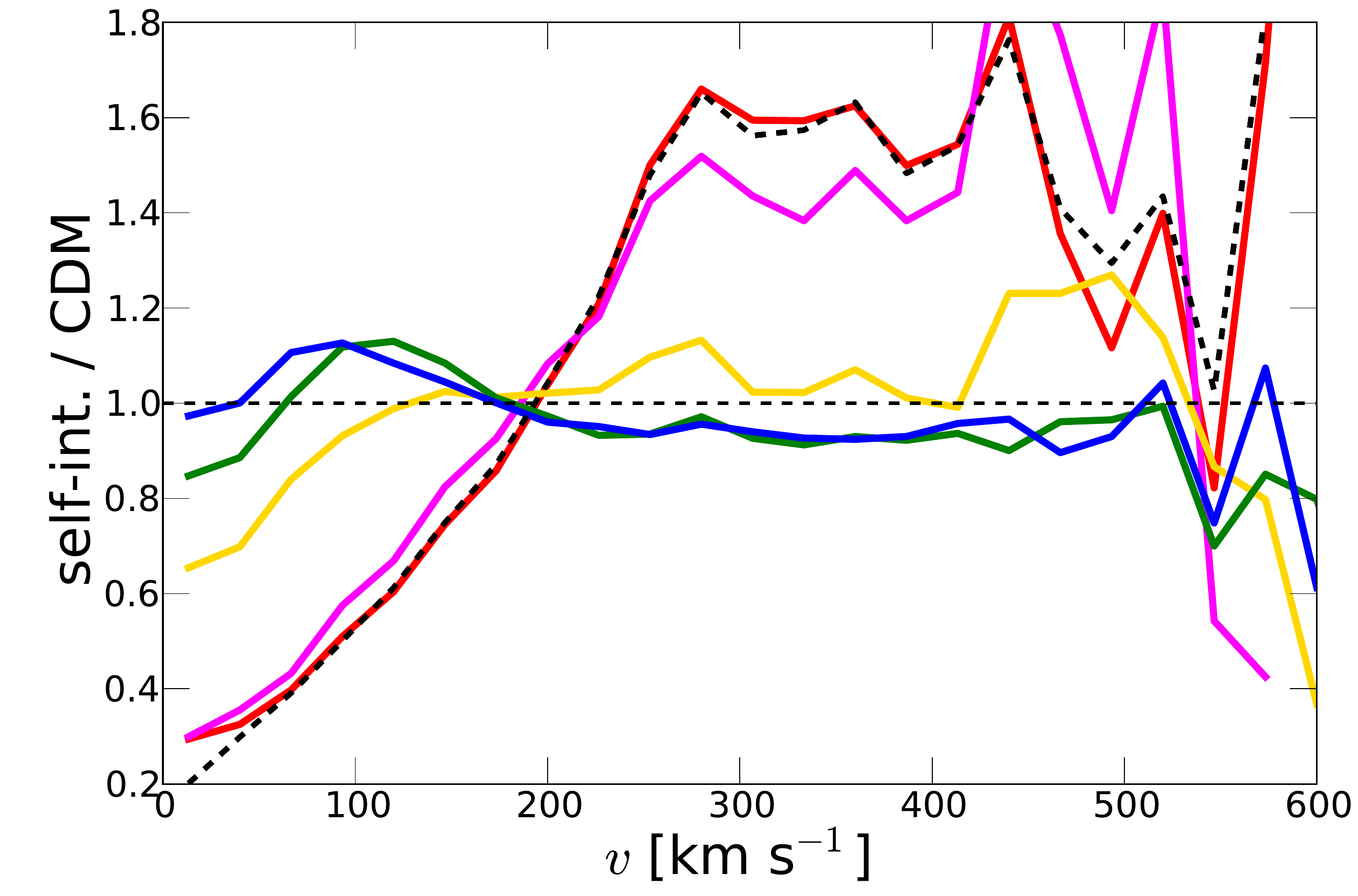}\\
\hspace{-0.5cm}\includegraphics[width=0.5\textwidth]{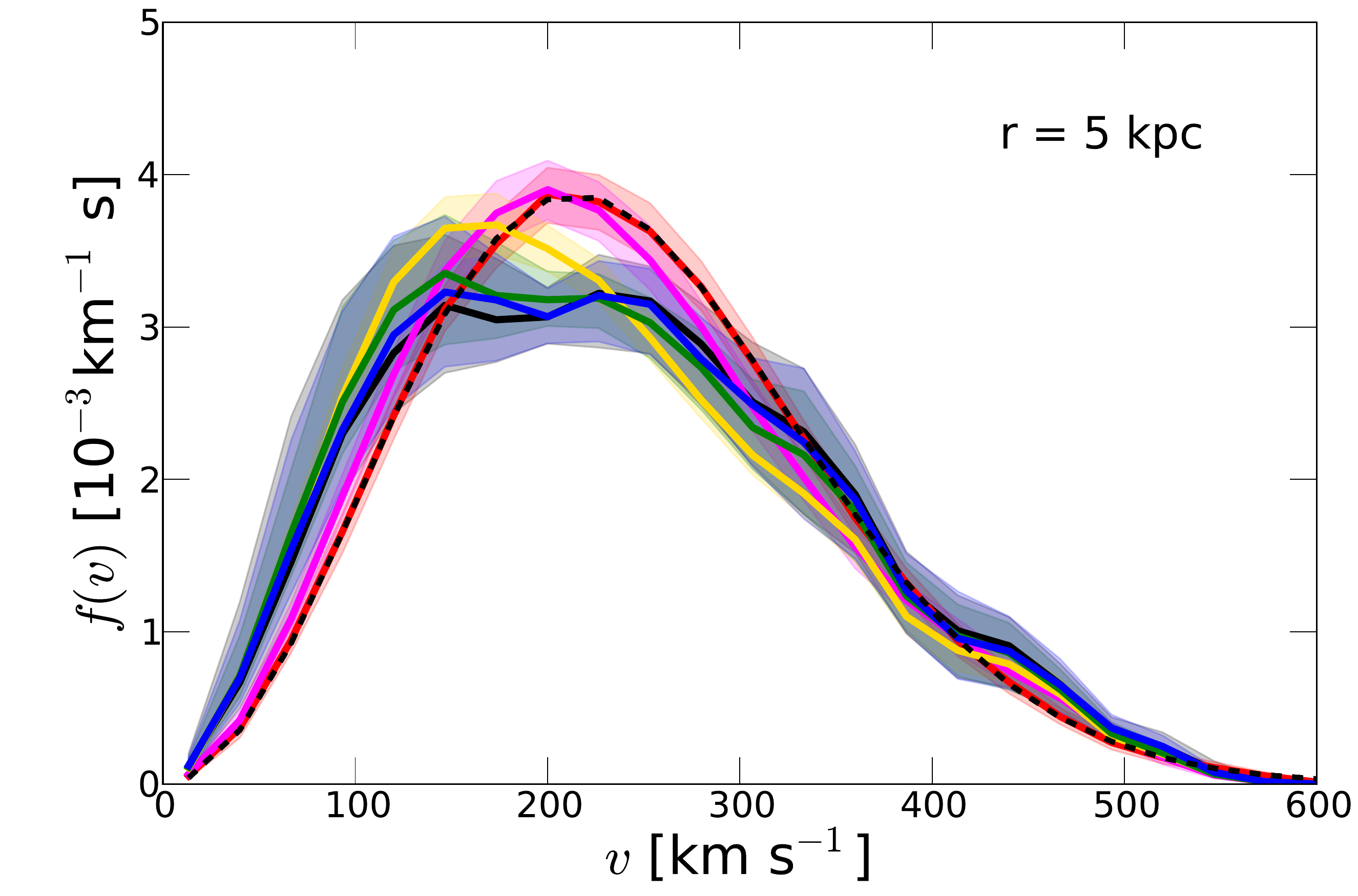}
\includegraphics[width=0.5\textwidth]{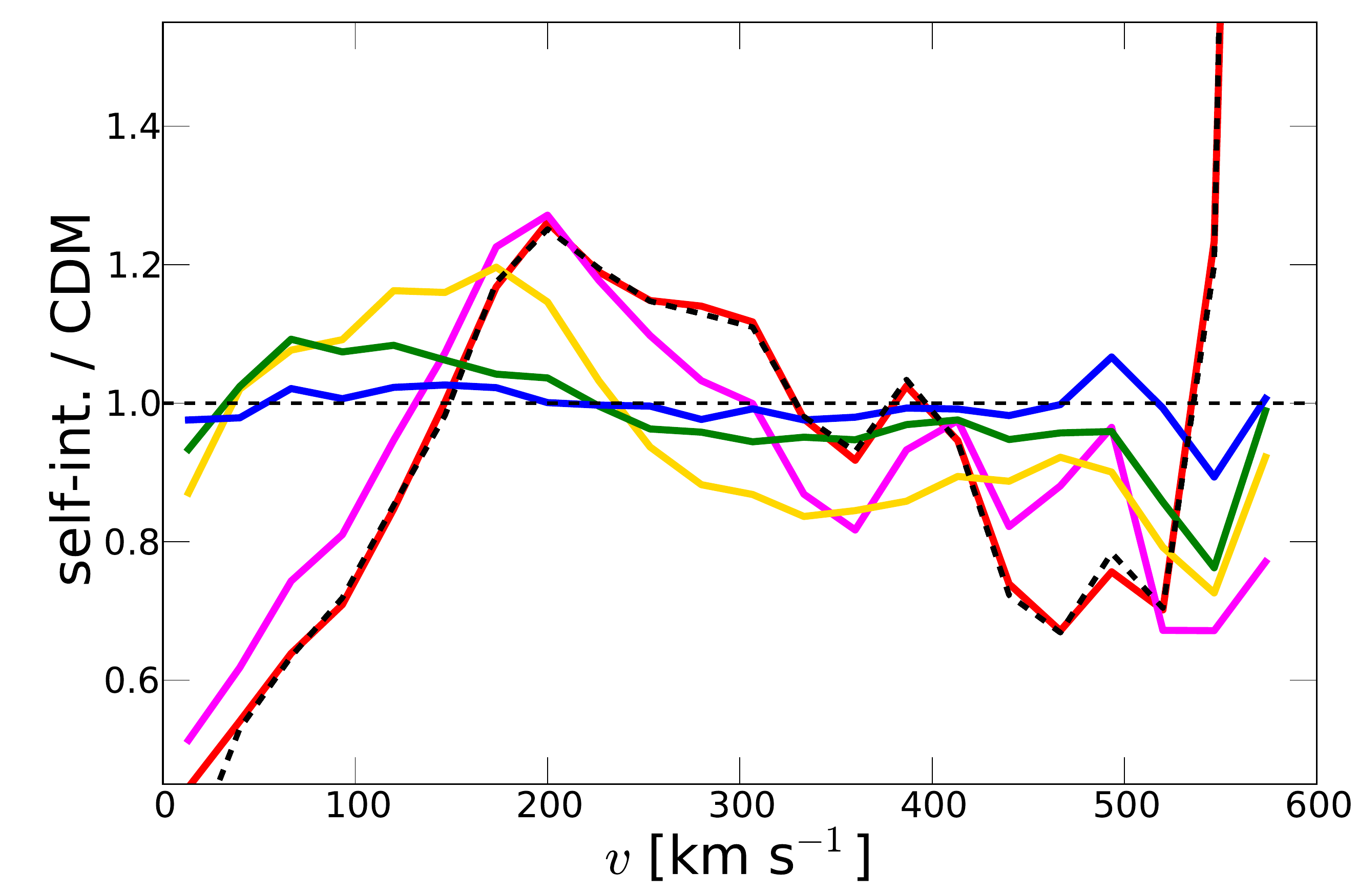}\\
\hspace{-0.5cm}\includegraphics[width=0.5\textwidth]{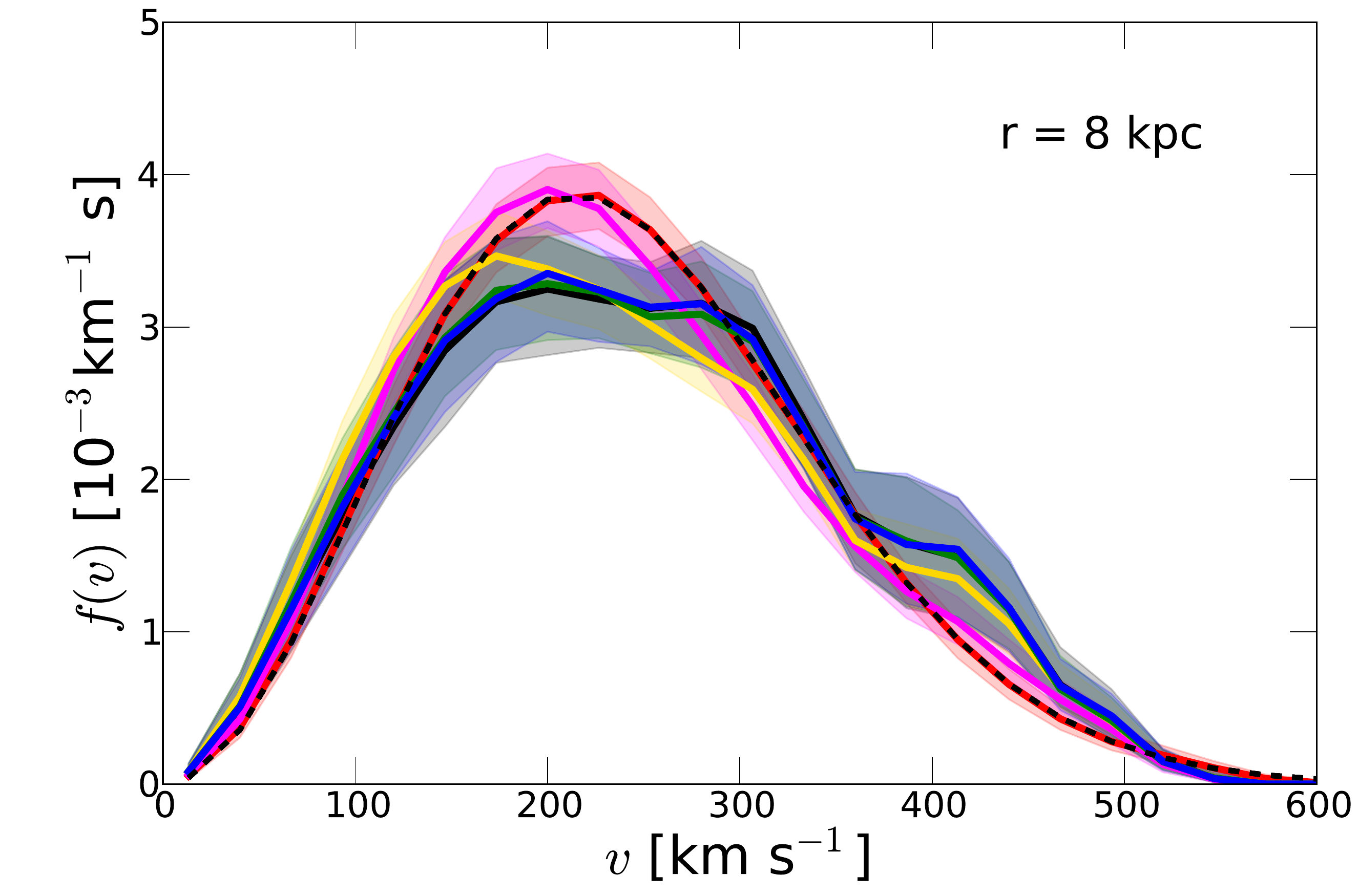}
\includegraphics[width=0.5\textwidth]{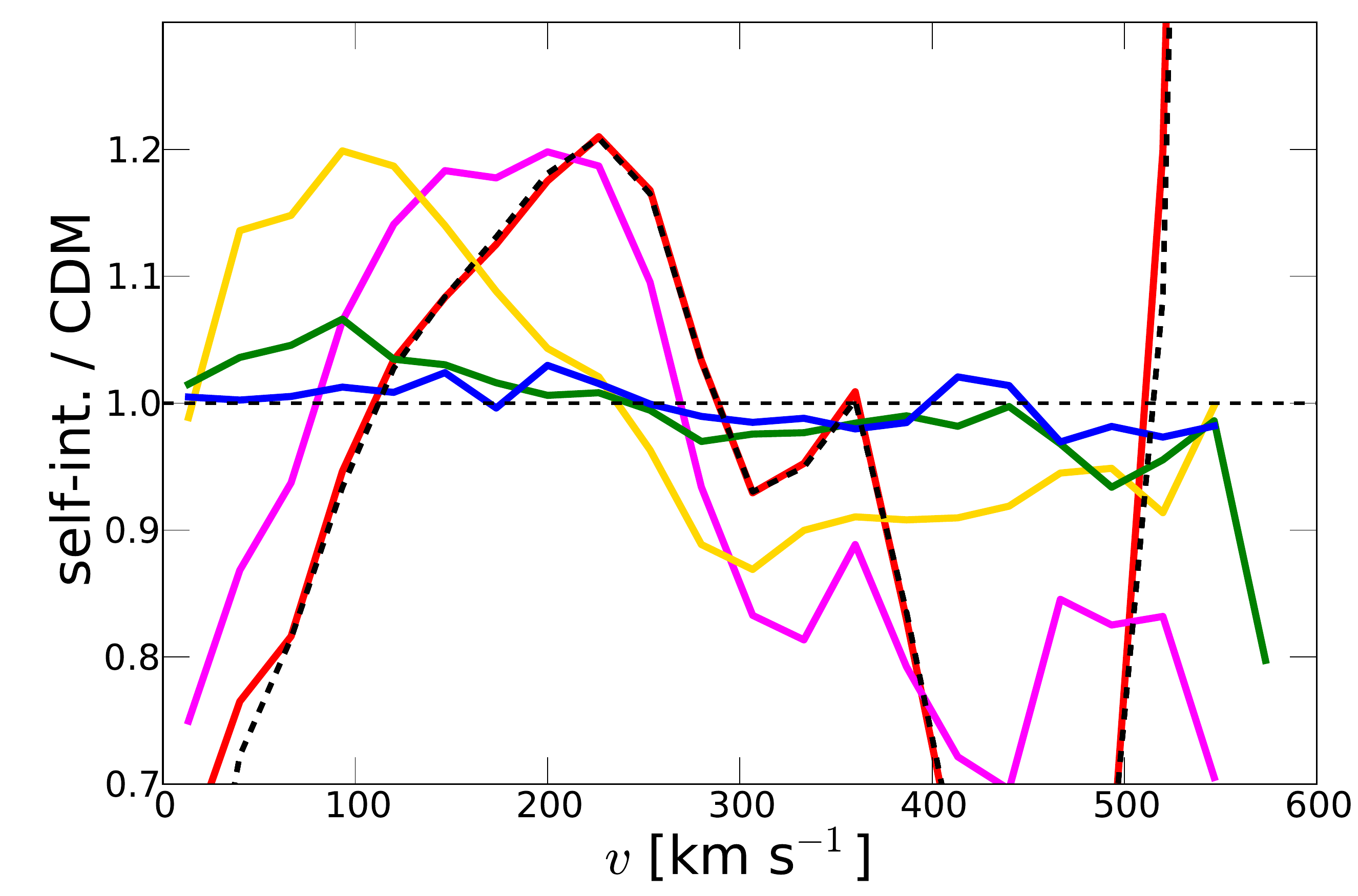}
\caption{Left panels: Distribution of the velocity modulus measured in spheres
of $1\kpc$ radius for $1000$ random positions at a given halocentric distance
(as indicated).  The median and $10-90\%$ quantiles of the distribution (over
all observers) are shown with solid thick lines and shaded regions,
respectively. Right panels: The median of the ratio of the velocity
distributions to the CDM case for each observer. Colours are as in
Fig.~\ref{fig:cross_section}.  The mean free path of the DM particles is
shortest in the halo center, creating the largest differences between the
self-interacting and CDM models.  For larger radii, the decrease in the density
diminishes the scattering rate and hence the local velocity distributions are
more similar.  The large and constant cross section of SIDM results in a
smooth and isotropic Maxwellian distribution at all radii within the core
region ($\sim10\kpc$). To lower extent, SIDM1 and SIDM0.1 with $10$ and $100$
times lower constant cross section, respectively, still show clear deviations
from CDM.  On the contrary, at $8\kpc$, the velocity-dependent SIDM cases
(vdSIDMa and vdSIDMb) deviate from the CDM model by at most $5\%$. The dashed black
line in the left panels shows a Maxwell-Boltzmann distribution with
$\sigma_{\rm 1D}=151.6\kms$, which fits the SIDM10 distribution very well. The
distribution of SIDM1 is still quite Maxwellian, whereas all other models
clearly have very different distributions} 
\label{fig:vdist} 
\end{figure*}

For the self-interacting models we can also weight the bins in the phase-space
distribution by the number of scattering events (until $z=0$), as we show in
the leftmost column of Fig.~\ref{fig:phasespace_scattering}.  Specifically, we
are plotting the mean mass-weighted number of scatter events for each
phase-space bin. We note that we do not use the same colour scales for the
different models in order to show the details in each case more clearly.  The
actual numbers of scatter events are indicated by the colour bars. As expected,
the centres of the main halo and its subhaloes have the highest numbers of
scatter events, although in the constant cross section cases (particularly in
SIDM10) individual subhaloes are not easily distinguishable in this phase-space
projection with most of the particles having encountered a large number of
scatter events until $z=0$.  vdSIDMa and vdSIDMb look very similar to each other
and significantly different from all constant cross section models. The central
main halo region, with large scatter counts, is more extended in the latter
whereas it is very small for the vdSIDM models since the velocity scaling of
the cross section compensates the increasing density towards the main halo
center leading to a suppression of scattering events. Subhaloes are clearly
visible in the vdSIDM models since self-scattering is enhanced in these cold
(low velocity) and dense structures. We note that this behaviour depends on the
details of the vdSIDM models. Specifically, it depends on the velocity at which
the cross section peaks. As shown in Fig.~\ref{fig:cross_section}, vdSIDMa and
vdSIDMb both peak at rather low velocities, which was chosen specifically to get
significant self-scatterings at the scale of dSphs (see VZL). Other choices of
the peak velocity will lead to different phase-space structures.

\begin{figure*}
\centering
\hspace{-0.5cm}\includegraphics[width=0.5\textwidth]{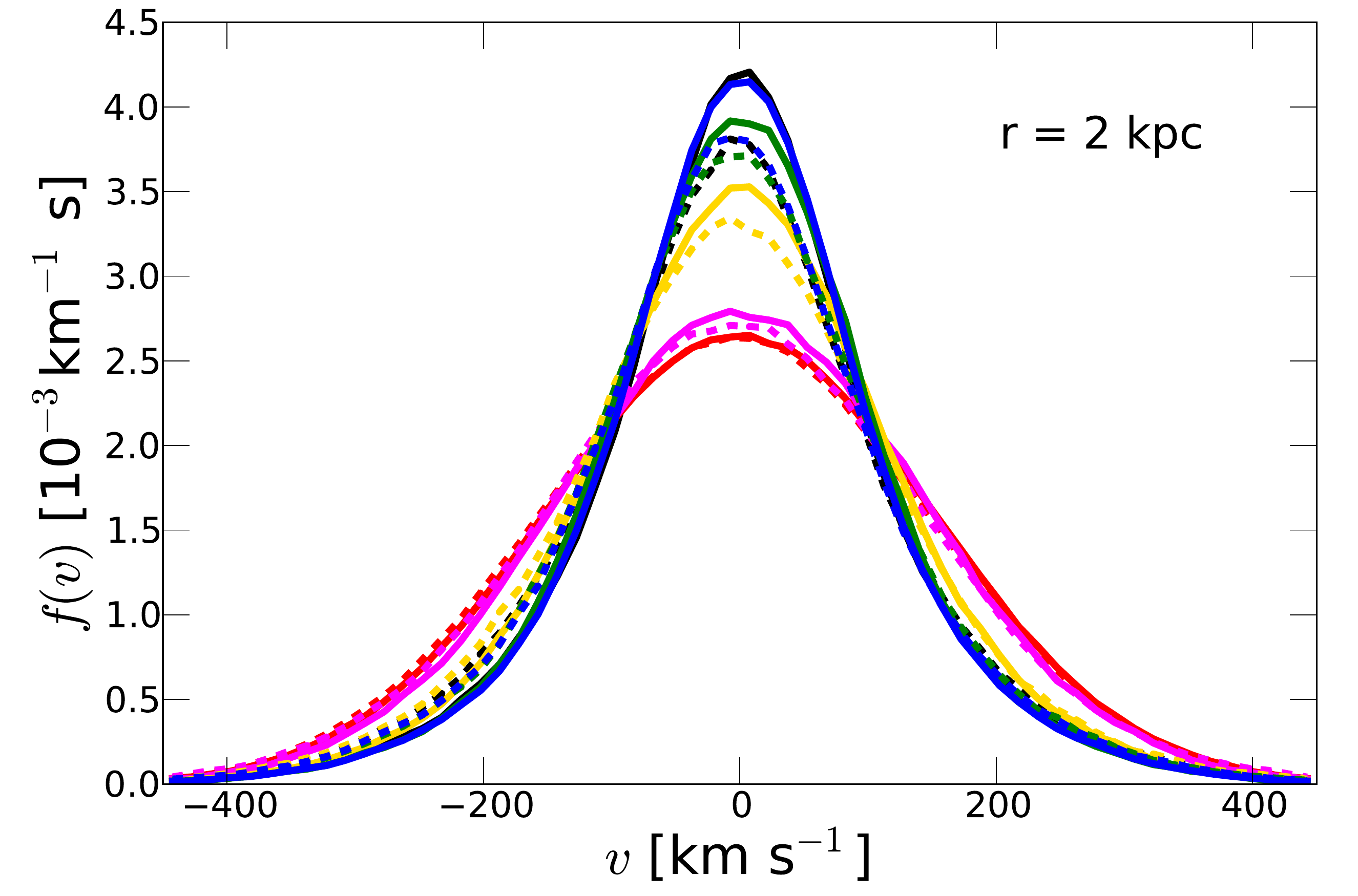}
\includegraphics[width=0.5\textwidth]{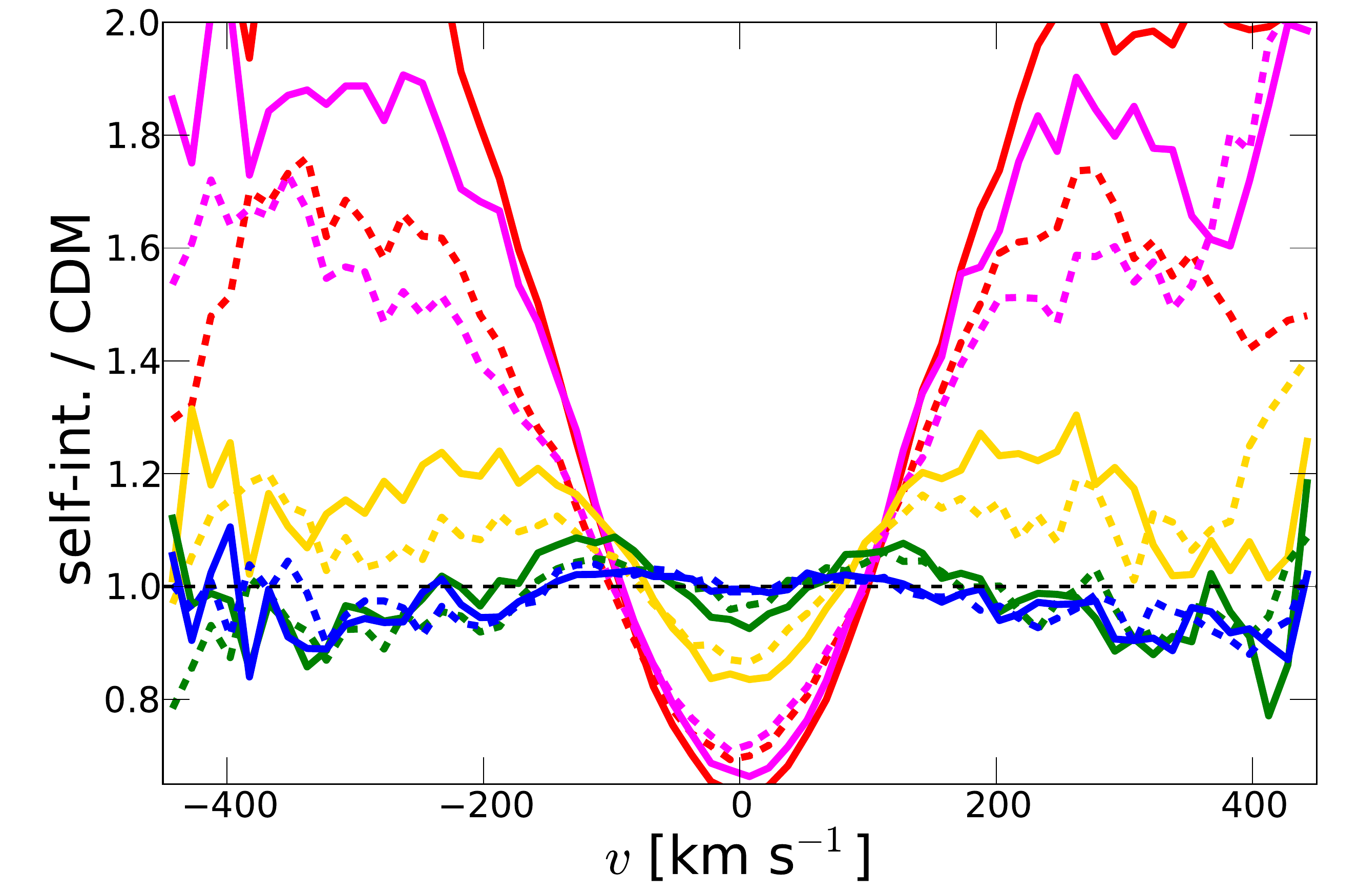}\\
\hspace{-0.5cm}\includegraphics[width=0.5\textwidth]{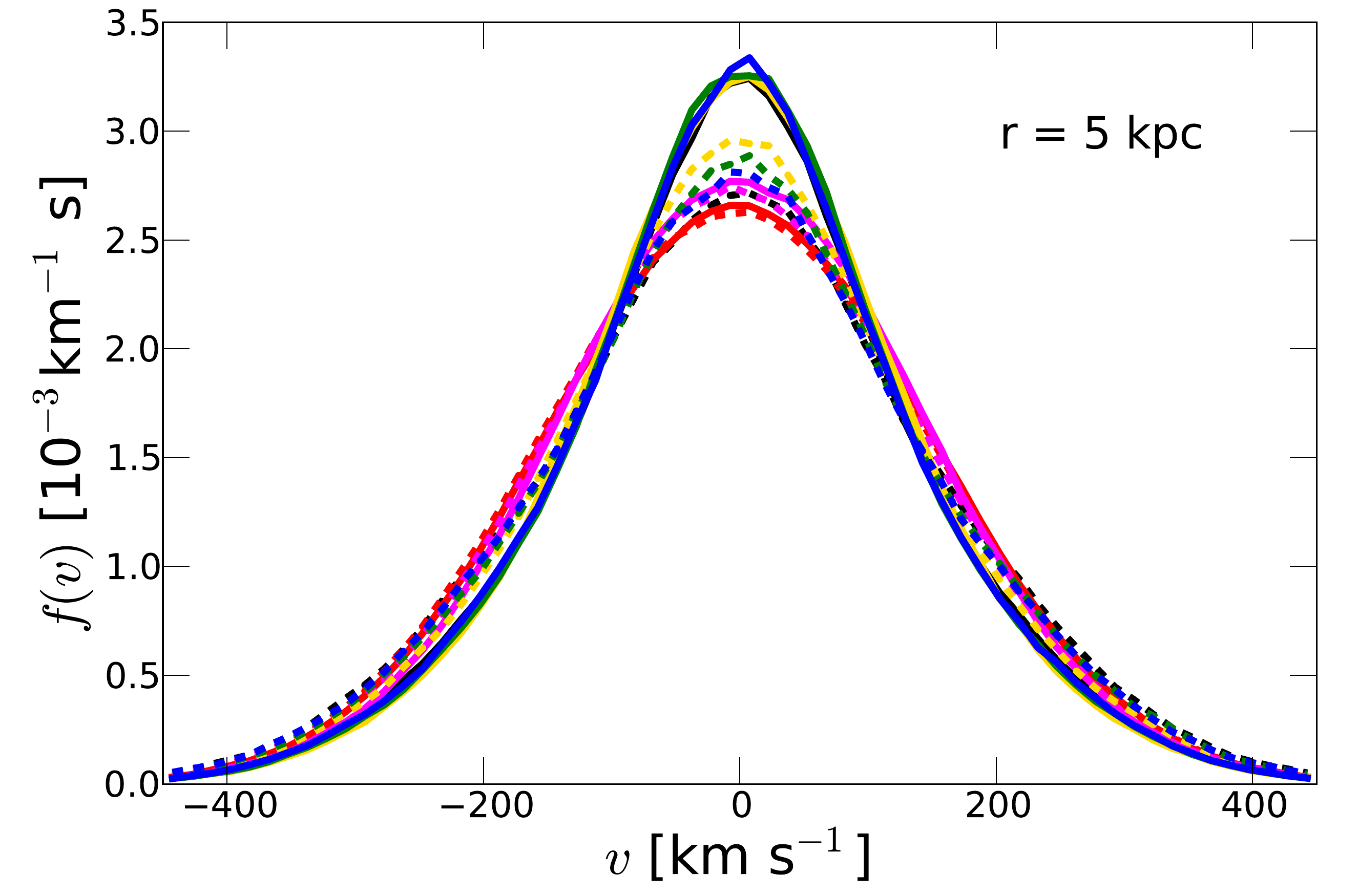}
\includegraphics[width=0.5\textwidth]{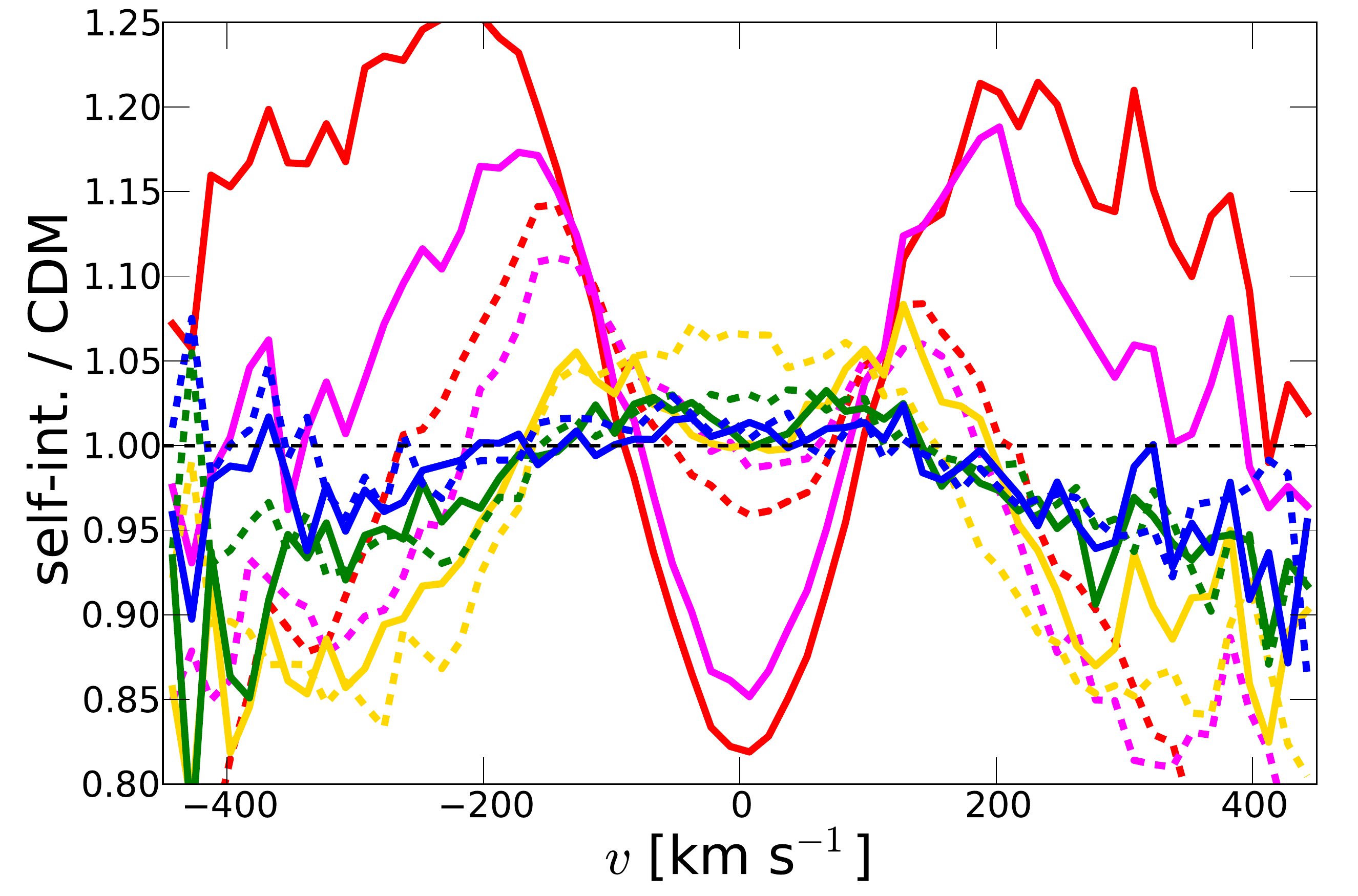}\\
\hspace{-0.5cm}\includegraphics[width=0.5\textwidth]{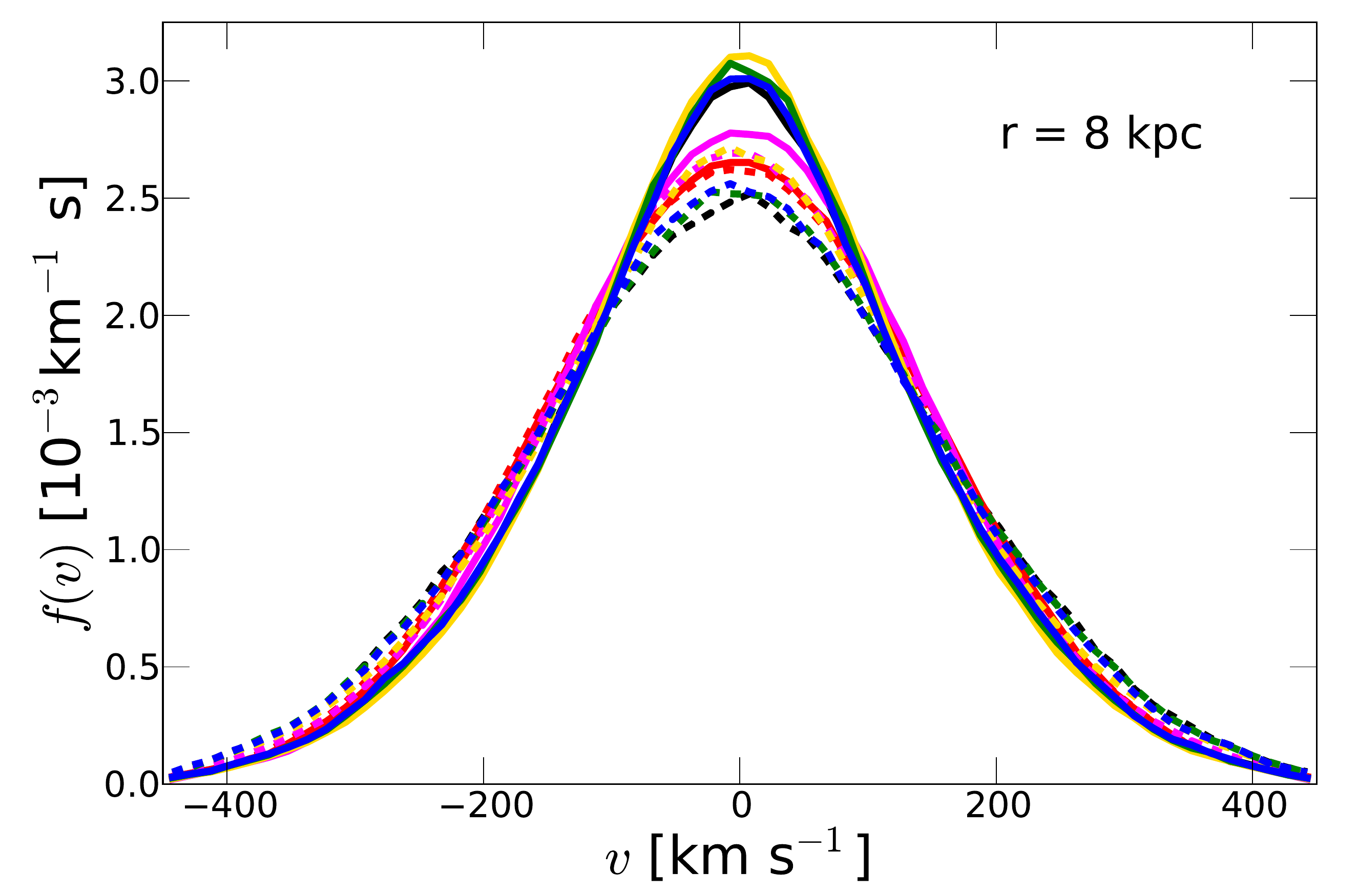}
\includegraphics[width=0.5\textwidth]{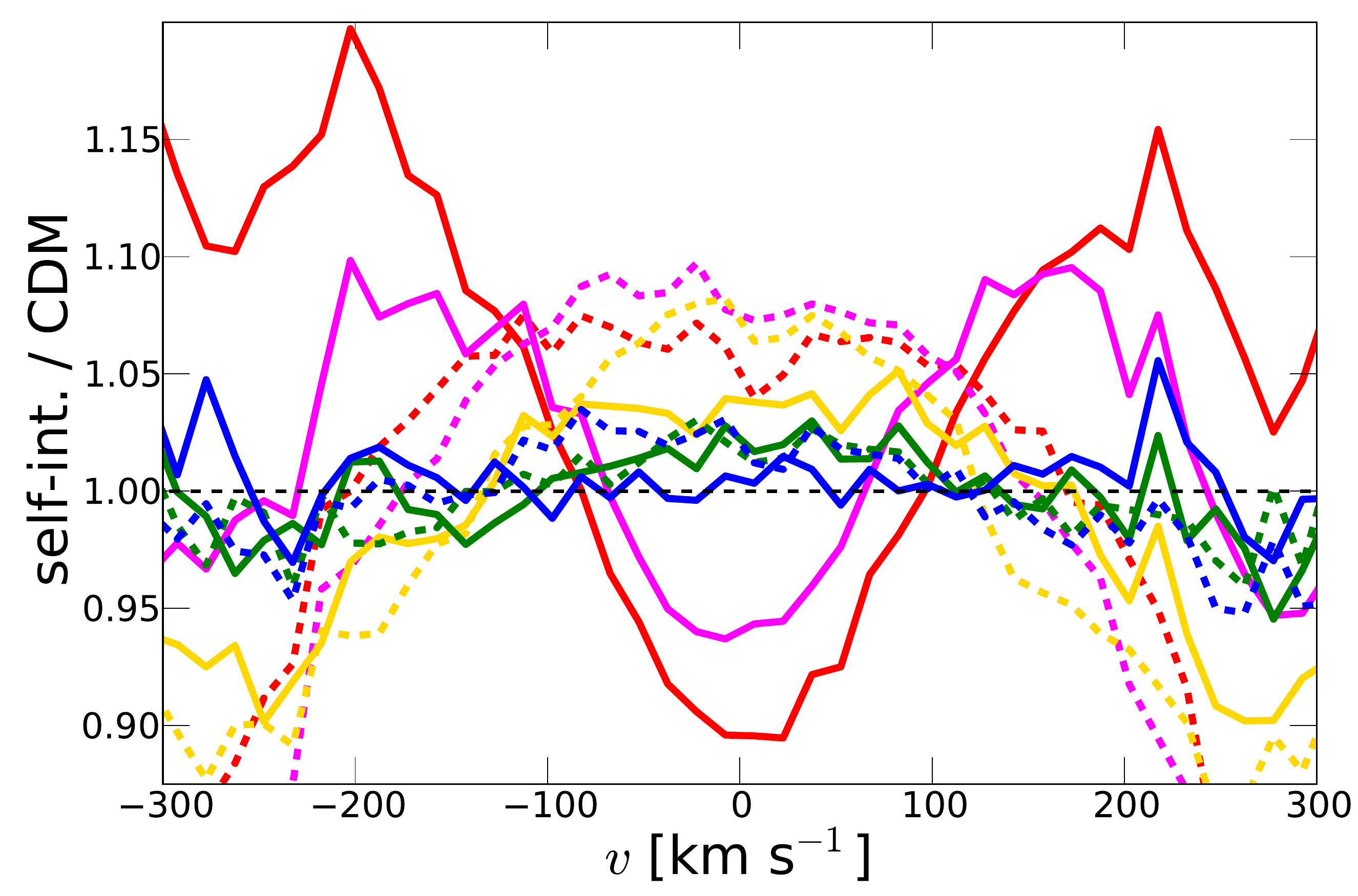}
\caption{Left panels: Velocity distribution along the largest (solid) and
smallest (dashed) components of the local velocity ellipsoid measured in
spheres of $1\kpc$ radius for $1000$ random positions at a given halocentric
distance (as indicated). Only the median of the distributions is shown. Right
panels: Median of ratio of each distribution to the corresponding CDM case.
Colours are as in Fig.~\ref{fig:cross_section}.  Compared to the SIDM models
CDM haloes are more anisotropic in the inner regions (see also
Fig.~\ref{fig:vel_axis_ratios}), whereas the SIDM models clearly show a
velocity ellipsoid which is much more spherical, particularly for the cases
with a large constant cross section. At the solar circle, the relative
deviations between the models and CDM is rather small, particularly for the
vdSIDM models, although the SIDM10 and SIDM1 cases still show some significant
differences.} 
\label{fig:vcompdist} 
\end{figure*}

To study the relation between scattering events and the local velocity
structure in more detail we also show in Fig.~\ref{fig:phasespace_scattering}
histograms of the local velocity modulus versus the number of scatterings
(until $z=0$) at three halocentric distances: $2\kpc$ (second column), $5\kpc$
(third column) and $8\kpc$ (fourth column).  The largest constant cross section
case SIDM10 shows a very large number of scattering events within a $10\kpc$
region from the center. In this region, most of the particles have been
scattered multiple times creating a constant isothermal density core (see VZL).
This process of transforming the velocity distribution into a Maxwellian one is
incomplete for the SIDM1 and SIDM0.1 cases where the cross section is much lower.
On the other hand, for the vdSIDM cases most of the scatter events have
occurred within $1\kpc$ which is the region with the lowest velocity dispersion
and highest density.

We note that in our analysis we are not including the effects of baryonic
physics on the DM phase-space structure. Typically, predictions on direct
detection signals are obtained without considering these effects. It is
expected that the domination of the central gravitational potential by the
galactic disc might alter dramatically the inner DM distribution, flattening
the central cusp in CDM haloes due to repeated episodes of gas outflows driven
by supernovae \citep[e.g.][]{Navarro1996,Pontzen2012}. Another process that
might impact the inner DM phase-space structure are resonant interactions
between the DM particles and the stellar bar \citep[e.g.][]{Weinberg2002}. DM
subhaloes are also affected if they tidally interact with the galactic disc
\citep[e.g.][]{Donghia2010}. Also the DM phase-space structure will be affected
by baryons, which will also alter the detection signal. The impact of baryonic
processes would likely be quite different in SIDM than in CDM (especially for
subhaloes) and goes beyond the scope of the present study, since our goal is to
compare the differences with CDM driven purely by DM self-scattering.

\section{Velocity structure of the inner halo}

\begin{figure}
\centering
\includegraphics[width=0.475\textwidth]{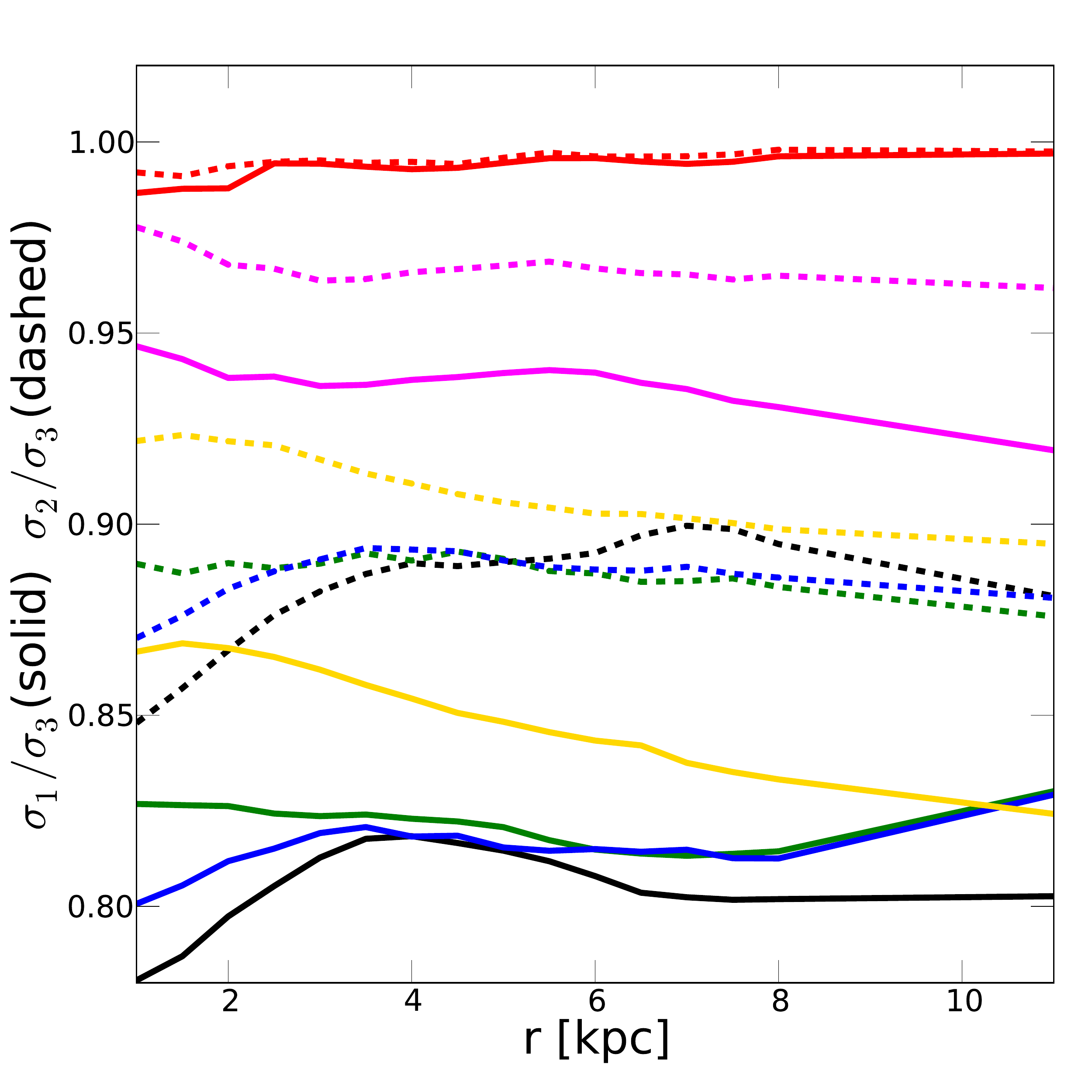}
\caption{Ratios of the components of the diagonal velocity dispersion tensor
$\sigma_1 < \sigma_2 < \sigma_3$ measured in different radial shells for the
different DM models.  We show the smallest-to-largest ($\sigma_1/\sigma_3$) and
intermediate-to-largest ($\sigma_2/\sigma_3$) ratios. Colours are as in
Fig.~\ref{fig:cross_section}. The velocity dispersion tensor in the SIDM10 case
is essentially isotropic in the inner halo with
$\sigma_1/\sigma_3\sim\sigma_2/\sigma_3\sim1$, while the other constant cross
section cases (SIDM1, SIDM0.1) show a gradual increase in the degree of
anisotropy with decreasing cross section. The vdSIDM models have
ratios very close to the CDM case and are clearly anisotropic and very
distinct from the models with a constant cross section.}
\label{fig:vel_axis_ratios}
\end{figure}

The local velocity structure is important for direct detection experiments
since the interaction rate is proportional to the $1/v$-weighted integral over
the local velocity distribution (see below). To explore the velocity structure
we measure the distribution of the local velocity modulus at different
halocentric distances ($2\kpc$, $5\kpc$, $8\kpc$). At each distance we
construct histograms in $v$ for all particles within spheres of $1\kpc$ radius
sampled at the three different radii for $1000$ randomly selected observers.
Fig.~\ref{fig:vdist} shows the resulting distributions for the different
halocentric distances. We show the median velocity distribution (thick lines)
and $10-90\%$ quantiles (shaded regions) to indicate the scatter at different
random positions at a given halocentric distance.  The left three panels of
Fig.~\ref{fig:vdist} show the actual distributions, whereas the right panels
show the median of the ratio of the distribution for the SIDM models to the
distribution of the CDM case over the sample of all observers.
Clearly, the distribution for the SIDM10 case (with the largest constant cross
section) is significantly smoother due to the large number of isotropic
scatterings through the halo history. The bumps and wiggles that are present in
the CDM case, and that reflect the assembly history of the halo
\citep{Vogelsberger2009}, are largely preserved for vdSIDMa, vdSIDMb and SIDM0.1, but
are essentially washed out for SIDM10 and significantly reduced for SIDM1.
Notice that despite the relatively low constant cross section in SIDM0.1, it
still produces a distinct velocity distribution from that of the CDM case.
For all models the largest difference with the collisionless case occurs for
the innermost radial distance where the density is higher leading to the
highest number of scattering events. For vdSIDMa, vdSIDMb and SIDM0.1, the
differences become smaller with increasing halocentric distance since
scattering events become less likely. In the former two cases the scattering
rate decreases more rapidly with radius because of the steep velocity
dependence of the cross section.  We note that also the scatter of the
distributions agrees very well between all models at large distances, which
indicates a very similar velocity structure in the halo at these halocentric
distances. Only the models SIDM and SIDM1 have non-negligible deviations for
$r\sim8\kpc$ since they still produce a relatively large scattering rate even
at these distances. The dashed black line in the left panels of
Fig.~\ref{fig:vdist} shows a Maxwell-Boltzmann distribution with $\sigma_{\rm
1D}=151.6\kms$, which fits the SIDM results well, demonstrating that
self-interactions lead in that case to a significant thermalisation of the
inner halo.  The distribution of SIDM1 (with a ten times smaller cross section)
cannot be fitted exactly by a Maxwell-Boltzmann distribution, but is still
quite close, whereas all other models clearly have very different
distributions.

\begin{figure}
\centering
\hspace{-0.5cm}\includegraphics[width=0.5\textwidth]{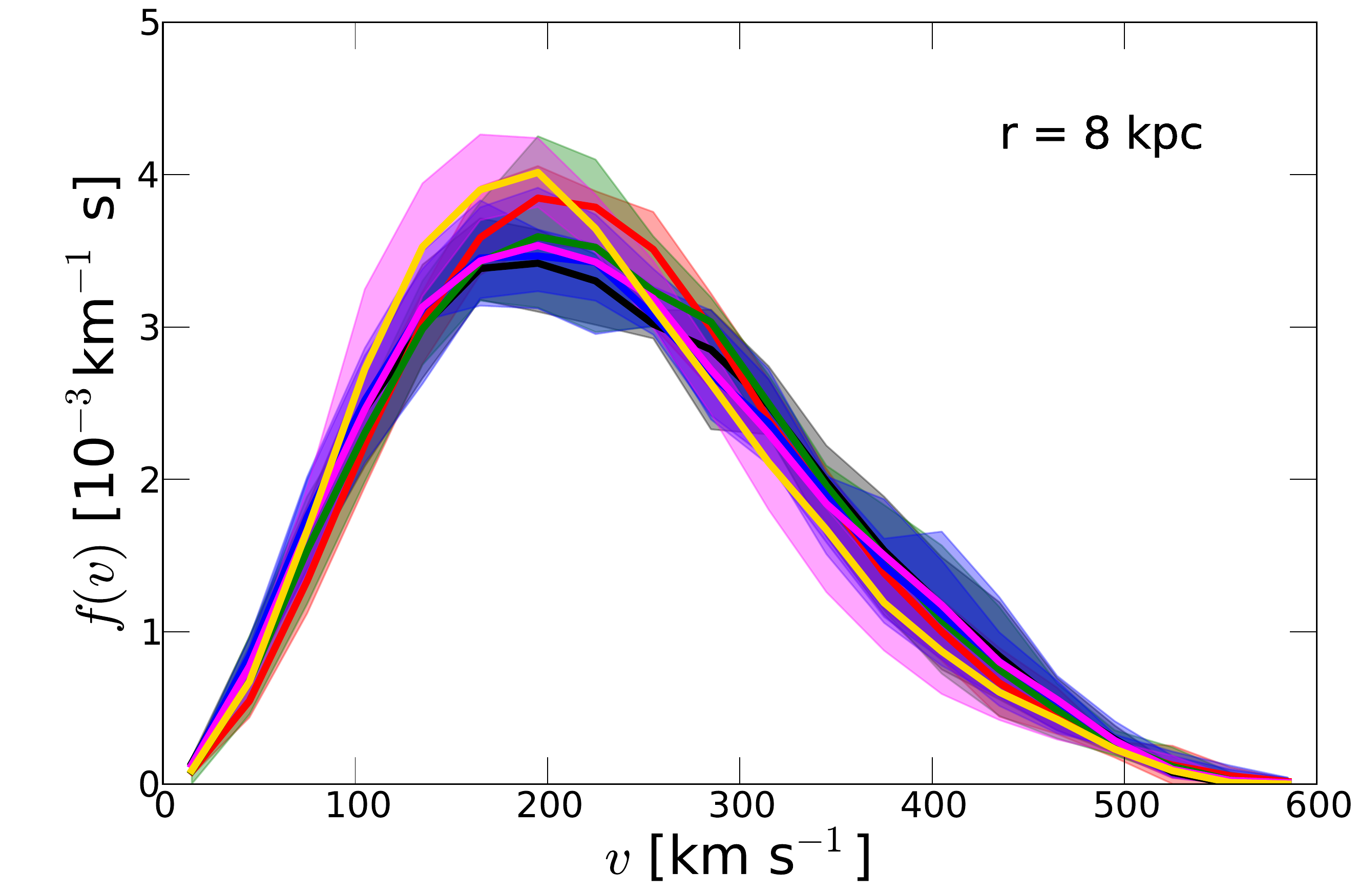}\\
\hspace{-0.5cm}\includegraphics[width=0.5\textwidth]{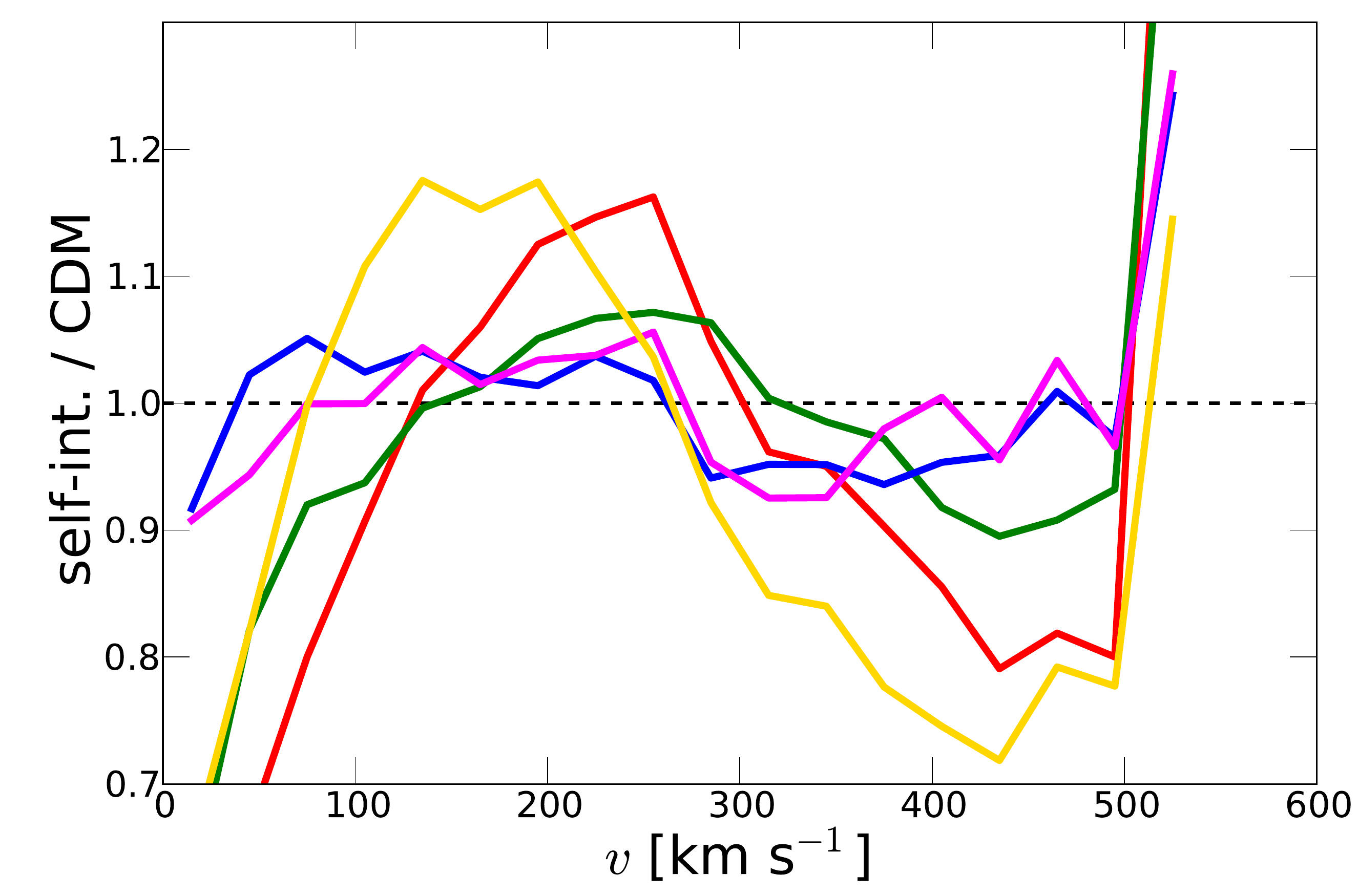}
\caption{Velocity distribution as in Fig.~\ref{fig:vdist}, but for a sample of
different MW-size haloes (Aq-A-4 to Aq-F-4).  Top panel: Distribution of the
velocity modulus measured at $8\kpc$ from the halo centre.  For each model and
halo, we first measure the velocity distribution in $1000$ randomly selected
spheres of $1\kpc$ radius at the indicated distances, and then calculate the
median over all observers for each model and halo. We then take the mean of
these median distributions over all haloes to derive the average distribution
for each model (solid lines).  The shaded regions show the range of
distributions that is spanned by the individual medians of all haloes for a
given model, i.e., this shows the halo-to-halo scatter of the medians. Bottom
panel: Ratio of the medians of the different models to the CDM case.
The two panels demonstrate that the trends found for the halo sample (Aq-A-4 to
Aq-F-4) are essentially the same as those found for the single halo in
Fig.~\ref{fig:vdist}. Thus, these findings are robust with respect to cosmic
variance and should hold for generic haloes in the mass range of our own MW
halo.}
\label{fig:vdistvar}
\end{figure}

When there are several collisions per particle, the velocity distribution tends
to be thermalised (i.e., it becomes Maxwellian). This clearly happens at small
radii for the cases of large constant cross section, as we mentioned in the
previous section. In this process, particles from the low and high velocity
tails are transferred towards intermediate velocities.  When the number of
collisions is low, however, the distribution is not fully thermalised and it is
modified only slightly in a more discrete way.  At the high-end of the velocity
distribution, the cases with a constant cross section preferentially scatter
these high velocity particles with those with lower velocities (since their
number density is higher). Since scatter happens elastically, the high velocity
particles end up with lower velocities and hence, the particle number density
in the velocity tail is lower than in the CDM case. For the vdSIDM cases on the
other hand, there is a selection effect in the scattering process for particles
that are moving roughly in the same direction and speed (in order to have low
relative velocities). Thus, particles in the high-end velocity tail collide
with others in the tail as well and this leaves the velocity distribution
essentially unchanged.

CDM haloes have a velocity structure that is anisotropic
\citep[e.g.][]{Navarro2010}. Since we are only considering isotropic
scattering, i.e., the velocities are redistributed on the unit sphere after a
scatter event, a large number of scattering events should lead to a velocity
structure that is less anisotropic than in the CDM case. The difference is more
important near the center where the scattering rate is higher. To test this we
show in Fig.~\ref{fig:vcompdist} the distribution of the largest and smallest
principal components of the velocity ellipsoid: specifically, for each of the
observer's $1\kpc$ spheres, we calculate the local velocity ellipsoid and
project the velocity along the principal axes. As in Fig.~\ref{fig:vdist}, the
left panels show the distribution at different halocentric distances, whereas
the right  panels show the deviations of each of the SIDM cases relative to the
CDM case. Most strikingly, the large number of scatter events in the SIDM10 case
produces an almost fully isotropic distribution at all radii within $\sim8\kpc$
of the main halo (which is roughly the size of the core induced by
self-scattering, see VZL).  The components of the velocity ellipsoid are
Gaussian with a nearly identical variance which leads to a nearly Maxwellian
distribution for the modulus of the velocity vector as we have shown in
Fig.~\ref{fig:vdist}. The ratios of the dispersions of the principal components
$\sigma_1 < \sigma_2 < \sigma_3$ for each case are shown in
Fig.~\ref{fig:vel_axis_ratios} as a function of radius. This plot shows more
clearly the degree of anisotropy in the velocity distributions. SIDM10 is
essentially isotropic in the inner halo with $\sigma_1/\sigma_3$ and
$\sigma_2/\sigma_3$ being close to unity, while the other constant cross
section models (SIDM1, SIDM0.1) show a gradual increase on the anisotropy;
particularly towards large radii where the scattering rate is lower. On the
other hand, the vdSIDM models are just slightly more isotropic
than CDM.

So far we have focused our analysis only on one specific halo. We will now
explore how the velocity results presented above change if we consider
different haloes.  Fig.~\ref{fig:vdistvar} shows for each DM model the
distribution of the velocity modulus for different haloes measured at $8\kpc$
halocentric distance. For each halo the measurement is done exactly the same
way as described above for Aq-A. The thick lines show the mean over all haloes
(Aq-A-4 to Aq-F-4) for each model. The shaded region encloses the distribution
of all haloes for each model. The trends found for this halo sample are
essentially the same as those found for the single halo in
Fig.~\ref{fig:vdist}. I.e., these trends are robust with respect to cosmic
variance and should hold for generic haloes in the mass range of our own Milky
Way halo. We have also checked that this is true for different halocentric
distances. Based on this, we conclude that the results found for Aq-A-3 should
hold for generic MW-sized haloes.

\begin{figure*}
\centering
\hspace{-0.5cm}\includegraphics[width=0.5\textwidth]{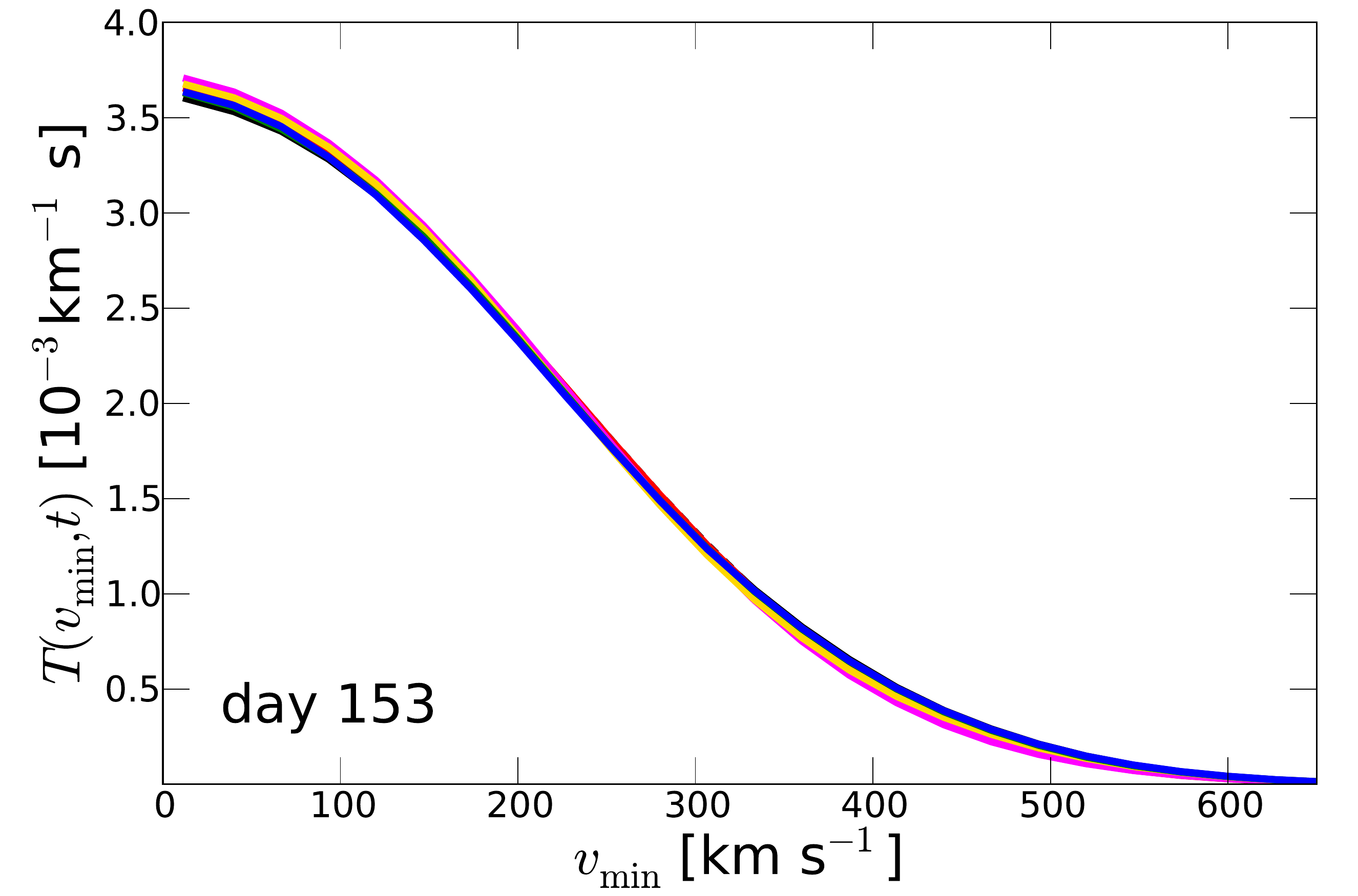}
\includegraphics[width=0.5\textwidth]{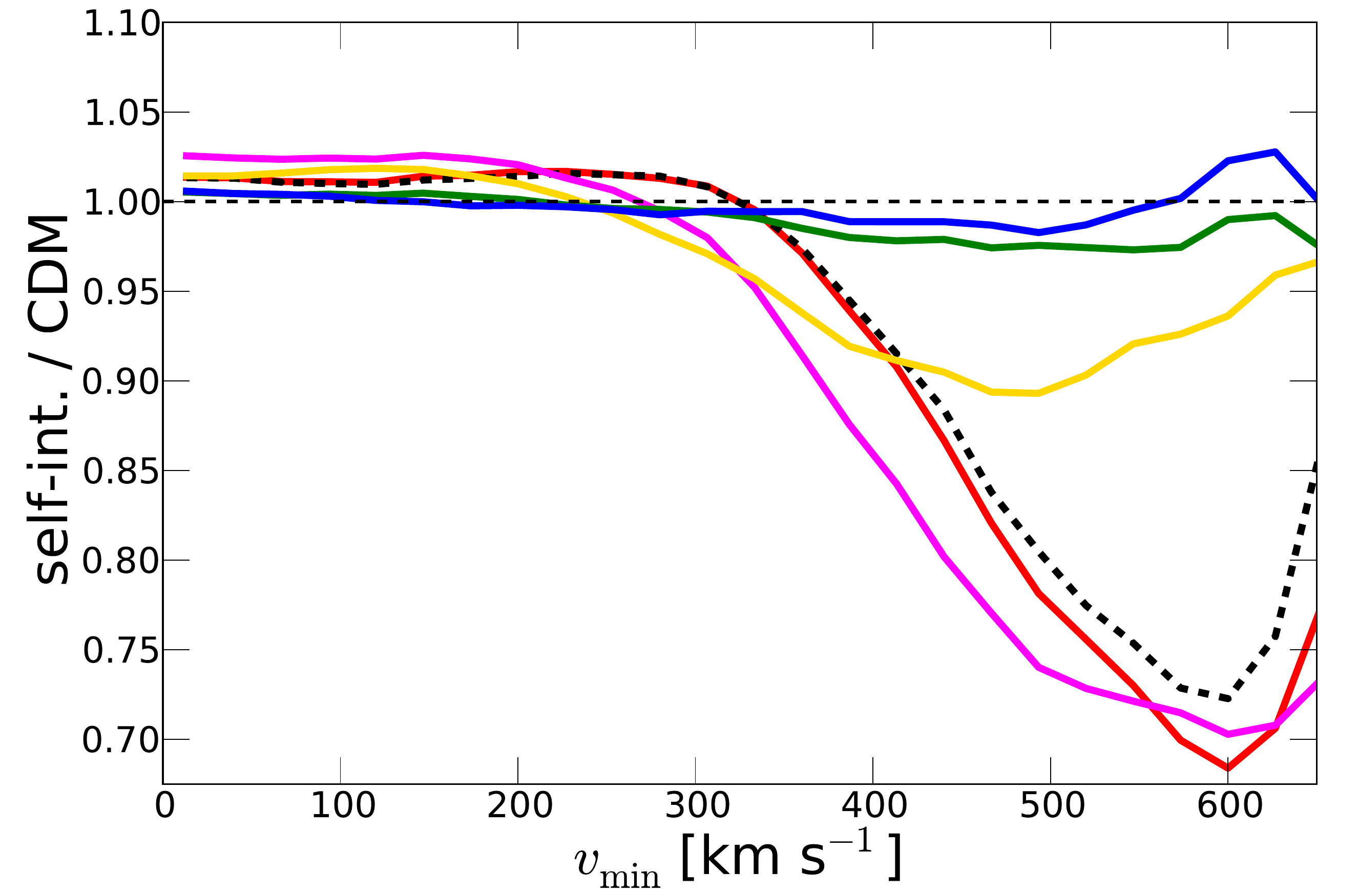}\\
\hspace{-0.5cm}\includegraphics[width=0.5\textwidth]{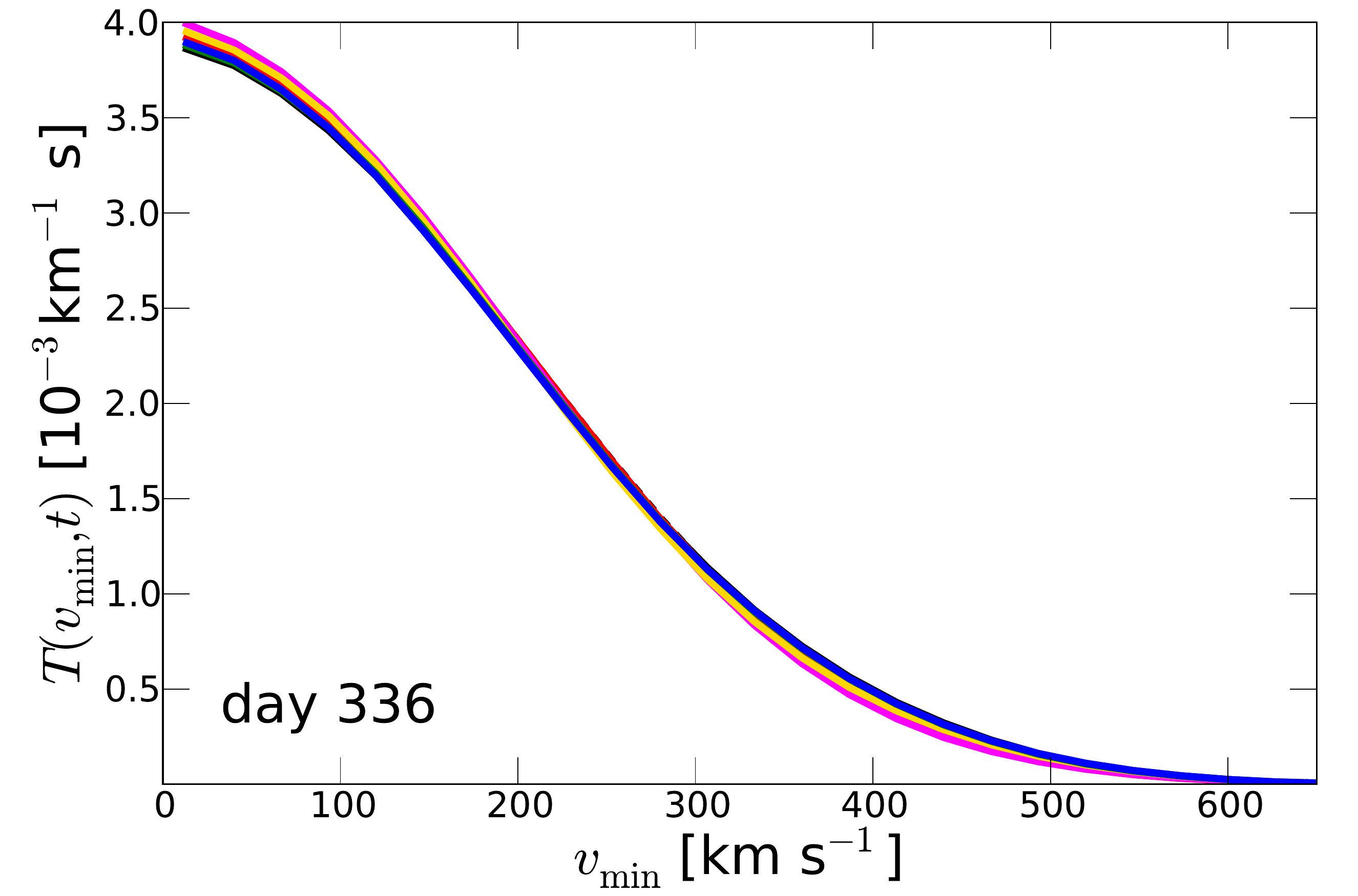}
\includegraphics[width=0.5\textwidth]{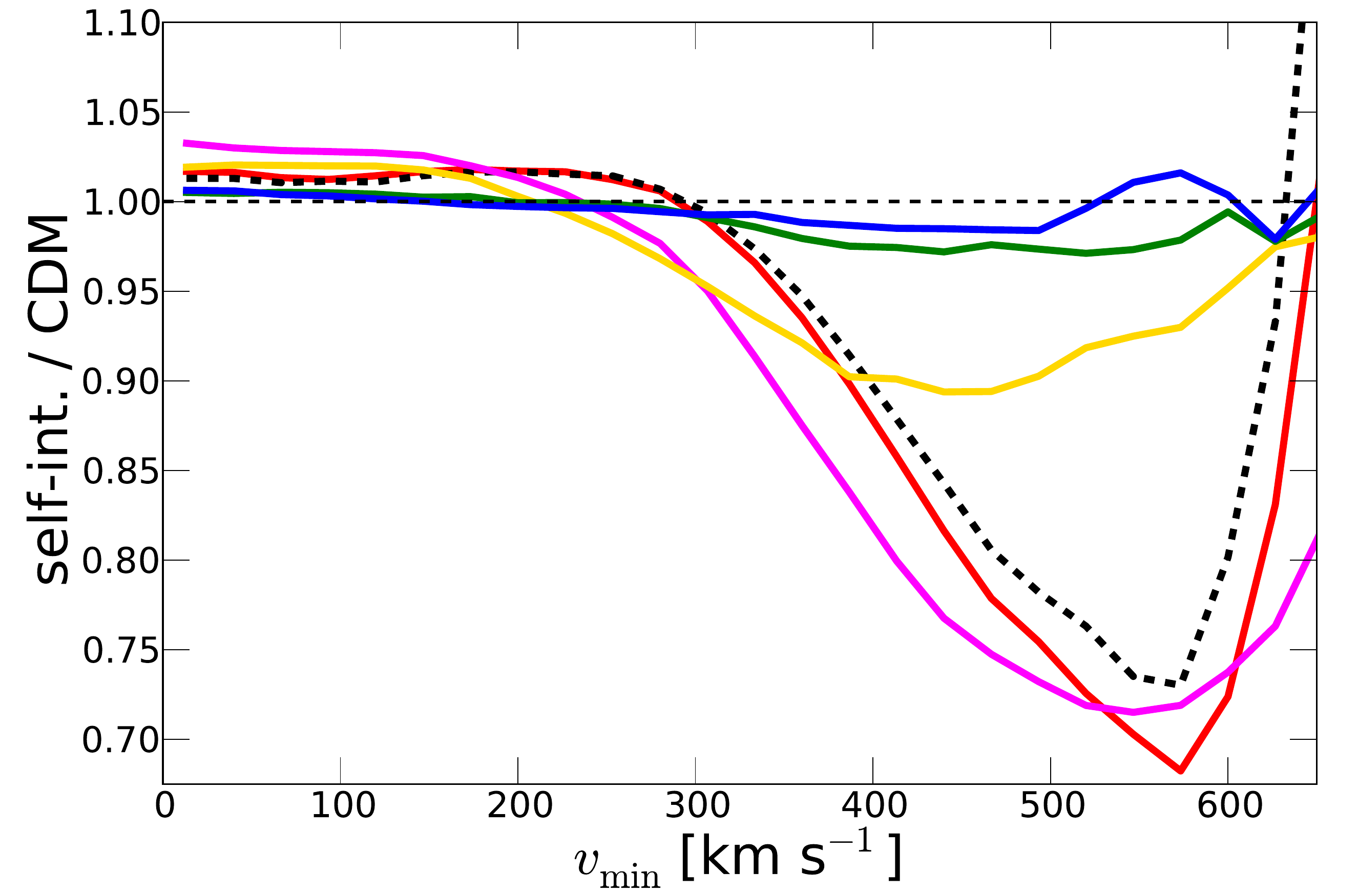}\\
\caption{Left panels: Detector recoil rates ($T(v_{\rm min},t)$) for the
different DM models.  We show the median signal for $1000$ randomly selected
observers at the solar circle ($8\kpc$ halocentric distance), where the
detector signal is sampled within spheres of $1\kpc$ radius. The selected two
days correspond to June 2nd (day $153$) and December 2nd (day $336$) of 2013. Right panels: Median of
ratios of the SIDM models relative to the local CDM case for all observers.
Colours are as in Fig.~\ref{fig:cross_section}. The low energy distribution
shows only minor changes with respect to the CDM case for all models ($\lesssim
5\%$) with the vdSIDM models showing essentially no deviation (percent level)
from the CDM case. Above $v_{\rm min}\!\sim\!  250\kms$ the constant cross
section models start to deviate significantly from CDM showing strongly
reduced recoil rates.  For SIDM0.1, the ratio is smallest around $v_{\rm
min}\!\sim\!  450\kms$, where the SIDM rate is about $10\%$ smaller than the
CDM prediction.  SIDM10 and SIDM1 both have their largest deviation from the CDM
case at about $v_{\rm min}\!\sim\! 600\kms$.  Although these two models differ
by a factor of $10$ in their cross section, they both have a recoil rate which
is approximately $30\%$ smaller than that of CDM around that velocity. Black
dashed lines in each panel show the expected detector signal for the
Maxwell-Boltzmann distribution ($\sigma_{\rm 1D}=151.6\kms$) shown in
Fig.~\ref{fig:vdist}, which fits the velocity distribution of SIDM10 best and is
quite similar to the standard halo model.} 
\label{fig:vmindist} 
\end{figure*}

\section{Direct detection signal}

The DM phase-space structure directly affects the detection rate for
experiments that are looking for DM particles signals. As demonstrated above,
significant DM self-scattering leads to a different DM velocity distribution
and we examine now the expected impact on the recoil rates. 

The DM-nucleon scattering event rate is given by \citep[e.g.][]{Jungman1996}:
\begin{equation}
R(E, t) = \mathcal{R} ~ \rho_0 ~ T(E,t) = \mathcal{R} ~ \rho_0 ~ T(v_{\rm min}(E),t),
\end{equation}
where $\mathcal{R}$ encapsulates the particle physics parameters (mass and
cross section of the DM particle; form factor and mass of the target nucleus),
$\rho_0$ is the local dark matter density, and the local velocity distribution
enters in an integrated form as:
\begin{equation}
T(v_{\rm min}(E),t) \! = \!\!\!\!\!\!\!\!\!\int\limits_{v_{\rm min}(E)}^\infty \!\!\!\!\!\!\!\!\frac{f_v(t)}{v}{\rm d} v, \,v_{\rm min}(E) \!\!=\!\! \left ( \frac{E~(m_\chi + m_N)^2}{2 m_\chi^2 m_N} \right )^{1/2}\!\!\!\!\!\!\!\!, 
\label{eq:Tint}
\end{equation}
where $f_v$ is the DM speed distribution in the rest frame of the detector
integrated over the angular distribution; $v_{\rm min}(E)$ is the minimum DM
particle speed (detector-dependent) that can cause a recoil of energy $E$. This
threshold velocity depends on the DM particle mass $m_\chi$ and nuclear mass of
the target $m_N$.  Since we are primarily interested in the impact of different
velocity distributions, we focus in the following on $T(v_{\rm min},t)$ rather
than $R(E,t)$ which is more sensitive to the details of the detector and
particle physics.

\begin{figure*}
\centering
\hspace{-0.5cm}\includegraphics[width=0.5\textwidth]{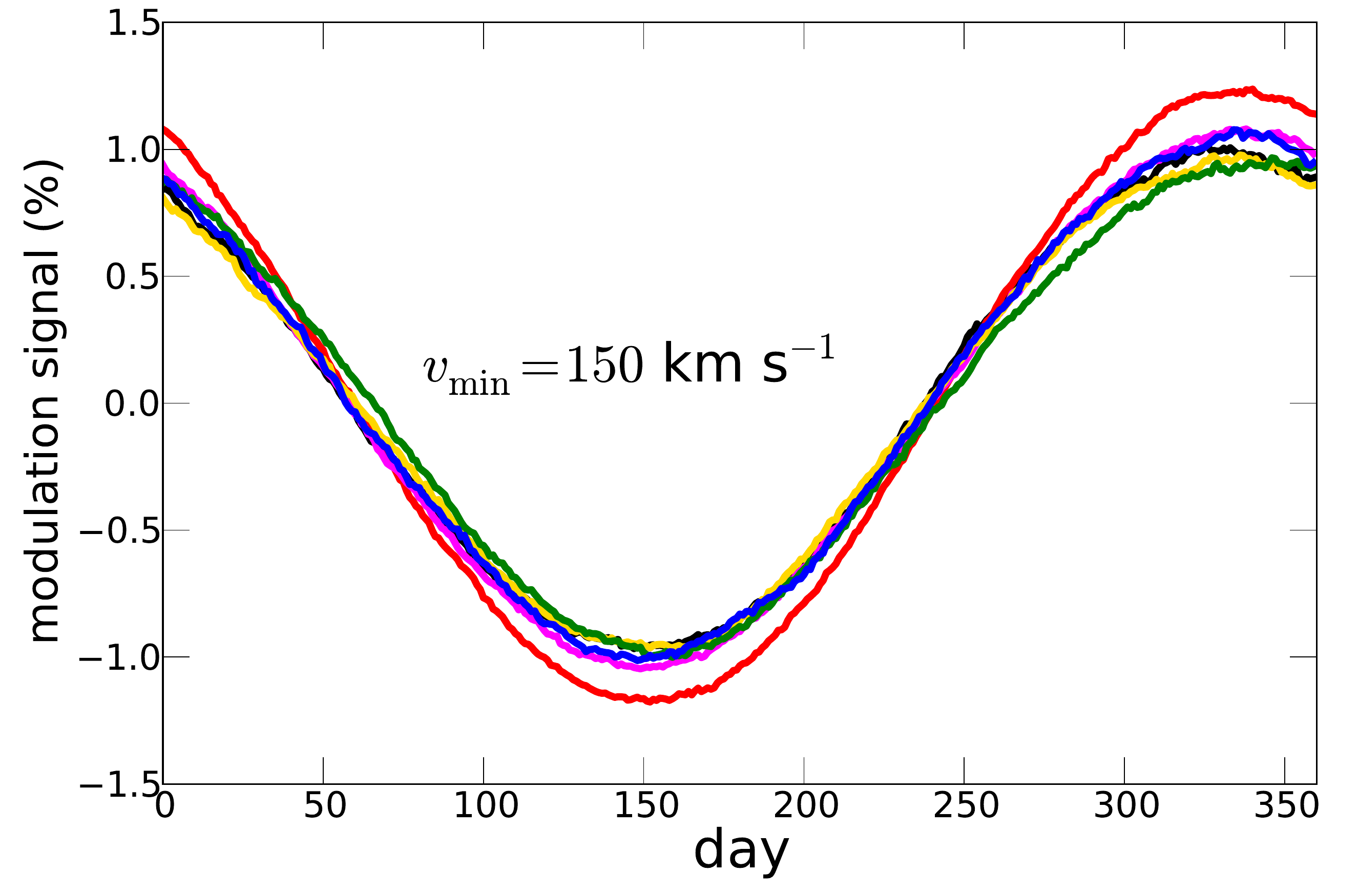}
\includegraphics[width=0.5\textwidth]{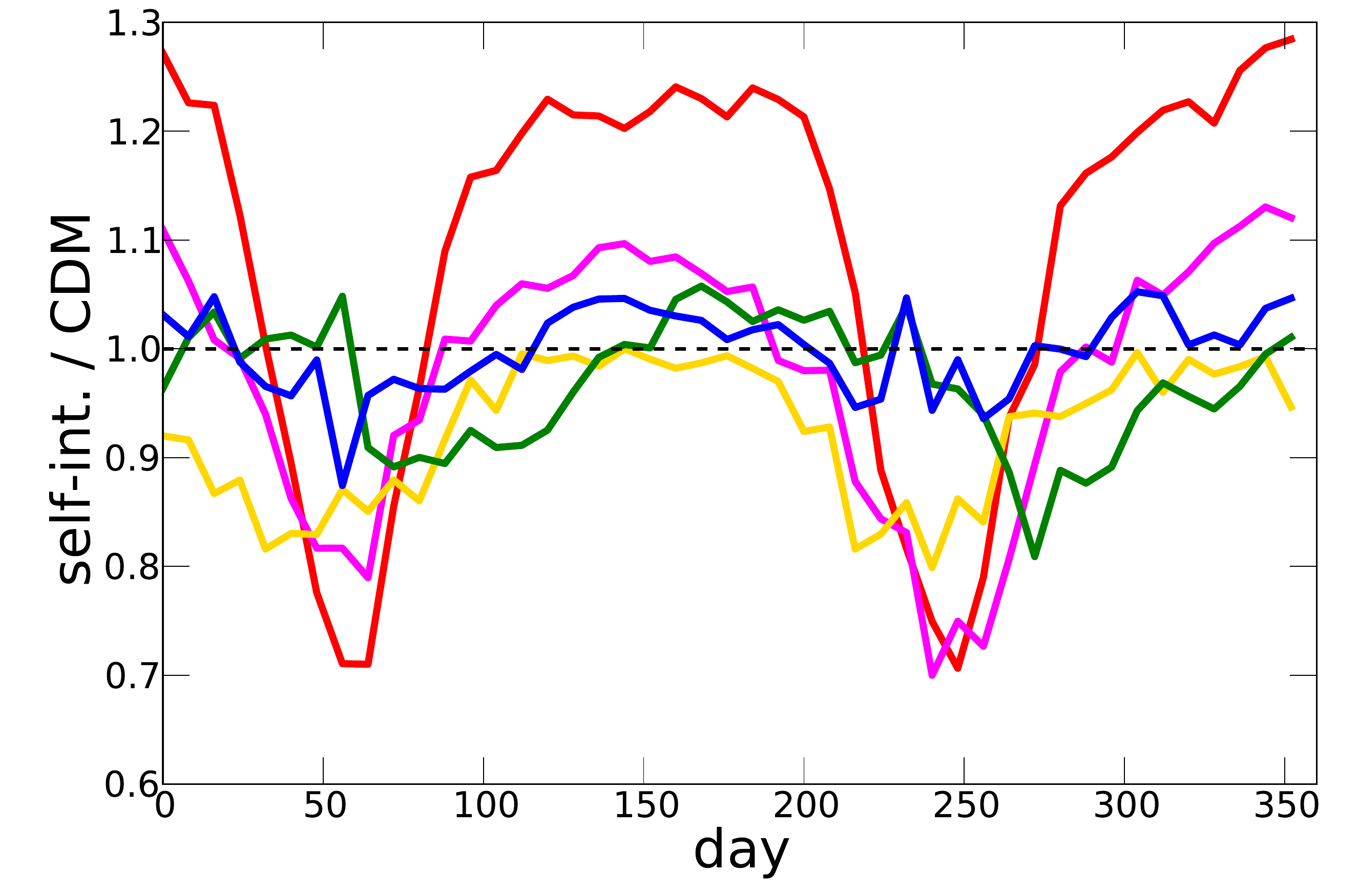}\\
\hspace{-0.5cm}\includegraphics[width=0.5\textwidth]{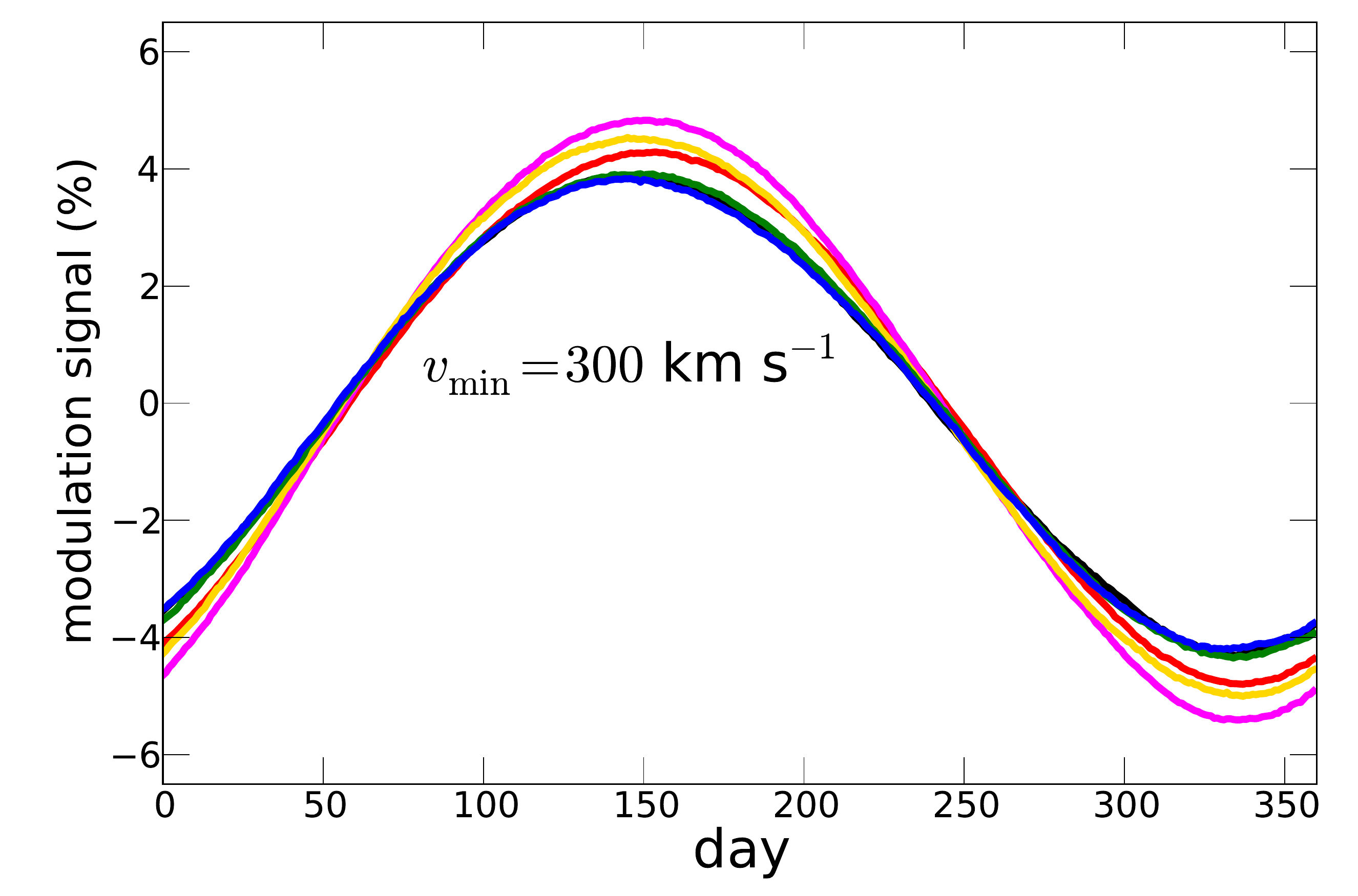}
\includegraphics[width=0.5\textwidth]{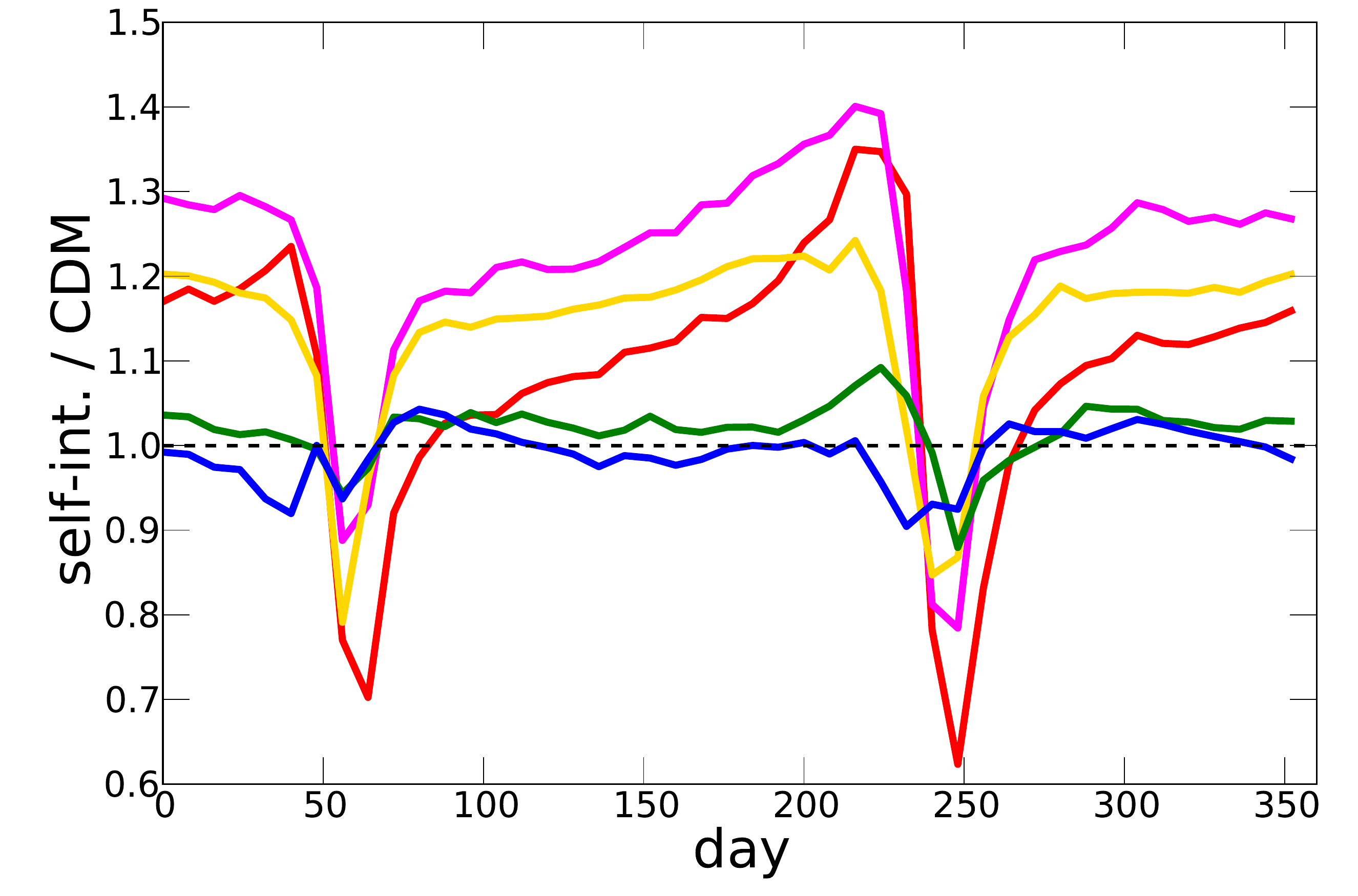}\\
\hspace{-0.5cm}\includegraphics[width=0.5\textwidth]{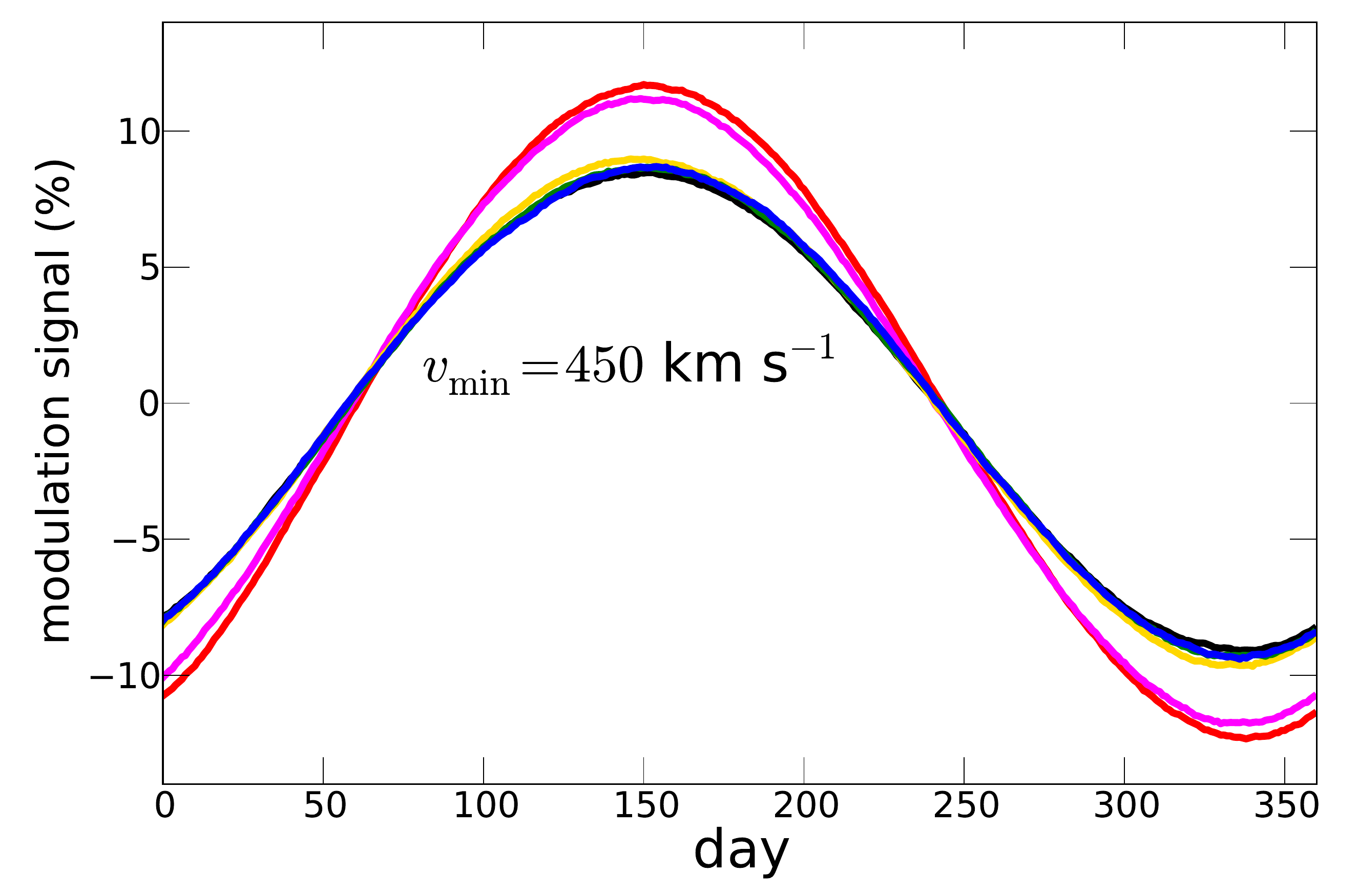}
\includegraphics[width=0.5\textwidth]{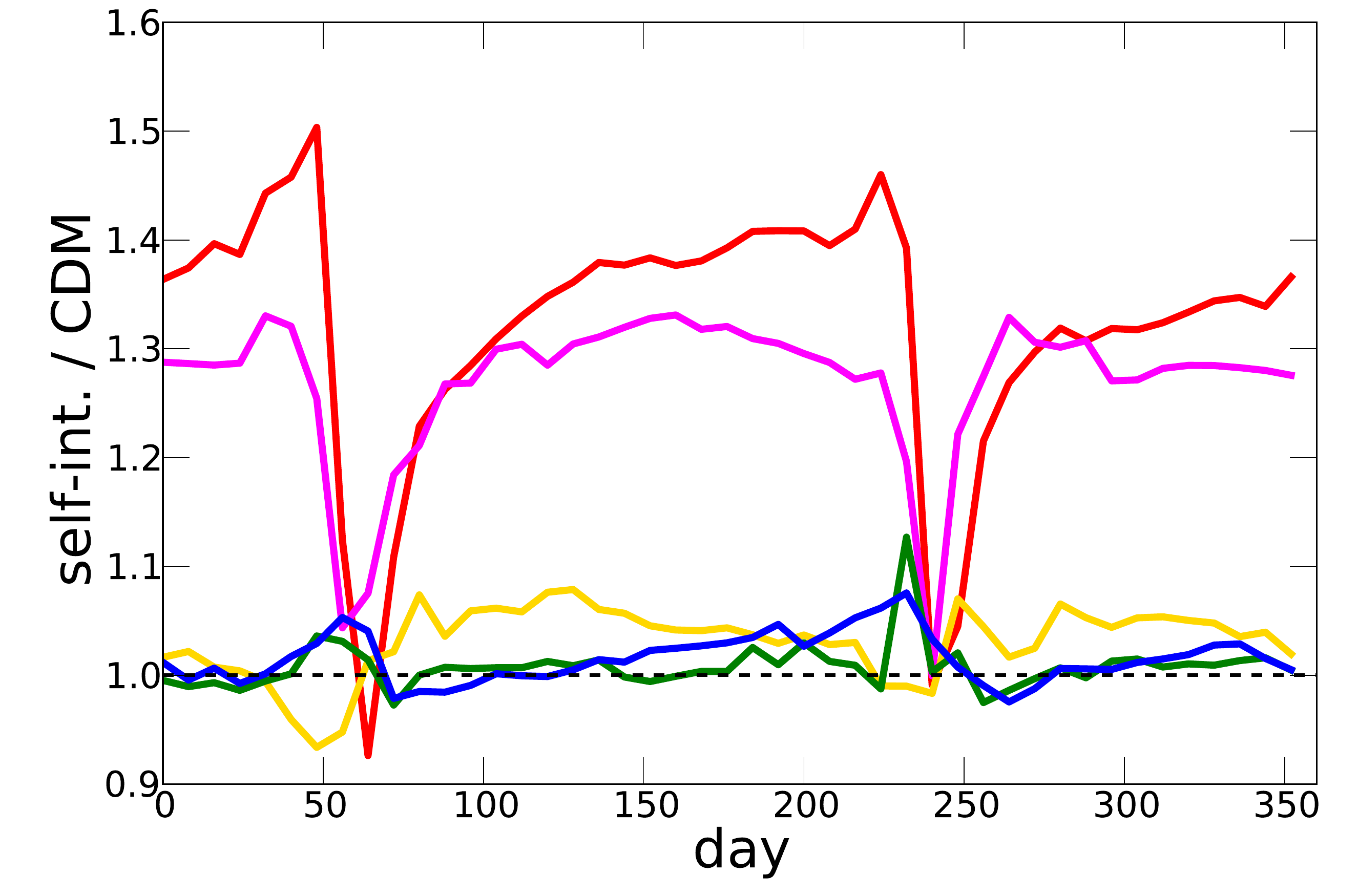}
\caption{Left panels: Median annual modulation signal of $1000$ randomly
selected spheres ($1\kpc$ radius) at $8\kpc$ halocentric distance for different
recoil energies as indicated by the corresponding $v_{\rm min}$ velocities.
Colours are as in Fig.~\ref{fig:cross_section}. Right panel: Median of the
ratio of the SIDM models over the CDM prediction. Depending on the DM model,
the modulation amplitude can differ from the CDM prediction by up to $40\%$.
Furthermore, self-interaction induces a change in the day of the peak of the
modulation such that this day varies by about two weeks for the different
models.}
\label{fig:modulation}
\end{figure*}

The velocity distribution function depends on time because of the motion of the
Earth with respect to the galactic halo. This produces an annual modulation in
the recoil rate. We parametrise the motion of the Earth following \citep{Kerr1986, Lewin1996, Dehnen1998}, and refer the reader to
\cite{Vogelsberger2009} for the detailed parametrisation. For definitiveness,
we fix the value of $t$ to two days during the course of a year when
calculating the recoil rate spectrum $T(v_{\rm min},t)$. Specifically, these
are day 153 (June 2nd) and 336 (December 2nd) of 2013 (left panels of
Fig.~\ref{fig:vmindist}). These days are close to the extrema of the modulation
signal.  The right panels of Fig.~\ref{fig:vmindist} show the ratio of the
$T(v_{\rm min},t)$ recoil rates relative to the vanilla CDM model for
the different SIDM models. We note that these rates and the ratios were
calculated at $8\kpc$ halocentric distance and, as above, for $1000$ randomly
selected spheres of $1\kpc$ radius. Black dashed lines in each panel show the
expected detector signal for the Maxwell-Boltzmann ($\sigma_{\rm
1D}=151.6\kms$) distribution shown in Fig.~\ref{fig:vdist}, which fits the
velocity distribution of SIDM10 best and is quite similar to the standard halo
model.

The relative variation between the different models does not seem to depend
strongly on the day of the year.  For all models, the low energy distribution
shows only very minor changes with respect to the CDM case. These differences
are at maximum $5\%$ for the most extreme constant cross section models (SIDM10,
SIDM1), while there is essentially no deviation (percent level) for the vdSIDM
models.  Above $v_{\rm min}\!\sim\! 250\kms$ all constant cross section models
start to deviate significantly from CDM showing lower recoil rates.  For
SIDM0.1 the ratio is smallest around $v_{\rm min}\!\sim\!  450\kms$, where the
SIDM rate is about $10\%$ smaller than the CDM prediction.  SIDM10 and SIDM1,
which have $100$ and $10$ times larger constant cross sections than SIDM0.1, both
have their largest deviation from the CDM case at about $v_{\rm min}\!\sim\!
600\kms$.  Although these two models differ by a factor of $10$ in their cross
section, they both have a recoil rate which is approximately $30\%$ smaller
than that of CDM around that velocity. Interestingly, self-interaction only
leads to a reduction of the recoil rates with respect to the CDM case. I.e.,
except for the very small increase towards lower recoil energies,  we do not
find a regime where the recoil rate is increased significantly.

\begin{table}
\begin{center}
\begin{tabular}{ccc}
\hline
Experiment         & $E_\mathrm{thres}$ [${\rm keV}$] & $v_\mathrm{min,thres}^{m_\chi = 10\,\mathrm{Gev}}$ [${\rm km}\,{\rm s}^{-1}$]\\ 
\hline
CDMS-Si            & 7   &  398                                                 \\
CRESST-O-band      & 15  &  527                                                \\
COGENT-Ge          & 1.9 &  282                                                 \\
DAMA-Na            & 6.7 &  371                                                 \\
Xenon              & 2   &  361                    \\
\hline
\end{tabular}
\caption{Typical threshold energies and corresponding $v_{\rm min}$ values for
current experiments with different detector materials (Si, O, Ge, Na, Xe).
Taken from \citet{Fox2011} (see their Fig. 1). For such a particle (and heavier
ones), most of the current experiments have $v_{\rm min}$ thresholds  below the
value where we expect to see a significant difference in the recoil rates due
to self-scattering (see Fig.~\ref{fig:vmindist}). This means that most
experiments should be sensitive to the effects of self-interacting DM.}
\end{center} 
\label{table:recoil_energies} 
\end{table}

To make connection with experimental efforts, Table~3 
lists the threshold recoil energies ($E_\mathrm{thres}$) for a few experiments
and the corresponding $v_\mathrm{min,thres}$ values for a $m_\chi =
10\,\mathrm{Gev}$ DM particle.  For such a particle (and heavier ones), most of
current experiments have $v_{\rm min}$ thresholds  below the value where we
expect to see a significant difference in the recoil rates due to
self-scattering (see Fig.~\ref{fig:vmindist}). This implies that these
experiments should actually be sensitive to the effects of self-scattering and,
in principle, they should be able to distinguish different constant and
velocity-dependent models. However, elastic scattering rates are dominated by
the lowest recoil energies, thus, 
experiments with $v_\mathrm{min}\lesssim 300\kms$, which is where SIDM effects 
start to be important,
are mostly dominated by events for which the scattering rate is indistinguishable from
CDM. 
Because of this, experiments with a larger threshold energy, in the range
where SIDM effects are maximised, 
are better suited to detect the imprints of self-interacting DM. Of the experiments listed in
Table~3 
CDMS-Si and CRESST (oxygen band) operate in this regime (this is of course
mass dependent, DM particles lighter than $m_{\chi}=10\,{\rm GeV}$ would have higher thresholds, and
vice versa). It is important to mention, however, that distinguishing different dark matter
models using a detector signal might be challenging in current experiments due to the associated 
errors in the recoil rates. For instance for CoGENT and CRESST, the relative error in the number
of events observed by the detectors near threshold is $\sim10-20\%$ \citep[see Figs. 3 and 4 of][]{Hooper2010}, 
which is of the order of the differences we expect between allowed SIDM models and CDM.

We note that the interpretation of most direct detection results assume a simple
standard halo model with a Maxwell-Boltzmann distribution to infer, for
example, limits on particle cross-sections and masses. Typically the
one-dimensional velocity dispersion is then either assumed to be $\sigma_{\rm
1D}=156\kms$  or taken directly from a fit to CDM simulations.  We find that
the most extreme SIDM model (SIDM10) establishes a Maxwell-Boltzmann
distribution quite close to the one assumed in the standard halo model: we
found a best fit with $\sigma_{\rm 1D}=151.6\kms$, just slightly lower than the
standard assumption. However, the SIDM cases of interest, those not ruled out
by other observations, show significant departures from the standard halo,
which implies that the simplified assumptions about the halo velocity structure
are not valid for self-interacting DM.

The present DAMA/LIBRA experiment and the former DAMA/NaI one have reported an
annual modulation signal from results corresponding to a total exposure of
$1.17$ ${\mathrm{ton}\times\mathrm{yr}}$ over $13$ annual cycles with a
confidence level of $8.9\sigma$ \citep[][]{Bernabei2012}.  Such a signal is
expected from DM due to the motion of the Earth through the galactic halo.
Since self-scattering changes the DM velocity distribution and, hence, the
detector signal, it is expected that the annual modulation signal will also be
sensitive to self-interactions of DM particles.  In Fig.~\ref{fig:modulation},
we show the annual modulation signal of the different DM models for three
different recoil energies given by three different minimum velocities, $v_{\rm
min}=150\kms{\rm (top)},300\kms{\rm (middle)},450\kms{\rm (bottom)}$. The left
panels show the actual modulation signal, whereas the right panels show the
median ratios of the non-CDM models to the CDM case.  To avoid too much
noise in this ratio, we show it only for every 4th day, whereas the actual
modulation signal is sampled for each day.  

The qualitative difference  between the $v_{\rm min}=150\kms$ modulation signal
and the other two values of $v_{\rm min}$ is due to the phase reversal effect
\citep[e.g.][]{Primack1988,Lewis2004}, which is consistently observed for all
DM models.  This phase reversal occurs below a critical recoil energy, which is
a function of the mass of the DM particle, as well as the nuclear mass of the
target. We find that it also depends on how collisionless is DM.  For
definiteness, we take the time of reversal as the first time when the day of
maximum amplitude is not close to day $\sim 300$, but rather jumps to about
$>100$ days earlier.  With this definition we find for the time of phase
reversal (CDM, SIDM10, SIDM1, SIDM0.1, vdSIDMa, vdSIDMb): 177, 185, 177, 172,
176, 180, i.e., SIDM0.1 changes first around day $172$ and SIDM10 changes phase
about $13$ days later.  The time of phase reversal of the the different DM
models therefore spans about two weeks, which provides a good probe to
distinguish between different (non-) self-interacting DM candidates.  

The modulation signal clearly depends on the particular DM model. As expected,
based on the results presented in the previous Sections, SIDM10 and SIDM1 lead
to the largest deviations from the CDM modulation signal.  
For all recoil energies, these modulation signals deviate by about $30\%$ from the CDM case.
(we note that the
best fit amplitude in the modulation signal reported by DAMA/NaI and DAMA/LIBRA experiments
has a relative error of $\sim10\%$, \citealt{Bernabei2012}).
For $v_{\rm min}=300\kms$, not only does the amplitude changes, but also a
clear shift of the peak is visible for the constant cross section models. This
shift is largest for SIDM10, but still clearly present for SIDM1 and to some
degree also for SIDM0.1.  The days with the indicated amplitudes for CDM,
SIDM10, SIDM1, SIDM0.1, vdSIDMa, vdSIDMb are: 141 ($3.8\%$), 156 ($4.3\%$), 142
($4.8\%$), 151 ($4.6\%$), 146 ($3.9\%$), 137 ($3.8\%$).  The vdSIDM models lead
only to variations of the order of $10\%$ in the modulation amplitude.  For the
recoil energy corresponding to $v_{\rm min}=450\kms$, the vdSIDM models
and SIDM0.1 lead only to very small differences ($<10\%$) from the CDM case. In
most of the SIDM models, the modulation amplitude is increased compared to the
CDM case for most of the days during the year, especially for $v_{\rm
min}=300\kms$, where all models show a rapid decrease in the otherwise higher
modulation signal. The days when this happens correspond to the days in the
left panels where the different modulation signals cross the CDM signal. 

\section{Summary and Conclusions}

Definitive proof for the existence of DM as a new particle beyond the Standard
Model of particle physics is expected to come from laboratories on Earth that
are looking for the collision of dark matter with nuclei in target detectors.
The hypothetical interaction rate depends on the intrinsic properties of the
dark matter (DM) particle and on the phase-space DM distribution at the scale
of the detector.  Thus, interpreting a positive signal or  
deriving a constraint on the DM properties based on a null result,
can only be done reliably through a detailed
knowledge of its local phase space distribution. The most powerful approach to
accomplish this is through the use of numerical simulations that follow the
formation and evolution of DM structures through cosmic time. These simulations
have already been used to report significant differences over the traditional
assumptions of a smooth density distribution and a Maxwellian velocity
distribution \citep{Vogelsberger2009,Kuhlen2010}.  To date, all predictions for
direct detection experiments from numerical simulations have been obtained
within the context of the Cold Dark Matter (CDM) paradigm. It is possible,
however, that DM has a non-negligible self-scattering cross section, large enough to
have a significant impact at galactic scales and yet, sufficiently small to
satisfy current astrophysical constraints
\citep{Vogelsberger2012,Rocha2012,Peter2012}. The interest in self-interacting
DM (SIDM) models has been renewed recently, since they are able to successfully
reduce the central densities of dark matter (sub)haloes and create dark matter
cores that are seemingly consistent with observations of dwarf galaxies
\citep{Boylan2011a, Vogelsberger2012, Rocha2012}.

In this paper we present the first study on the impact of DM self-scattering on
the velocity distribution of dark matter haloes and on the anticipated direct
detection signals. We imagine a particle physics scenario where dark matter
strongly interacts with itself but interacts with nuclei at the level probed by
current DM detectors (a possible scenario would be along the lines of the models
proposed in \cite{Fornengo2011} and \cite{Hooper2012}, see also \cite{Cyr-Racine2012}
for a different plausible model).  Although it is
expected that DM collisions will tend to modify the collisionless velocity
distribution towards a Maxwellian one, we quantify this in detail by
re-simulating the MW-size Aquarius haloes \citep{Springel2008} within the
context of SIDM using the algorithm described in \citet{Vogelsberger2012}
(VZL). We study five different SIDM models, three with a constant cross
section: SIDM10 ($\sigma_T/m_\chi=10\,{\rm cm}^2\,{\rm g}^{-1}$, ruled-out by observations),
SIDM1 ($\sigma_T/m_\chi=1\,{\rm cm}^2\,{\rm g}^{-1}$, likely ruled out) and SIDM0.1 
($\sigma_T/m_\chi=0.1\,{\rm cm}^2\,{\rm g}^{-1}$, consistent with observations), and two with a
velocity-dependent cross section (allowed by current observations and studied
in VZL within the model of \cite{LoebWeiner2011}). It was demonstrated in
VZL that the latter two models can also alleviate the tension between the MW's
dSphs and theoretical predictions.

We find that all SIDM models show a significant departure from the velocity
distribution of the CDM model in the center of the MW halo (within $\sim
2\kpc$), while at the solar circle the scattering effect is less significant
(see Fig.~\ref{fig:vdist}).  Only the cases with constant cross section still
show a significant difference relative to the CDM prediction.  Particularly,
those with the largest cross section ($\sigma_T/m_\chi>1\,{\rm cm}^2\,{\rm g}^{-1}$) result in
a velocity distribution that is essentially Maxwellian at the solar circle.  In
these cases, the distribution within $\sim 10\kpc$ can be described very well
by a Maxwell-Boltzmann distribution function with a one-dimensional velocity
dispersion of $\sigma_{\rm 1D}\sim152\kms$, which is quite close to the value
of $156\kms$ assumed in the standard halo model.  On the other hand, although
the case with a self-scattering cross section of $\sigma_T/m_\chi=0.1\,{\rm cm}^2\,{\rm
g}^{-1}$ also shows distinct deviations from the CDM distribution
($\lesssim20\%$), its distribution is clearly non-Maxwellian.  The
velocity-dependent SIDM cases show even smaller departures from CDM ($<5\%$).
We note that the latter three models (vdSIDMa, vdSIDMb, SIDM0.1) are all consistent
with current observational constraints.  In addition to thermalising the
velocity distribution, DM scatterings also isotropized the orbits, which we
clearly observe by looking at the principal components of the velocity
dispersion tensor (see Figs.~\ref{fig:vcompdist} and
\ref{fig:vel_axis_ratios}). However, only the constant cross section models
show considerable departures from an anisotropic distributions.  Models
with a velocity dependent cross section are very close to CDM in the
inner halo.

The differences in the velocity distribution impact the predicted recoil rates
in direct detection experiments. This depends on the recoil energy,
corresponding to a minimum velocity $v_{\rm min}$ of the DM particles necessary
to impart a measurable recoil in a given detector. We find that the recoil
rates begin to deviate significantly from the CDM predictions for $v_{\rm
min}>250\kms$, being lower by as much as $30\%$ for the largest constant cross
section cases, and only at the percent level for the vdSIDM cases (see
Fig.~\ref{fig:vmindist}). These changes occur in a recoil energy range
currently accessible by essentially all direct detection experiments.  Models
that are fully consistent with current observational constraints can lead to a
deviation of about $10\%$. 

We also find differences between the SIDM models and CDM in the amplitude and
peak days of the annual modulation signal (see Fig.~\ref{fig:modulation}). The
former is larger by as much as $40\%$ ($25\%$ for allowed models) while the
latter is shifted by two weeks at the most. Interestingly, we find a
significant shift in the day when the well-known phase reversal effect of the
modulation signal \citep[e.g.][]{Primack1988,Lewis2004} occurs. The time of
phase reversal of the the different DM models spans about two weeks, which
therefore provides a sensitive probe to distinguish between different (non-)
self-interacting DM candidates. Even the cases where the scattering cross
section is velocity-dependent show a distinct phase reversal shift by a couple of
days.

We conclude that direct detection experiments should in principle be sensitive
to currently allowed SIDM models. In general models with velocity dependent
cross sections peaking at the typical velocities of dwarf galaxies lead only to
minor changes in the detection signals, whereas allowed constant cross section
models lead to significant changes. 

\section*{Acknowledgements}

We thank Volker Springel for giving us access to {\sm GADGET-3},  Philippe Di
Stefano and Naoki Yoshida for initial discussions that inspired this work,
Adrian Jenkins for generating the initial conditions of Aq-F-4, Adrienne
Erickcek, Lars Hernquist, Michael Kuhlen, Avi Loeb, Josef Pradler, Kris
Sigurdson, Paul Torrey and Simon White for useful suggestions.  JZ is supported by the
University of Waterloo and the Perimeter Institute for Theoretical Physics.
Research at Perimeter Institute is supported by the Government of Canada
through Industry Canada and by the Province of Ontario through the Ministry of
Research \& Innovation.  JZ acknowledges financial support by a CITA National
Fellowship.  MV acknowledges support from NASA through Hubble Fellowship grant
HST-HF-51317.01. 

\bibliography{paper}

\label{lastpage}

\end{document}